\documentclass{aa}  

\usepackage{adjustbox}
\usepackage{threeparttable}
\usepackage{graphicx,url,twoopt,natbib}
\usepackage[varg]{txfonts}           
\usepackage{hyperref}   
\usepackage{longtable}
\usepackage{pifont}

\hypersetup{
  colorlinks=true,   
  urlcolor=blue,     
  linkcolor=blue,     
  citecolor=blue}

\usepackage{txfonts}
\usepackage{subcaption}

\newcommand{\xco}{\mbox{$X_{\rm CO}$}}
\newcommand{\irassesenta}{\mbox{IRAS\,60$\mu$m}}
\newcommand{\kms}{\mbox{km\,s$^{-1}$}}
\newcommand{\mloss}{\mbox{$\dot{M}$}}
\newcommand{\ppuls}{\mbox{$P$}} 
\newcommand{\rfuvnuv}{\mbox{$R_{FUV/NUV}$}}
\newcommand{\ls}{\mbox{$L_{\odot}$}}
\newcommand{\msun}{\mbox{$M_{\odot}$}}
\newcommand{\my}{\mbox{$M_{\odot}$~yr$^{-1}$}}

\newcommand{\kb}{\mbox{$k_{\rm B}$}}

\newcommand{\ta}{\mbox{$T^*_{\rm A}$}}  
\newcommand{\tmb}{\mbox{$T_{\rm MB}$}}
\newcommand{\teff}{\mbox{$T_{\rm eff}$}}
\newcommand{\eff}{\mbox{$\eta_{\rm eff}$}} 
\newcommand{\iram}{\mbox{IRAM-30\,m}}
\newcommand{\farc}{\mbox{$.\!\!^{\prime\prime}$}}

\newcommand{\lbol}{\mbox{$L_{\rm bol}$}}
\newcommand{\vexp}{\mbox{$V_{\rm exp}$}}

\newcommand{\tex}{\mbox{$T_{\rm ex}$}}
\newcommand{\tkin}{\mbox{$T_{\rm kin}$}}

\newcommand{\rsou}{\mbox{$\theta_{\rm s}$}}  
\newcommand{\rco}{\mbox{$R_{\rm CO}$}}  
\newcommand{\rs}{\mbox{$R_{\rm s}$}}  
\newcommand{\fsixty}{\mbox{F$_{\rm 60}$}}
\newcommand{\micron}{\mbox{$\mu$m}}

\newcommand{\doceuno}{$^{12}$CO\,($J$=1--0)}
\newcommand{\docedos}{$^{12}$CO\,($J$=2--1)}
\newcommand{\docetres}{$^{12}$CO\,($J$=3--2)}
\newcommand{\docecuatro}{$^{12}$CO\,($J$=4--3)}

\def\snu#1{\ifmmode {S_\nu\,\propto\,\nu^{#1}}
          \else \hbox{$S_\nu$\,$\propto$\,$\nu^{#1}$}\fi}
\def\cm#1{\ifmmode {\,{\rm cm^{-#1}}}                  
        \else \hbox{$\,${\rm cm$^{\rm -#1}$}}\fi}
\def\raw {\ifmmode\rightarrow\else$\rightarrow$\fi}
\def\ex#1{\ifmmode {\times 10^{#1}}         
        \else \hbox{{$\times 10^{\rm #1}$}}\fi}

\begin{document}

   \title{CO emission survey of asymptotic giant branch stars with ultraviolet excesses}

   \author{J. Alonso-Hernández
          \inst{1}
          \and
          C. Sánchez Contreras \inst{1} \and R. Sahai \inst{2}  }

   \institute{Centro de Astrobiologia (CAB), CSIC-INTA. Postal address: ESAC, Camino
Bajo del Castillo s/n, 28692, Villanueva de la Ca\~nada, Madrid, Spain\\
              \email{jalonso@cab.inta-csic.es}
         \and
             Jet Propulsion Laboratory, California Institute of
Technology, Pasadena, CA 91109, USA\\
             }


 
  \abstract
   {The transition from the spherically symmetric envelopes around asymptotic giant branch (AGB) stars to the asymmetric morphologies observed in planetary nebulae is still not well understood, and the shaping mechanisms are a subject of debate. Even though binarity is widely accepted as a promising option, it is limited by the complication of identifying binary AGB stars observationally. Recently, the presence of ultraviolet excesses in AGB stars has been suggested as a potential indicator of binarity.}
   {Our main goals are  to characterise  the properties of the circumstellar envelopes (CSEs) around candidate AGB binary stars, specifically those selected based on their UV excess emission, and to compare these properties with those derived from previous CO-based studies of AGB stars.}
   {We  observed the \doceuno\ and \docedos\  millimetre-wavelength emission in a sample of 29 AGB binary candidates with the \iram\ antenna. We measured the systemic velocities and the terminal expansion velocities from their line profiles. Population diagrams were used to interpret the results, enabling the estimation of excitation temperatures (\tex), mass-loss rates (\mloss), and the characteristic sizes of the envelope layers where the CO millimetre emission originates (\rs). We explored different trends between the envelope parameters deduced, multiwavelength flux measurements, and other properties of our sample, and compared them with those previously  derived from larger samples of AGB stars found in the literature. } 
  {We detected $^{12}$CO emission in 15 sources, of which 5 are first detections. We found relatively low expansion velocities (3\,\kms\,$\lesssim$\,\vexp\,$\lesssim$\,20\,\kms) in our sample. We derived the average excitation temperature and column density of the CO-emitting layers, which we used to estimate self-consistently the average mass-loss rate ($10^{-8}$\,\my\,$\lesssim$\,\mloss\,$\lesssim$\,$10^{-5}$\,\my) and the CO photodissociation radius (5 \ex{15} cm\,$\lesssim$\,$\rco$\,$\lesssim$ 2 \ex{17} cm) of our targets. We find a correlation between CO intensity and \irassesenta\, fluxes, revealing a CO-to-\irassesenta\ ratio lower than for AGB stars and closer to that found for pre-planetary nebulae (pPNe). An anti-correlation is observed between $^{12}$CO (and \irassesenta) and the near-ultraviolet (NUV), but no such correlation is observed with the far-ultraviolet (FUV). It is also worth noting that there is no correlation between bolometric luminosity and NUV or FUV.}
   {For the first time we have studied the mass-loss properties of UV-excess AGB binary candidates and estimated their main CSE parameters. Our sample of uvAGB stars shows similarities with the broader category of AGB stars, except for a distinct CO-to-\irassesenta\ trend suggesting enhanced CO photodissociation.  Our findings, based on single-dish low-$J$ CO line emission observations, support the dust-driven wind scenario and indicate that alternative mass-loss mechanisms are not necessary (in principle) to explain the $\sim$200-2000\,yr old mass-loss ejecta in uvAGBs. The different relationships between $^{12}$CO and \irassesenta, with NUV and FUV are consistent with an intrinsic origin of NUV emission, but potential dominance of an extrinsic process (e.g. presence of a binary companion) in FUV emission.}


   \maketitle
%

\section{Introduction}
\label{intro}

Most stars (i.e. with masses in the 1-8 \msun \, range) evolve from the main sequence to the Asymptotic Giant Branch (AGB) phase and develop dusty winds producing spherically symmetric CSEs expanding at low velocities (\vexp $\sim$ 5-15 \kms) \citep[see][]{ Castro-Carrizo_2010, Hofner_2018}.
As AGB stars transition into the pre-planetary nebula (PPN) stage, they undergo a remarkable transformation, exhibiting a wide range of non-spherical shapes such as elliptical, bipolar, or multipolar structures \citep[e.g.][]{Sahai_2007,Balick_2002}. Additionally, during this transitional phase, these stars display high-velocity outflows with expansion velocities typically exceeding 50-100\,\kms\ \citep{Bujarrabal_2001,sanchez-contreras_2012}.

In the past two decades, binarity has emerged as a widely accepted and decisive factor in shaping PPNe and planetary nebulae (PNe) \citep[e.g.][]{De_Marco_2009}. The prevalence of binarity in AGB stars can be attributed to its common occurrence among main-sequence stars, as established by \cite{Duquennoy_1991}.
In addition, recent observations have provided strong evidence for the presence of a close binary companion for the central star in many PNe \citep{Hillwig_2018}, reinforcing the idea that binarity should be common in the AGB phase. Nevertheless, identifying binarity is not possible for most AGB stars by classical methods (i.e. transits or radial velocities) due to their high luminosity and the intrinsic variability produced by their pulsations. Recent observational studies have identified asymmetries in AGB CSEs, including enhanced density equatorial structures and large-scale arcs or spiral patterns, that are consistent with theoretical predictions from binary models \citep[see e.g.][]{Soker_1994, Kim_2012, Decin_2020}.


\cite{Sahai_2008} were the first to hypothesise that UV emission excesses in AGB stars (i.e. UV emission considerably higher than the expected photospheric emission of single AGB stars) could provide an indication of binarity due to the fact that these excesses would be produced by photospheric and/or coronal emission from a hotter companion (\teff$\gtrsim$ 5500-6000\,K and/or accretion onto the companion. 
They identified UV emission as a common feature in a small sample of AGB stars in the Hipparcos catalogue with indications of binarity using the [GALEX] the Galaxy Evolution Explorer \citep[GALEX,][]{GALEX}. 
In their sample of 21 sources, 19 were detected with near-ultraviolet (NUV) excesses (hereafter nuvAGB stars); 9 of these 19 were detected with far-ultraviolet (FUV) excesses (hereafter fuvAGB stars) as well. In addition, \cite{Ortiz_2019} showed that these excesses correspond to continuum flux instead of emission lines. Hereafter we  refer to AGB stars detected in either the  NUV or FUV (or both) as uvAGBs. 

\cite{Sahai_2011, Sahai_2018} studied in deeper detail the case of Y Gem, an fuvAGB whose extreme UV flux, significant time variability and infall--outflow signatures seen in UV line spectra, confirmed the presence of an accretion disk around a companion star. 

An increasing number of fuvAGBs have been also observed to exhibit X-ray emission, which in principle could potentially be attributed to coronal activity or accretion processes involving a companion \citep{Ramstedt_2012}. The idea of accretion as a plausible mechanism for generating X-rays has gained support from recent investigations \citep[e.g.][]{Sahai_2015,Ortiz_2021}. 
 
 Previous studies have provided valuable insights into the nature of UV emission in AGB stars. \cite{Sahai_2016} suggested a possible connection between FUV and X-ray variability and accretion processes. Additionally, \cite{Ortiz_2016} supported the hypothesis of binarity as a primary source of UV emission, while not ruling out chromospheric activity. In contrast, \cite{Montez_2017}
 proposed that NUV emission in AGB stars could be attributed to chromospheric and/or photospheric activity rather than binarity, and the lack of detectability might be due to absorption by the interstellar medium (ISM) or the circumstellar medium (CSM). More recently, \cite{Sahai_2022}, based on modelling studies, concluded that AGB stars with \rfuvnuv\,$\gtrsim$\,0.06 are likely associated with accretion-dominated UV emission, while those with \rfuvnuv\,$\lesssim$\,0.06 may have UV emission mainly originating from chromospheric and/or low-level accretion processes.

Therefore, uvAGBs are ideal binary candidates and excellent sources to search for emerging asymmetries or other companion-induced changes in the properties of their envelopes. Moreover, uvAGBs may allow us to identify and characterise AGB companions, and thus    test the  binary evolution models.

In this paper we present the results from a study of the CSEs of a sample of uvAGB binary candidates based on single-dish observations of carbon monoxide (CO) at  millimetre-wavelengths. We compare the main properties of our sample of uvAGB stars with those derived from similar studies of larger samples of AGB stars.  
In Sect.\,\ref{sample} we introduce and describe our sample. The observations and main observational results are given in Sect.\,\ref{obs} and \ref{resul}, respectively. The analysis of the data, which includes population diagram analysis to derive the mass-loss rate and the mean excitation temperature of the envelopes around our sample of uvAGBs, as well as the search for correlations between different envelope properties, is presented in Sect.\,\ref{anal}. In Sect.\,\ref{trends} we make a comparison between stellar and envelope estimated parameters in order to identify correlations in this sample or deviations from well-known trends previously studied in common AGB stars. Our results are interpreted and discussed in Sect.\,\ref{dis} and a summary of our main conclusions is provided in Sect.\,\ref{summ}.

\section{The sample}\label{sample}

The sample of 29 uvAGB stars observed in this study are listed in Table~\ref{tab:astronomy} together with their equatorial coordinates and Gaia DR3 \citep{GAIA} parallaxes and distances ($D$). In addition, information about variability and spectral classification, pulsation period (\ppuls), chemical composition, luminosity (\lbol) and ISM reddening is presented in Table~\ref{tab:sources}.

\begin{table}[h!]

\renewcommand{\arraystretch}{1.3}
\small
\centering

\caption{Astronomical parameters of the sample from Gaia DR3.} 
\label{tab:astronomy}

\begin{adjustbox}{max width=\textwidth}
\begin{threeparttable}[b]

\begin{tabular}{l c c c c}
\hline\hline 
Source & RA  & DEC & parallax & $D$  \\  
       & (hh:mm:ss) & (\textbf{dd:mm:ss}) & $(\mathrm{mas})$ &  $(\mathrm{pc})$ \\  
\hline 

AT\,Dra & 16 17 15 & +59 45 18   & $4.76 \pm 0.16$ & $210 \pm 7$ \\

BC\,Cmi & 07 52 07 & +03 16 38   & $5.74 \pm 0.07$ & $174 \pm 2$ \\

BD\,Cam & 03 42 09 & +63 13 00   & $4.3 \pm 0.3$ & $234 \pm 14$ \\

BY\,Boo & 14 07 56 & +43 51 16   & $6.01 \pm 0.16$ & $166 \pm 4$ \\

CG\,Uma & 09 21 43 & +56 41 57   & $4.41 \pm 0.16$ & $227\pm 8$ \\

DF\,Leo & 09 23 30 & +07 42 49   & $3.03 \pm 0.14$ & $330 \pm 15$ \\

EY\,Hya & 08 46 21 & +01 37 56   & $2.34 \pm 0.07$ & $427 \pm 7$ \\

FH\,Vir & 13 16 24 & +06 30 17   & $2.75 \pm 0.10$ & $364 \pm 13$ \\

IN\,Hya & 09 20 37 & +00 10 54   & $3.03 \pm 0.08$ & $330 \pm 9$ \\

OME\,Vir & 11 38 28 & +08 08 03   & $5.6\pm 0.3$ & $180 \pm 10$ \\

R\,Lmi & 09 45 34 & +34 30 43   & $3.45 \pm 0.14$ & $290 \pm 12$ \\

R\,Uma & 10 44 38 & +68 46 33   & $1.75 \pm 0.09$ & $570 \pm 30$ \\

RR\,Eri & 02 52 14 & -08 16 01   & $2.63 \pm 0.12$ & $400 \pm 18$ \\

RR\,Umi & 14 57 35 & +65 55 57   & $9.9 \pm 0.5$ & $102 \pm 5$ \\

RT\,Cnc & 08 58 16 & +10 50 43   & $3.62 \pm 0.11$ & $276 \pm 8$ \\

RU\,Her & 16 10 15 & +25 04 14   & $1.40 \pm 0.08$ & $710 \pm 40$ \\

RW\,Boo & 14 41 13 & +31 34 20   & $3.94 \pm 0.10$ & $254 \pm 7$ \\

RZ\,Uma & 08 10 60 & +65 13 22   & $1.91 \pm 0.10$ & $520 \pm 30$ \\

ST\,Uma & 11 27 50 & +45 11 07   & $3.06 \pm 0.11$ & $327 \pm 11$ \\

SV\,Peg & 22 05 42 & +35 20 55   & $2.59\pm 0.17$ & $390 \pm 30$ \\

T\,Dra & 17 56 23 & +58 13 07   & $1.02 \pm 0.08$ & $980 \pm 80$ \\

TU\,And & 00 32 23 & +26 01 46   & $0.94 \pm 0.06$ & $1060 \pm 70$ \\

UY\,Leo & 10 29 22 & +23 03 44   & $1.35 \pm 0.04$ & $740 \pm 20$ \\

V\,Eri & 04 04 19 & -15 43 30   & $3.35 \pm 0.16$ & $298 \pm 14$ \\

VY\,Uma & 10 45 04 & +67 24 41   & $2.38 \pm 0.08$ & $420 \pm 14$ \\

W\,Peg & 23 19 51 & +26 16 44   & $2.94\pm 0.13$ & $340 \pm 15$ \\

Y\,Crb & 15 46 44 & +38 19 21   & $1.43 \pm 0.06$ & $700 \pm 30$ \\

Y\,Gem & 07 41 09 & +20 25 44   & $1.53 \pm 0.09$ & $654 \pm 40$ \\

Z\,Cnc & 08 22 25 & +14 59 32   & $2.34 \pm 0.06$ & $428 \pm 11$ \\

\hline
\end{tabular} 

\end{threeparttable}
\end{adjustbox}

\renewcommand{\arraystretch}{1.0}

\end{table}

\begin{table*}[h]

\renewcommand{\arraystretch}{1.3}
\small
\centering

\caption{Observational parameters of the sources.} 
\label{tab:sources}

\begin{adjustbox}{width=0.8\textwidth}
\begin{threeparttable}[h]

\begin{tabular}{l c c c c c c}
\hline\hline 
Source & Variability$^{(a)}$   & \ppuls$^{(a)}$   & Spectral$^{(b)}$  & Composition$^{(b)}$   & \lbol$^{(c)}$ & E(B-V) (ISM)$^{(d)}$ \\ 
 &  type &  $(\mathrm{days})$ &  type &  type &  $(\mathrm{L_{\odot}})$ & (mag) \\
\hline 

AT\,Dra & LB     &       ---    &      M4III D ~    &   O  & $1900 \pm 400$  & $<0.0018$  \\

BC\,Cmi & SRB     &      34.13  &       M4/5III D   &    O & $720 \pm 130$  & $0.0$  \\

BD\,Cam & LB   &  ---   & S3.5/2 B  &  S & $2600  \pm  600$ & $<0.04$  \\

BY\,Boo & LB     &   ---    &      M4.5III B     &  O & $1800 \pm 400$  & $0.0$  \\

CG\,Uma & LB   &         ---    &      M4IIIa D &   O & $2000 \pm 400$  & $0.13 \pm 0.05$  \\

DF\,Leo & SRB     &      70        &    M4III C ~    &   O & $900 \pm 200$  & $0.040 \pm 0.019$  \\

EY\,Hya &  SRA       &    182.7     &    M7 D        &    O & $6400 \pm 1900$  & $0.080 \pm 0.007$ \\

FH\,Vir & SRB      &     70    &        M6III C ~    &   O & $1500 \pm 400$  & $0.090 \pm 0.017$  \\

IN\,Hya & SRB         &  65        &    M3/4III D    &   O & $1200 \pm 200$  & $0.040 \pm 0.005$  \\

OME\,Vir & LB     &       ---   &       M4.5:III C &     O & $1900 \pm 500$  & $<0.005$ \\

R\,Lmi & M       &      372.19    &    M6.5-9e B  & O & $3100 \pm 900$  & $<0.014$ \\

R\,Uma & M     &  301.62 & M5-8e B & O  &  $5400  \pm  1300$ & $0.090 \pm 0.009$ \\

RR\,Eri & SRB    &       97         &   M5III C      &   O & $4000 \pm 1000$  & $0.060 \pm 0.012$ \\

RR\,Umi & SRB          & 43.3     &     M4.5III C    &   O  & $1300 \pm 300$  & $0.0$  \\

RT\,Cnc &  SRB     &      89.8 &         M5III C     &    O & $2900 \pm 600$  & $0.06 \pm 0.016$ \\

RU\,Her & M      &   484.83  &      M6-7e B       &  O & $20000 \pm 7000$  & $0.110 \pm 0.007$ \\

RW\,Boo & SRB       &    209  &         M5III: D ~   &   O & $2000 \pm 400$  & $<0.0018$ \\

RZ\,Uma & SRB  &  115    & M8 D ~  &  O &  $5000  \pm  1300$ & $0.110 \pm 0.005$ \\

ST\,Uma &  SRB      &     110 &          M4III C ~    &   O & $2800 \pm 600$  & $0.0$  \\

SV\,Peg & SRB       &    144.6  &       M7 D ~       &   O & $8000 \pm 2000$  & $0.130 \pm 0.008$ \\

T\,Dra & M        &     421.62  &      C6,2e C       &  C & $9000 \pm 3000$  & $0.020 \pm 0.007$ \\

TU\,And & M    &  316.8  & M6e B  &  O & $6600  \pm  1900 $ & $0.020 \pm 0.007$   \\

UY\,Leo &  LB       &     ---   &       M7III: D ~  &    O & $2400 \pm 500$  & $< 0.018$  \\

V\,Eri & SRC       &    97.0     &     M7+II: C      &  O & $5800 \pm 1500$ & $0.070 \pm 0.018$ \\

VY\,Uma & LB   &  ---   & C-N5 B  & C  &  $5300  \pm  1100$ & $<0.007$ \\

W\,Peg & M           &  345.5        & M6.5-7.5e B   &  O & $5300 \pm 1400$  &  $0.120 \pm 0.007$ \\

Y\,Crb & SRB      &     300       &    M8III: D ~    &  O & $4100 \pm 1000$ & $<0.005$ \\

Y\,Gem & SRB     &      160      &     M8 D ~      &    O & $7000 \pm 2000$  & $0.040 \pm 0.005$ \\

Z\,Cnc & SRB        &   95.4         & M6III C ~     &  O & $3700 \pm 800$ & $0.040 \pm 0.005$ \\

\hline
\end{tabular} 
\begin{tablenotes}
\item \textbf{Notes.} Column (1): source, Col (2): variability type, Col (3): expansion period when the source has a stable period, Col (4): spectral type, Col (5): composition (O for O-rich stars or C for C-rich stars), Col (6): bolometric luminosity, Col (7): reddening produced by the ISM.
\item[(a)] From the General Catalogue of Variable Stars \citep[GCVS,][]{GCVS}.
\item[(b)] From the SIMBAD astronomical database \citep{SIMBAD}.
\item[(c)] From integration of the extinction-corrected spectral energy distributions (SEDs) (see appendix~\ref{sources fluxes}).
\item[(d)] Interstellar medium reddening has been obtained with GALEXtin \citep[see][]{amores_2021} with the extinction map from Bayestar19 \citep[see][]{Green_2019}.
\end{tablenotes}

\end{threeparttable}
\end{adjustbox}

\renewcommand{\arraystretch}{1.0}

\end{table*}

Accurate distance estimation is a complicated subject for AGBs. While Very Long Baseline Interferometry (VLBI) parallax measurements of maser emissions remain the most reliable technique to date, they are limited to a small subset of AGBs \citep[see][]{Andriantsaralaza_2022}. Gaia DR3 offers a broader alternative, yet it is crucial to note that Gaia's distance uncertainties may be underestimated by up to a factor of five, as demonstrated by \cite{Andriantsaralaza_2022}. To enhance the statistical reliability of our analysis, we ensured that over 50\% of our sample met stringent criteria for reliable parallax measurements, including low brightness (G<8.0) and low RUWE (<1.4), as outlined in \cite{El-Badry_2021} and \cite{Andriantsaralaza_2022}. Nevertheless, it is important to acknowledge that uncertainties associated with distances (Table~\ref{tab:astronomy}) and derived luminosities (Table~\ref{tab:sources}) are likely underestimated.

Our sample of AGB binary candidates was derived from a larger sample of AGBs with NUV detections found in the GALEX MAST archive \citep[][]{Conti_2011} following a similar approach as in previous studies by \cite{Sahai_2008} and \cite{Sahai_2015}. 
The GALEX sample of ultraviolet excess AGB stars is substantial, comprising over 700 objects \citep{Sahai_2022}, many of which have multiple observations revealing significant variability in the ultraviolet \citep{Sahai_2022, Montez_2017}. To identify the most promising binary AGB star candidates, we have included a large fraction of stars exhibiting excesses in the FUV band (25 sources). Additionally, six of our sources have been detected in X-rays: RR Umi \citep{Hunsch_1998}, R\,Uma and T\,Dra \citep{Ramstedt_2012}, EY\,Hya and Y\,Gem \citep{Sahai_2015}, and BD\,Cam \citep{Lima_2022}.

In this section, we compare the properties of our sample of 29 uvAGBs with those of the complete uvAGBs sample identified to date as well as with the initial sample of 18381 galactic AGB stars from the catalogue published by \cite{Suh_2021}, which represents one of the most comprehensive collections of AGB stars published to date.

A complete sample of uvAGB stars has been derived after crossmatching the GALEX MAST catalogue with that by \cite{Suh_2021}. We used the TOPCAT tool \citep{topcat} and used a searching window of 5\arcsec\ radius. We identified that 53.5\% of the sources in the \cite{Suh_2021} catalogue are included in the GALEX NUV sky coverage, from which 10.4\% were detected in this band. On the other hand, the number of sources from the \cite{Suh_2021} catalogue that are in the GALEX FUV sky coverage is 8.0\%, with a detection rate in this band of only 4.5\%.

It is important to note that the statistical analysis of nuvAGBs and fuvAGBs is affected by the incomplete sky coverage of GALEX observations. Specifically, there is a lack of GALEX observations near the galactic centre, where a significant portion of AGB stars are located. This incomplete coverage hinders comprehensive statistical studies on the prevalence of nuvAGBs and fuvAGBs within the entire AGB population. In fact, 46.5\% of the sample was not observed in the NUV band, and this ratio increases to 92.0\% in the FUV band.

To assess the representativeness of our sample within the uvAGB class, we conducted a comparison of their properties with the complete uvAGB sample obtained by cross-matching the GALEX catalogue with the catalogue by \cite{Suh_2021}. This analysis enables us to identify similarities and differences between AGBs, nuvAGBs, and fuvAGBs, providing insights into the characteristics of these subgroups.

\begin{figure}[h!]
    \centering
    \includegraphics[width=\linewidth]{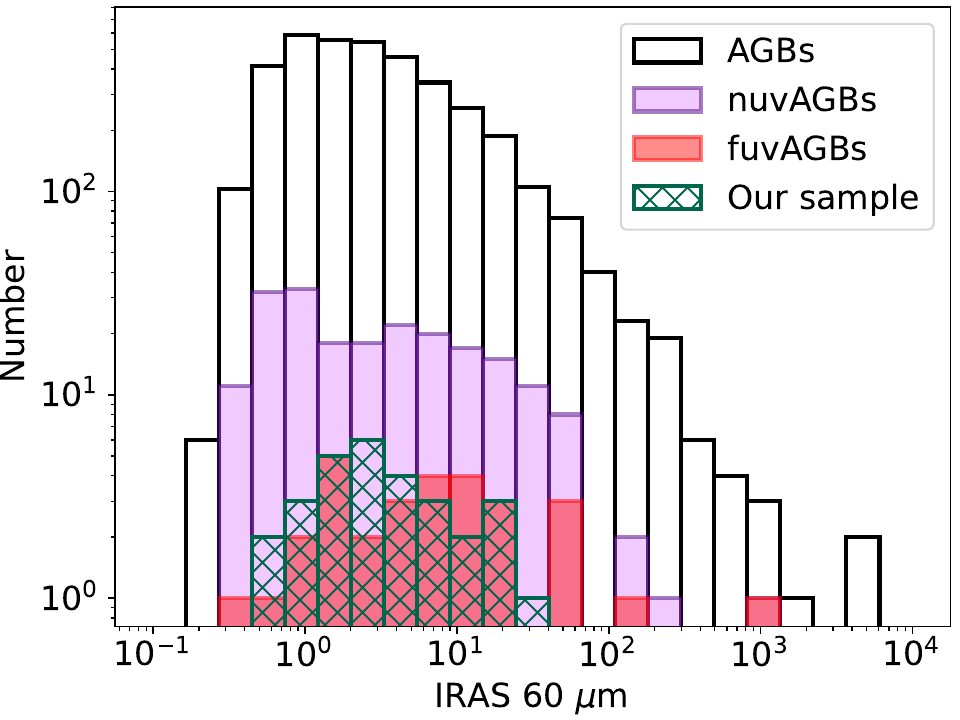}
    \caption{\irassesenta \, flux distribution of our targets (green) and of AGB stars (including nuvAGBs and fuvAGBs) from \cite{Suh_2021}, see Sect.~\ref{sample}. The  colours   represent the NUV/FUV emission;  the sample is indicated in the  top right corner  for reference.}
    \label{fig:IRAS60_histo}
\end{figure}

The distribution of \irassesenta\ flux for nuvAGBs and fuvAGB sources closely resembles that of AGB stars in the Suh catalogue (Fig.~\ref{fig:IRAS60_histo}). The main distinction is that the uvAGB subclass has fewer stars, and there is a relative deficiency of objects with extremely high \irassesenta\ fluxes within this subclass. Our sample of 29 uvAGBs, which has a higher proportion of fuvAGBs compared to nuvAGBs, exhibits a distribution similar to that of fuvAGBs. Thus, it appears that our sample can be regarded as representative of fuvAGBs and covers low and intermediate \irassesenta\ fluxes ($<$40\,Jy). However, it is worth noting that our sample lacks objects with high \irassesenta\ fluxes ($>$40\,Jy). A discussion of the \irassesenta\ luminosity (i.e. taking into account the distances to our targets) is presented in Sect.~\ref{luminosities}. 

\begin{figure*}[h!]
    \centering
    \includegraphics[width=0.8\linewidth]{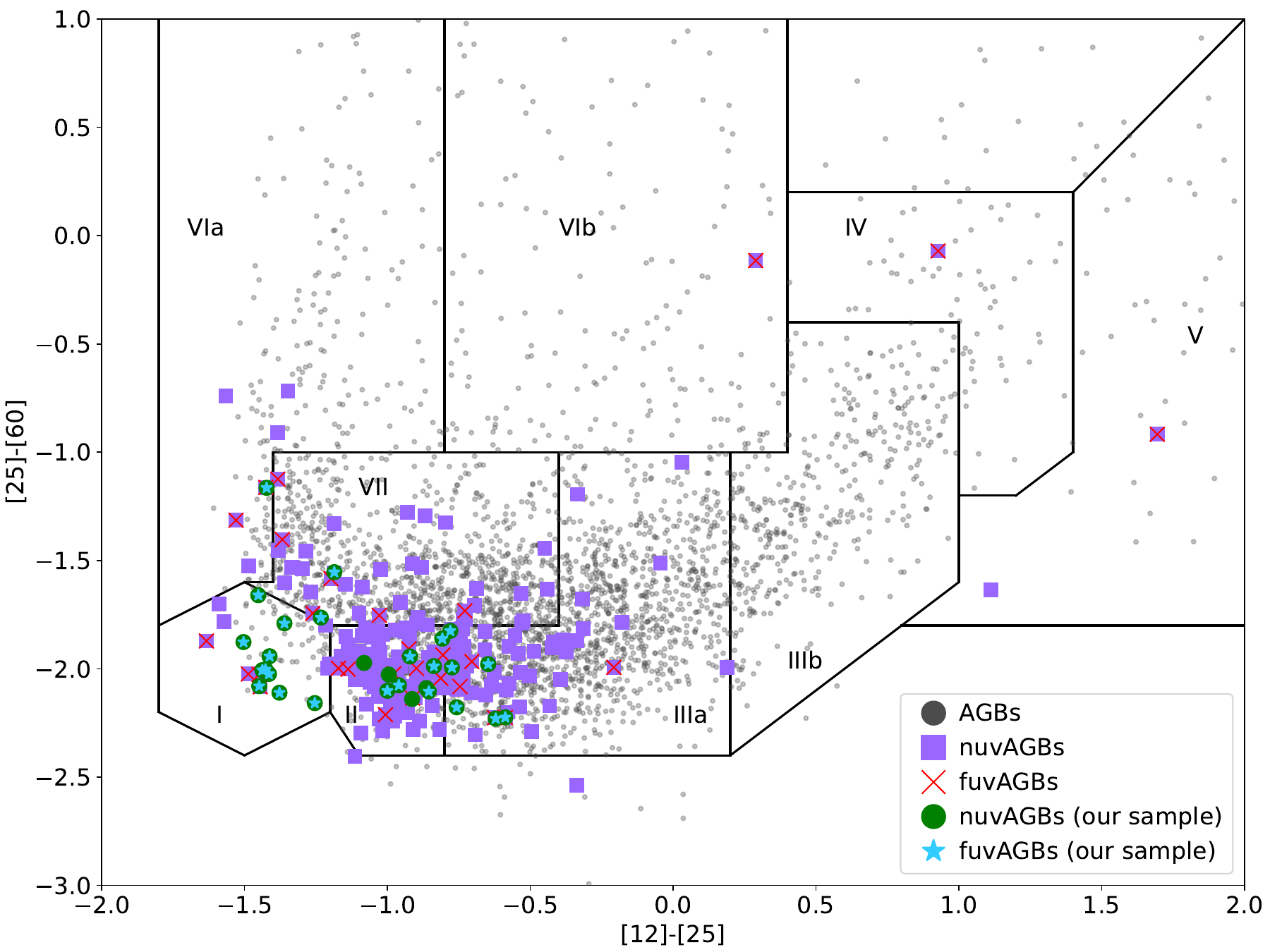}
    \caption{IRAS [25]-[60] vs [12]-[25] colour-colour diagram showing the location of our sample of 29 uvAGBs and the  AGB stars (with and without UV excess) from the catalogue by \cite{Suh_2021}, see Sect.~\ref{sample}. The colours of the markers represent the NUV--FUV emission;   the sample is indicated in the  bottom right corner  for reference.}
    \label{fig:IRAS_colour_diagram_NUV+FUV}
\end{figure*}

Figure \ref{fig:IRAS_colour_diagram_NUV+FUV} shows the  IRAS [25-60] versus [12-25] two-colour diagram of the catalogue sources. The black lines present regions that contain sources with similar CSE properties and in a near evolutionary stage \citep[for a detailed description, see][]{van_der_veen_1988}.

The majority of nuvAGBs and fuvAGBs are found in regions I, II, IIIa, or VII. Specifically, fuvAGBs tend to be concentrated in areas characterised by low values of [25]-[60] and [12]-[25] IRAS colours, primarily in regions I and II. This indicates that UV emissions are more prevalent in AGB stars with thin envelopes \citep[see e.g.][]{van_der_veen_1988}. Given that our uvAGB sample primarily consists of fuvAGBs, their distribution also aligns with the concentration in regions I and II observed in the \cite{Suh_2021} catalogue of fuvAGBs.

The sample used in this work includes objects with different types of variability according to the General Catalogue of Variable Stars \citep[GCVS,][]{GCVS} classification: 21\% are Miras (M), 55\% are semi-regular (SRs), and 24\% are irregulars (LBs) as shown in table~\ref{tab:sources}. Since most of the objects in our sample (25/29) have FUV+NUV emission, we compare the variability properties of our FUV+NUV sample with the FUV+NUV emitting AGB stars of the Suh (2021) catalogue, and find that the proportions of M, SRs and LBs are similar. However, for AGB stars with only NUV emission in the Suh (2021) catalogue, we find the proportions of SRs and LBs are much lower (23\% and 13\% respectively), and that Miras are most abundant (62\%). FUV+NUV emission is more likely associated with binarity (and related accretion activity) than only NUV emission as shown in Sahai et al. (2022). Thus, our finding above that there are larger fractions of SRs and LBs in a sample of AGB stars with FUV+NUV emission compared to one with only NUV emission, suggests that SR and LB variability in AGB stars is an indicator of binarity.
 
However, further investigations using larger samples are needed to explore this tentative result more thoroughly, taking into account the observational bias that large samples likely cover a larger distance range. This makes it  more difficult to detect and classify variability as well as to detect UV emission.

There are some targets in our sample that show low luminosities $\lesssim 3000 L_{\odot}$ (see table~\ref{tab:sources}). These low values are not expected for stars on the tip of the AGB. Therefore, it is possible that some of these low-luminosity sources are still on the Red Giant Branch (RGB) or in the early-AGB phase, and therefore have relatively low mass-loss rates with the result that they have not developed a detectable CSE (in millimetre-wave CO lines) as yet. Some of these low luminosities could potentially be attributed to distance inaccuracies.

\section{Observations}\label{obs}

The observations presented in this paper were carried out with the \iram\ radiotelescope (Pico Veleta, Granada, Spain) using the Eight MIxer Receiver \citep[EMIR,][]{Carter_2012} spectral line receiver. Observations were taken in two observational campaigns in 2009 and 2010.

We simultaneously observed the \docedos (230.538\,GHz) and \doceuno (115.271\,GHz) rotational transitions with E090 and E230 receivers operated in dual sideband (2SB) mode and dual (H+V) polarisation. 
Observations were performed with different backends simultaneously. In this work, we present only the spectra taken with the VErsatile SPectrometer Array (VESPA), which provides the best spectral resolution of $\sim$0.3\,MHz ($\sim$0.4 and 0.8\,\kms\ at 1 and 3\,mm, respectively) over a spectral bandwidth of $\sim$216\,MHz.  

Observations were performed in wobbler-switching mode with a wobbler throw of 120\arcsec\ and frequencies of 0.5\,Hz. Calibration scans on the standard two load system were taken every $\sim$10-15 min. Pointing and focus were checked regularly on nearby continuum sources. After pointing corrections, the typical pointing accuracy was $\sim$2\arcsec-5\arcsec. 
On-source integration times are in the range 1-7.5\,h (with a mean value of 1.7\,h) per target. 

The observation were reduced using CLASS\footnote{CLASS is a world-wide software used to process, reduce and analyse spectroscopic single-dish observations developed and maintained by the Institut de Radioastronomie Millimétrique (IRAM) and distributed with the GILDAS software, see {\tt \url{http://www.iram.fr/IRAMFR/GILDAS}}} following standard procedures, which include killing bad channels (if any), subtracting baseline, and averaging individual good-quality scans to produce the final spectra. 
For the sources observed in both the 2009 and 2010 observation campaigns (Y\,Crb, RW\,Boo, RU\,Her, DF\,Leo, OME\,Vir), the two epoch spectra were averaged after weighting by the inverse square of the rms noise of each individual spectrum. We present the spectra in the corrected antenna-temperature (\ta) scale, which can be converted to main-beam temperature (\tmb) applying the well-known relation \tmb=\ta/\eff, where \eff\ is the main-beam efficiency. For the \iram\ antenna \eff\ $\sim$0.83 at 115\,GHz and \eff\ $\sim$0.65 at 230\,GHz. The point source sensitivity (S/\ta, i.e. the K-to-Jy conversion factor) is $\sim$6.01 and $\sim$7.89 at 115 and 230\,GHz, respectively. The half-power-beam width (HPBW) of the \iram\ antenna is 
21\farc4 at 3mm and 11\farc7 at 1mm\footnote{The \iram\ efficiencies and beam widths can be found at {\tt \url{https://publicwiki.iram.es/Iram30mEfficiencies.}}}.

The noise of the spectra for a velocity resolution of 0.8\,\kms\ is in the range rms=7.7-95\,mK (with a median value of 12\,mK) at 1mm and rms=5-30\,mK (with a median value of 11\,mK) at 3mm. The uncertainty of the relative calibration of our observations was estimated by observing CW Leo, a well-studied AGB star with intense CO emission that is commonly used as a line calibration standard. The relative calibration uncertainty has also been checked by comparing the \docedos\ and \doceuno\ spectra of five of our targets observed in 2009 as well as in 2010. Based on this, we estimate total line flux uncertainties of $\sim$20-25\% at 1mm and of $\sim$10-15\%\ at 3mm.

\section{Observational results}\label{resul}

\begin{figure*}[ h!]
     \centering
     \begin{subfigure}[b]{0.31\linewidth}
         \centering
         \includegraphics[width=\linewidth]{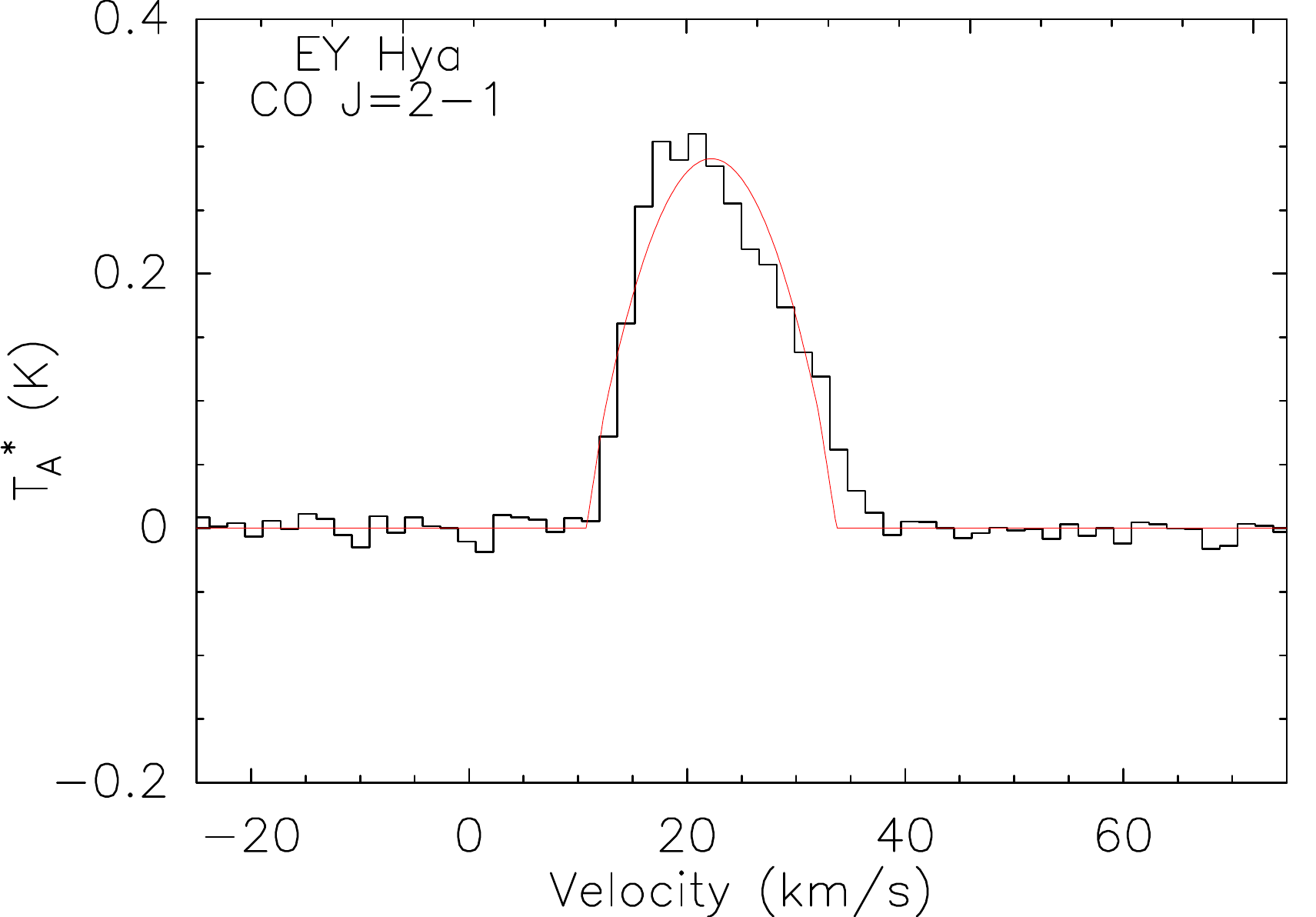}
     \end{subfigure}
     \begin{subfigure}[b]{0.31\linewidth}
         \centering
         \includegraphics[width=\linewidth]{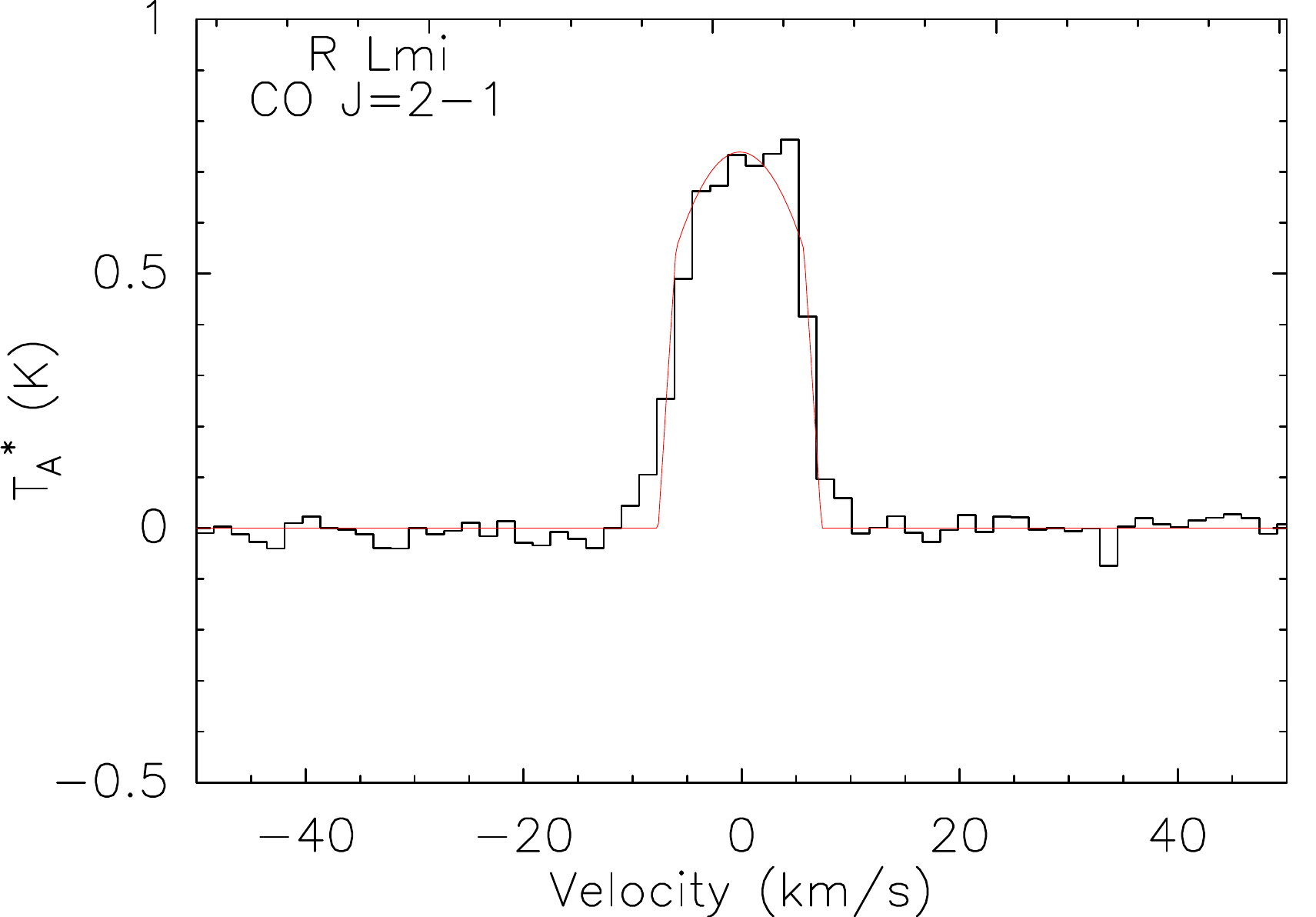}
     \end{subfigure}
     \begin{subfigure}[b]{0.31\linewidth}
         \centering
         \includegraphics[width=\linewidth]{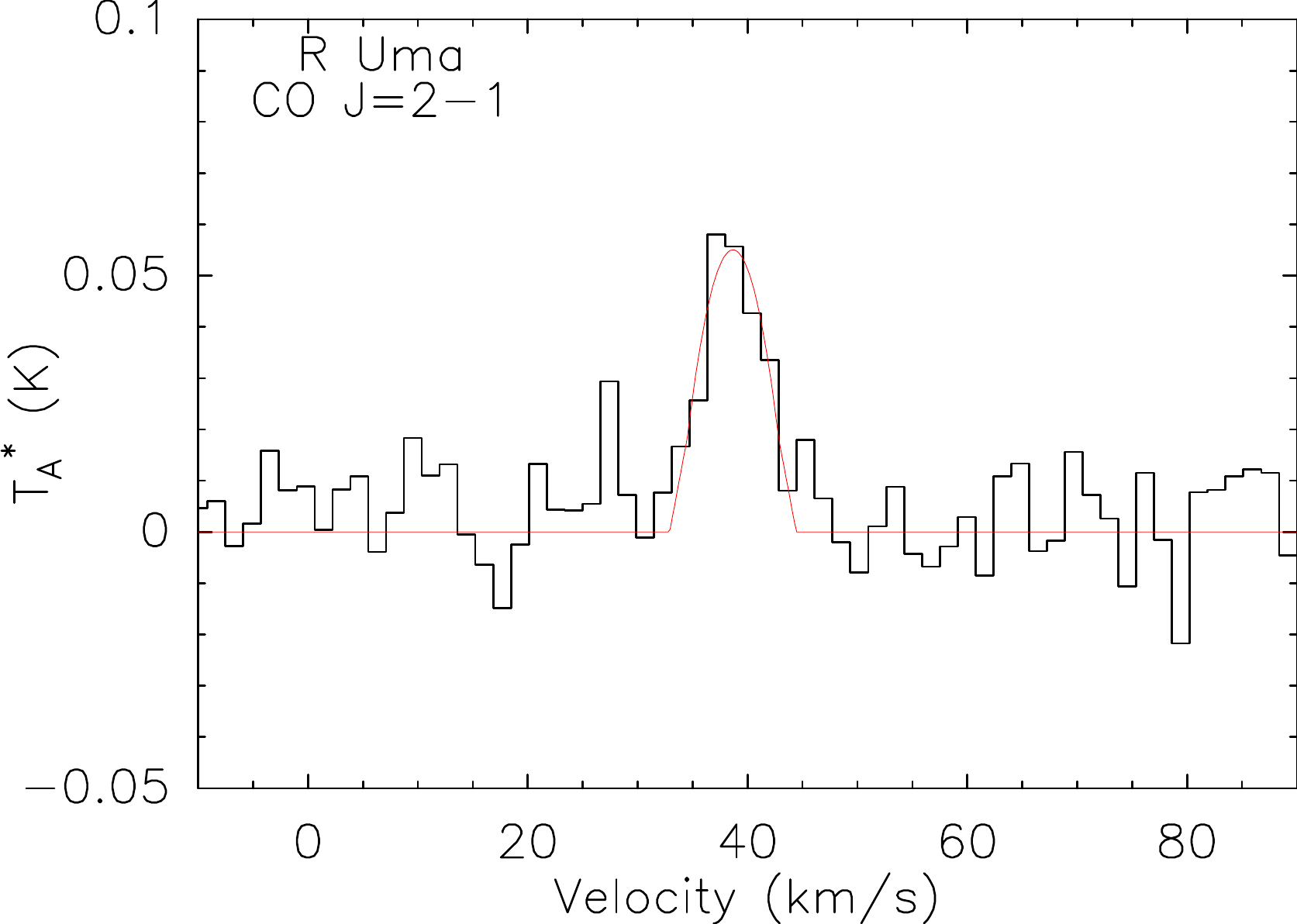}
     \end{subfigure}
     \par\bigskip
     \begin{subfigure}[b]{0.31\linewidth}
         \centering
         \includegraphics[width=\linewidth]{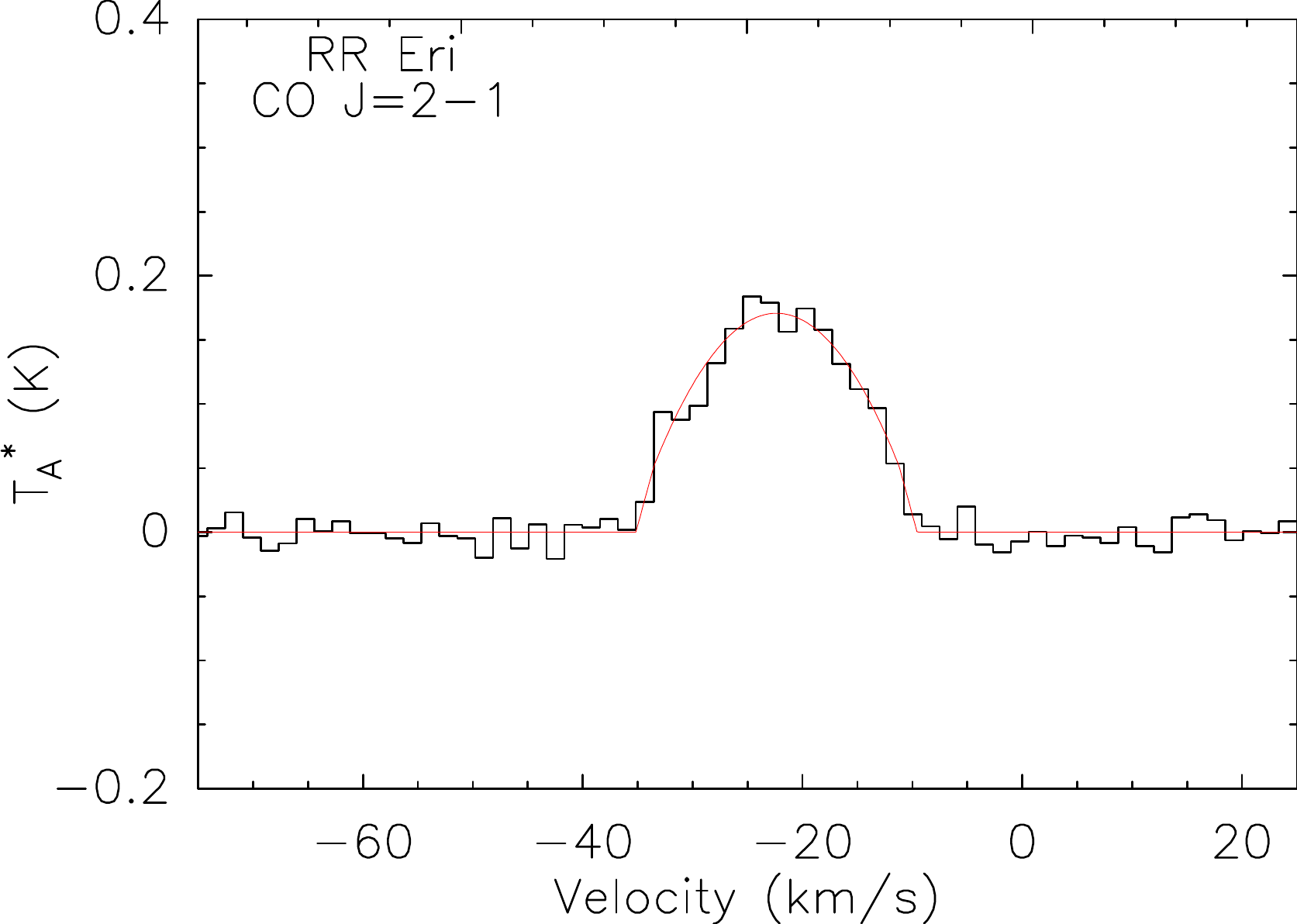}
     \end{subfigure}
     \begin{subfigure}[b]{0.31\linewidth}
         \centering
         \includegraphics[width=\linewidth]{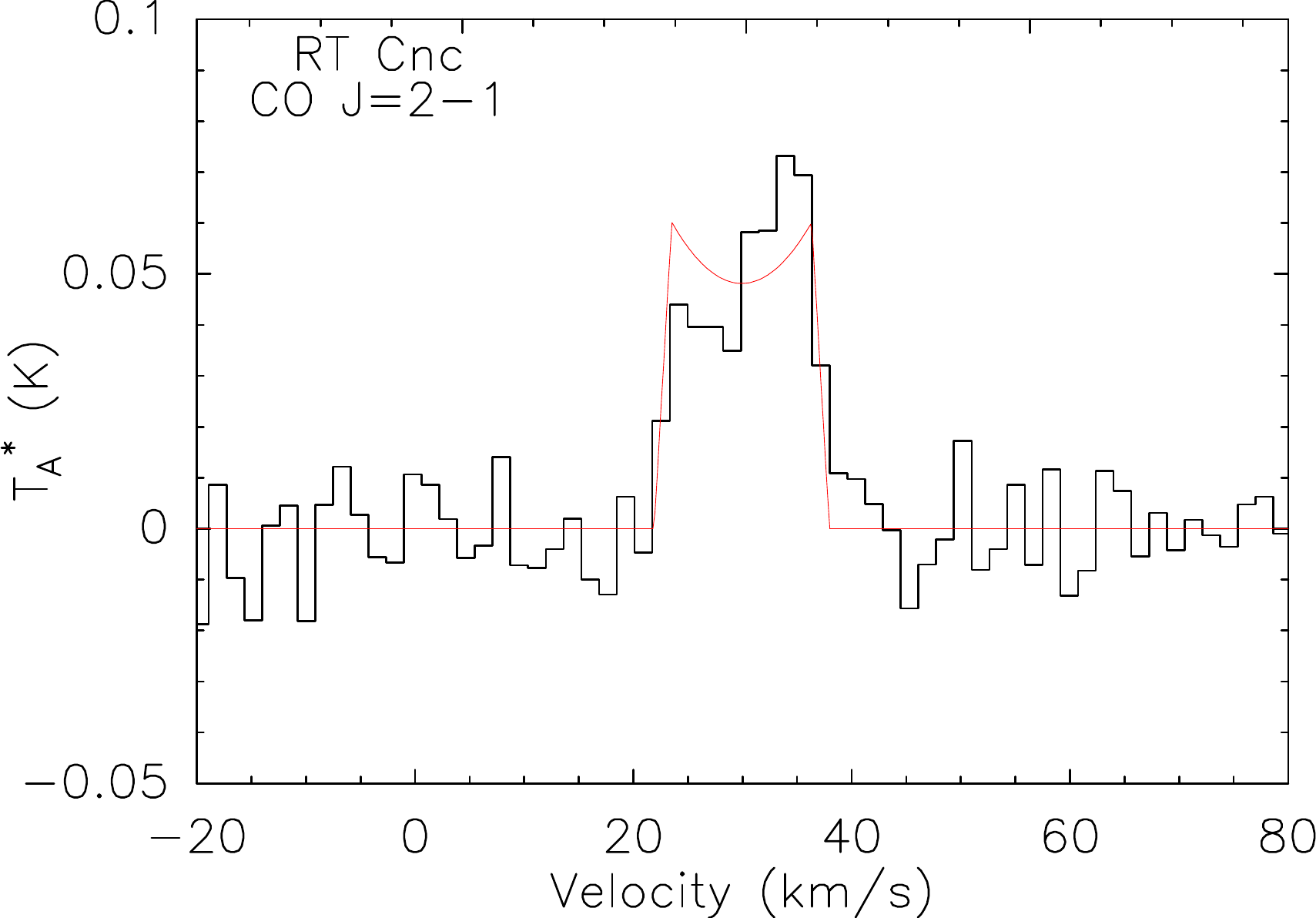}
     \end{subfigure}
     \begin{subfigure}[b]{0.31\linewidth}
         \centering
         \includegraphics[width=\linewidth]{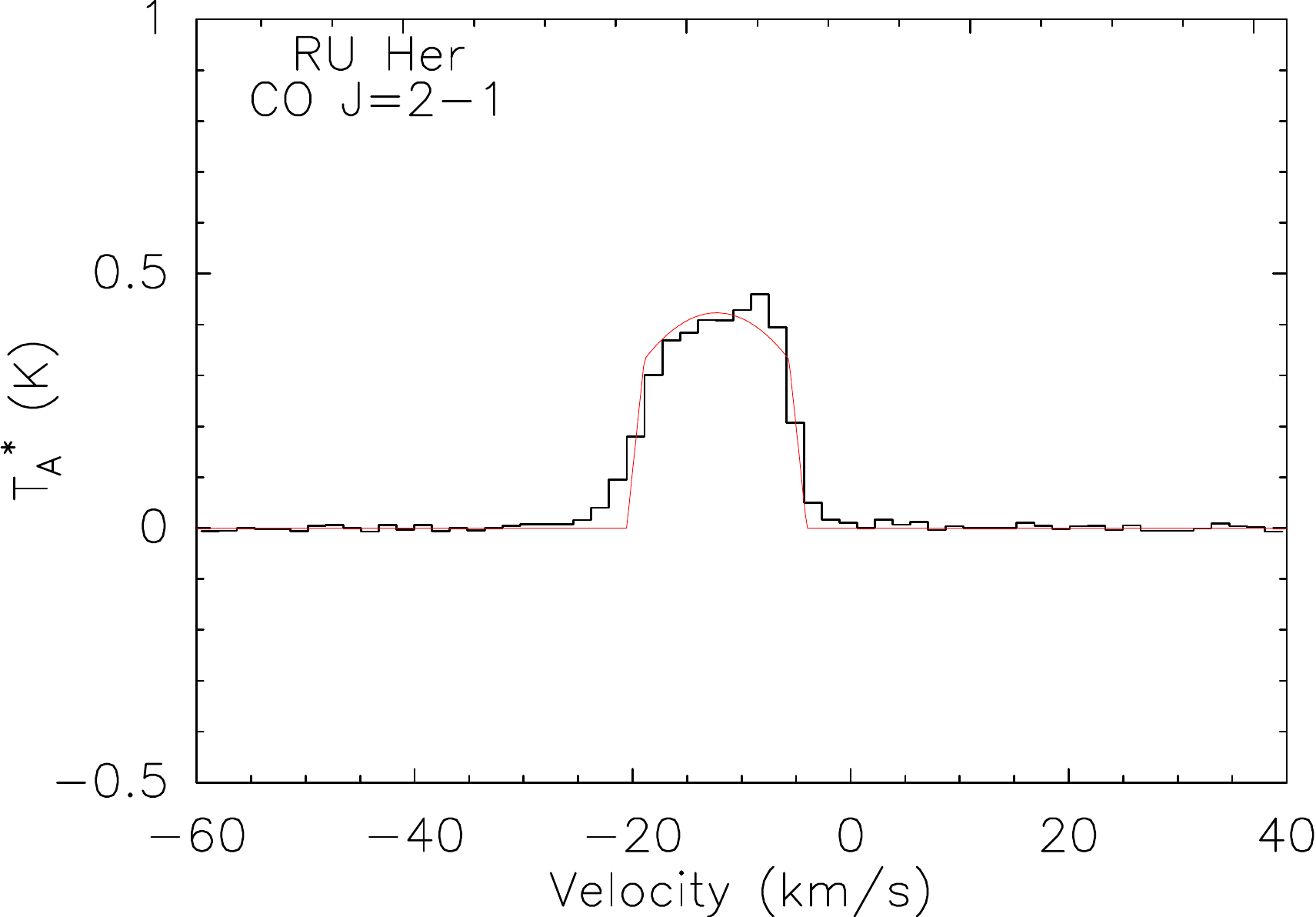}
     \end{subfigure}
     \par\bigskip
     \begin{subfigure}[b]{0.31\linewidth}
         \centering
         \includegraphics[width=\linewidth]{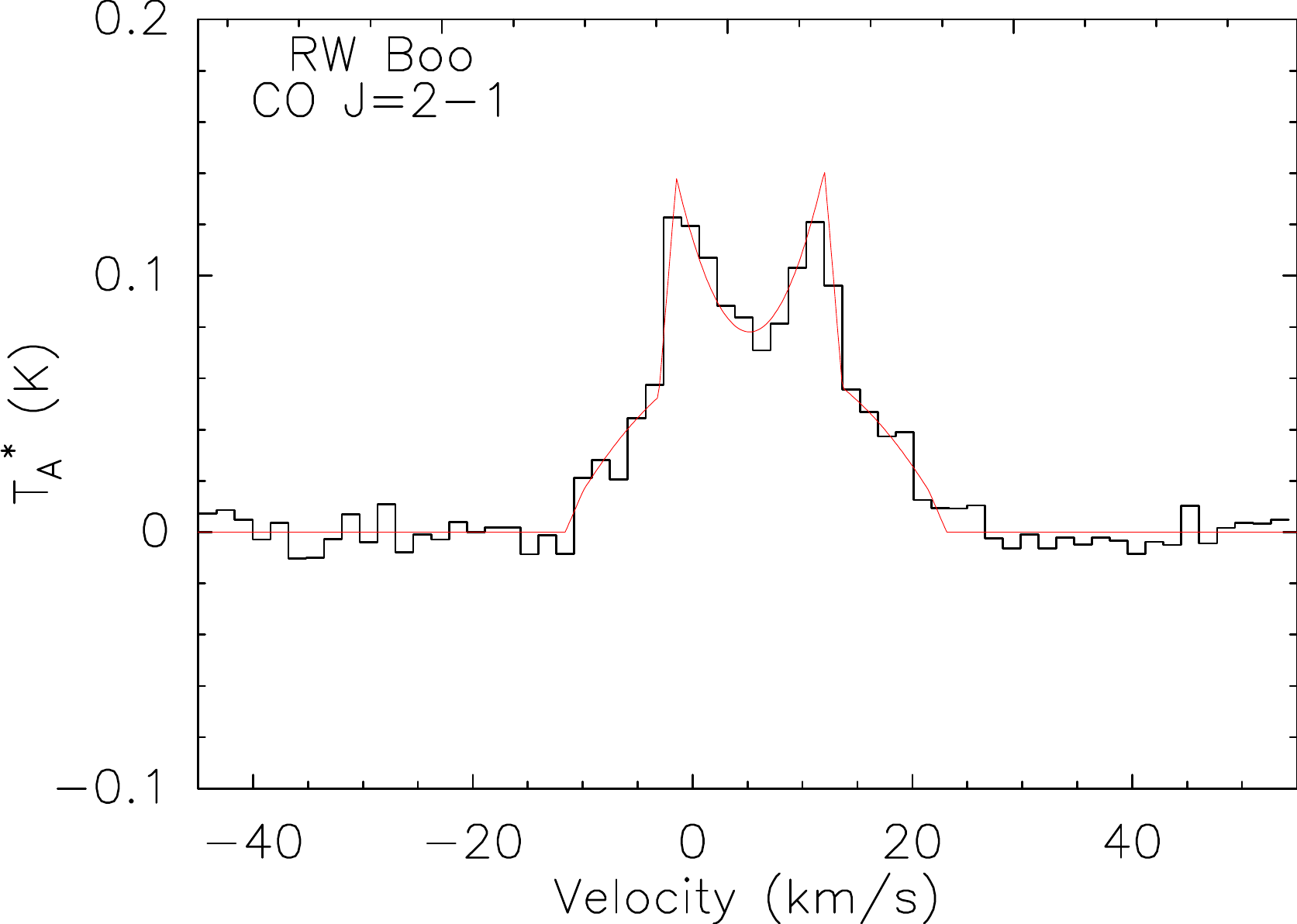}
     \end{subfigure}
     \begin{subfigure}[b]{0.31\linewidth}
         \centering
         \includegraphics[width=\linewidth]{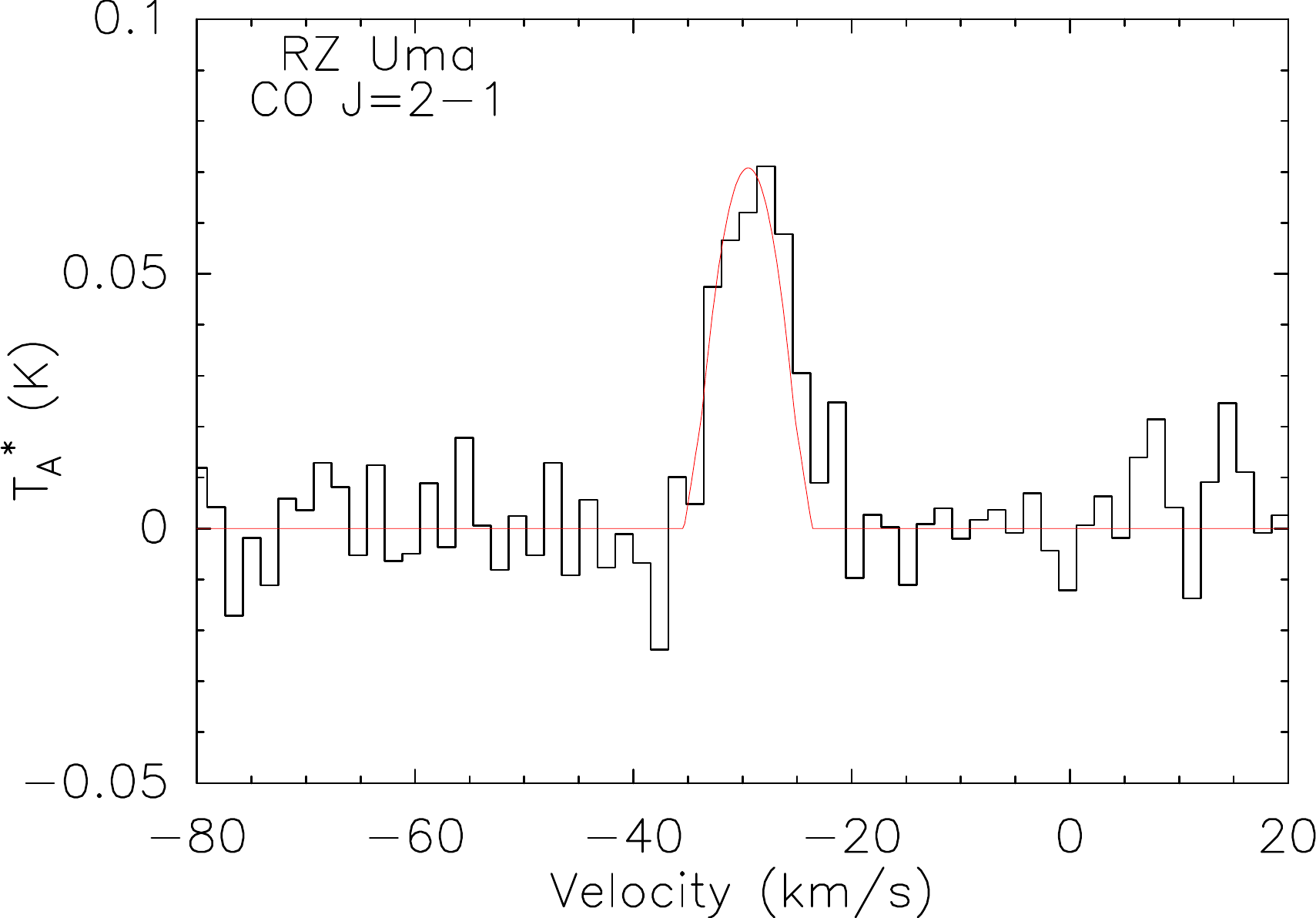}
     \end{subfigure}
     \begin{subfigure}[b]{0.31\linewidth}
         \centering
         \includegraphics[width=\linewidth]{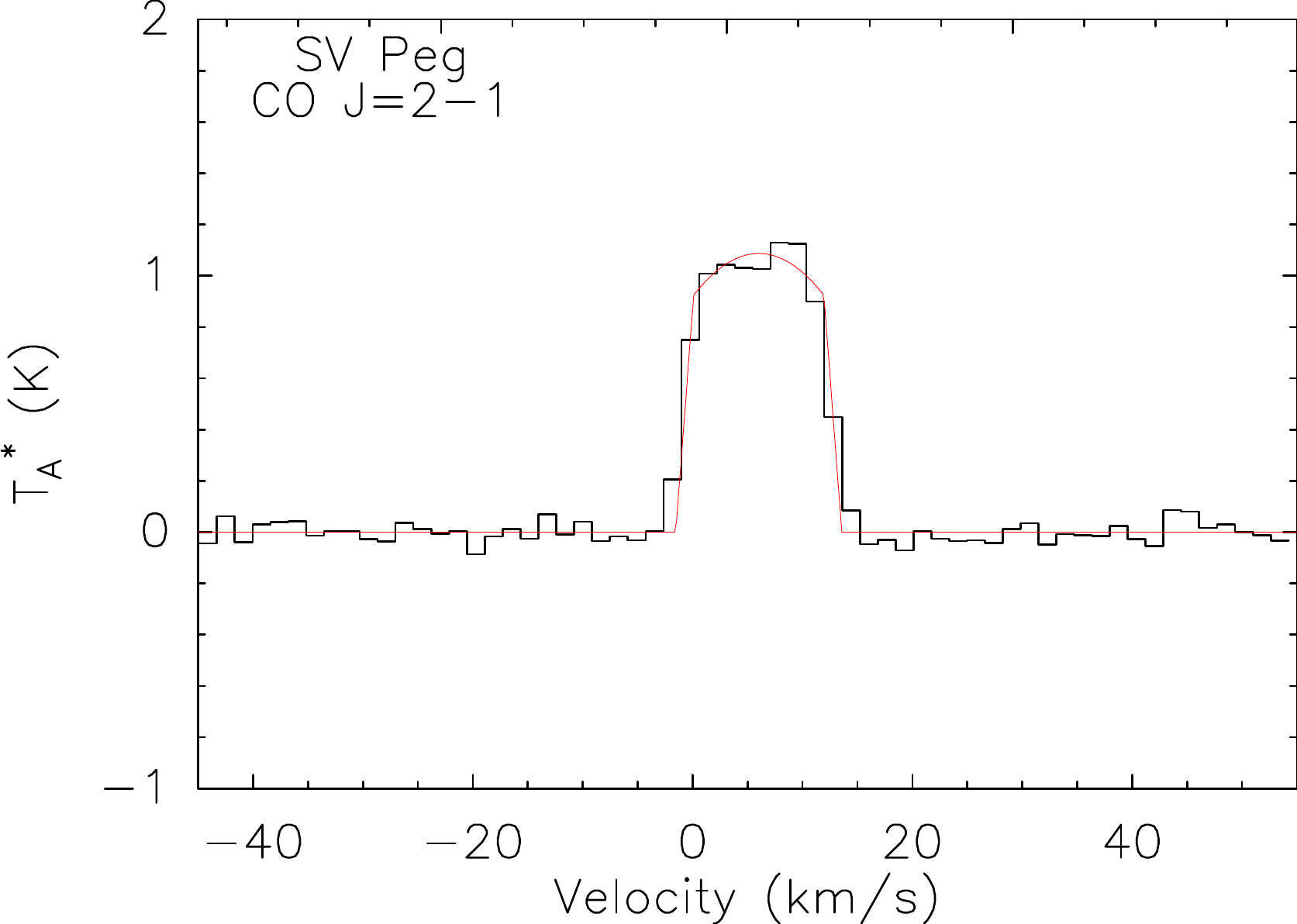}
     \end{subfigure}
     \par\bigskip
     \begin{subfigure}[b]{0.31\linewidth}
         \centering
         \includegraphics[width=\linewidth]{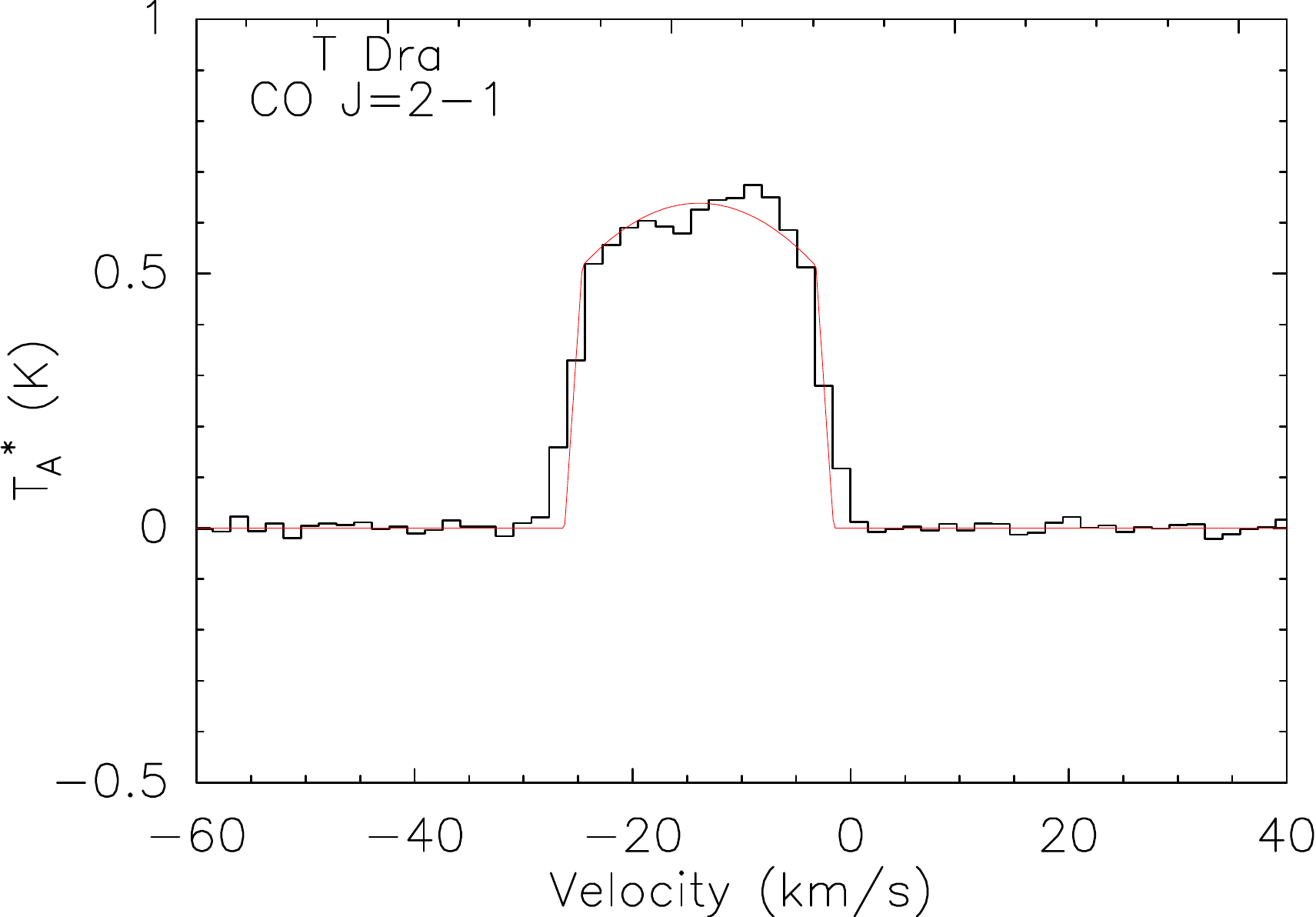}
     \end{subfigure}
     \begin{subfigure}[b]{0.31\linewidth}
         \centering
         \includegraphics[width=\linewidth]{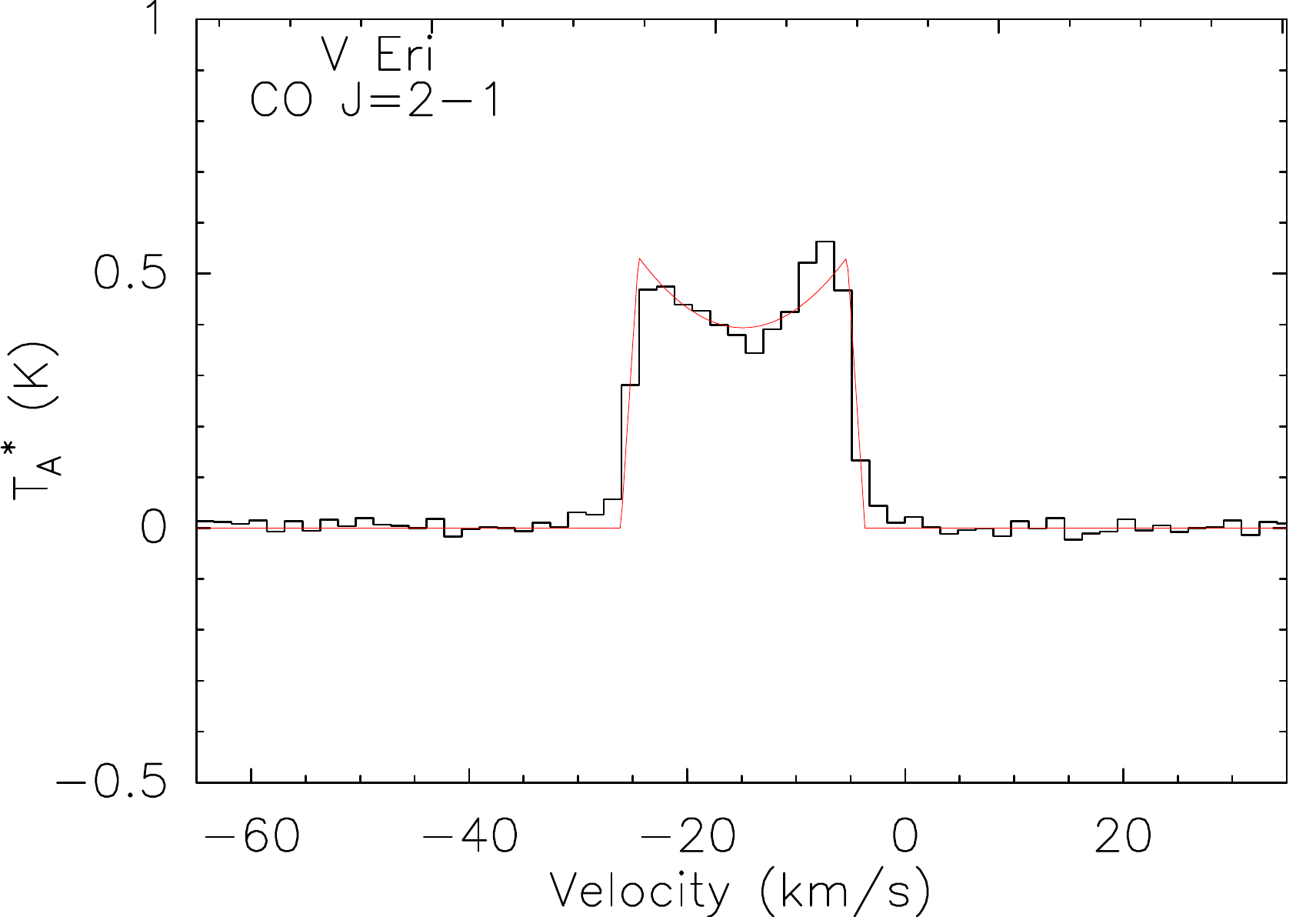}
     \end{subfigure}
     \begin{subfigure}[b]{0.31\linewidth}
         \centering
         \includegraphics[width=\linewidth]{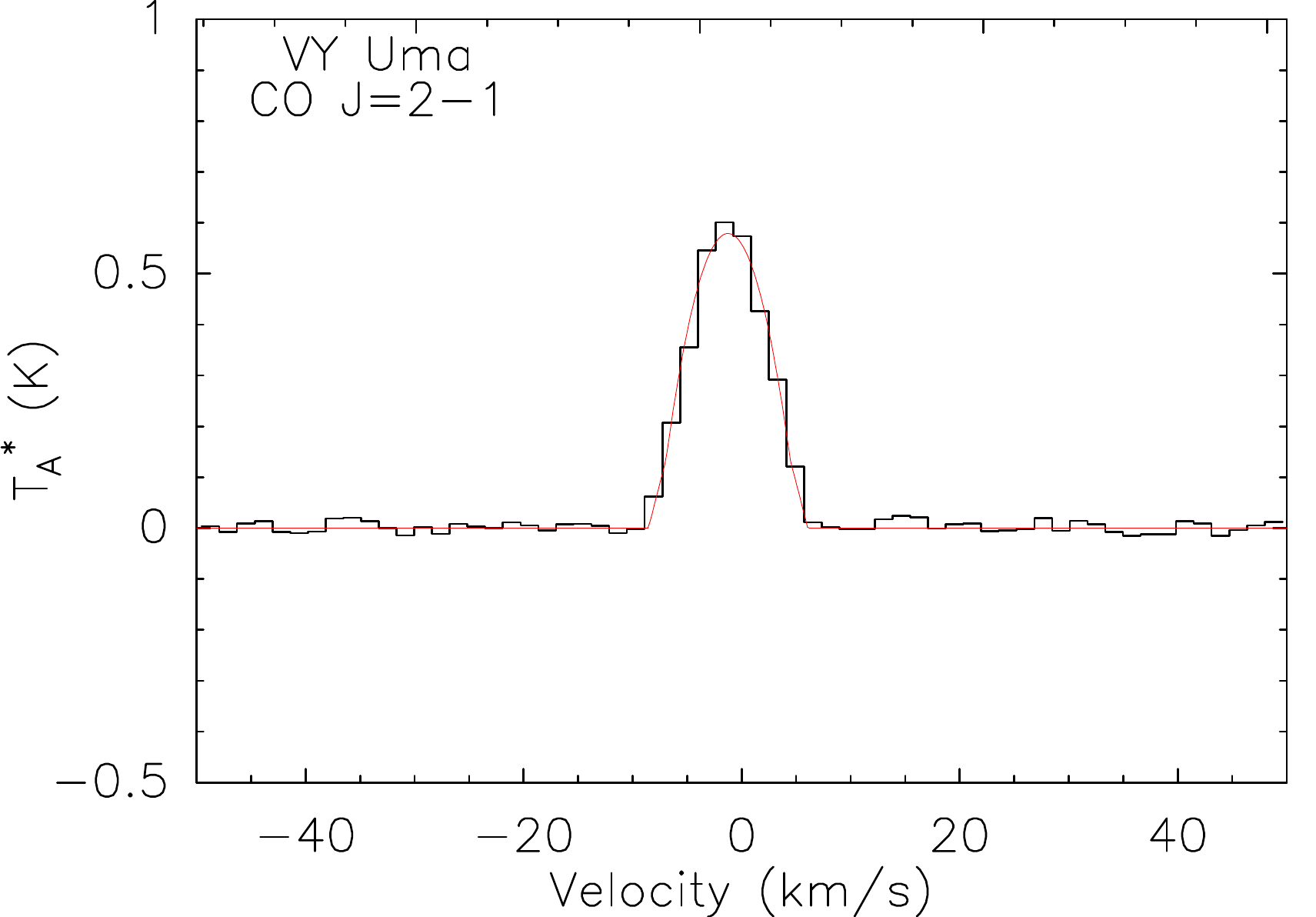}
     \end{subfigure}
     \par\bigskip
     \begin{subfigure}[b]{0.31\linewidth}
         \centering
         \includegraphics[width=\linewidth]{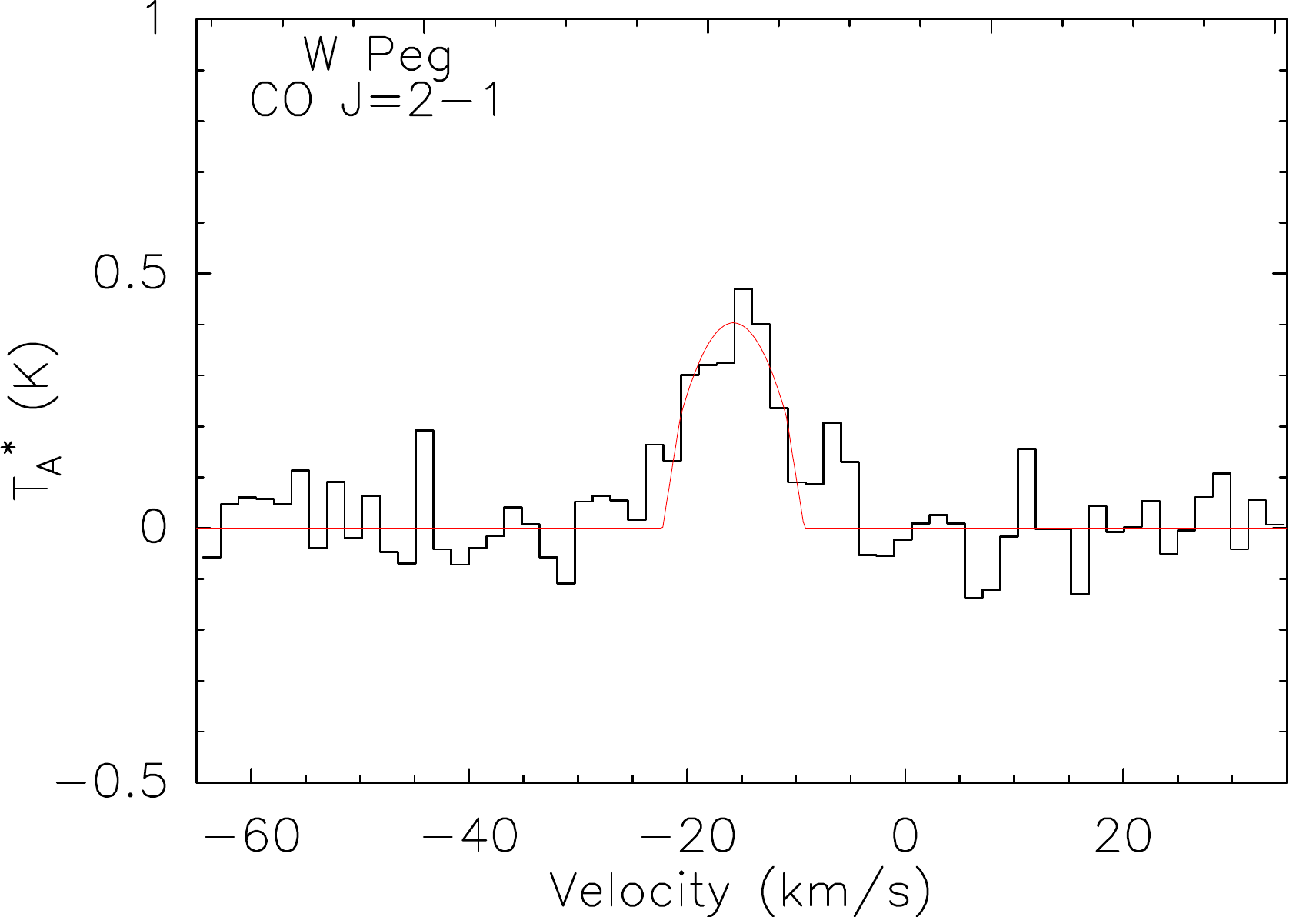}
     \end{subfigure}
     \begin{subfigure}[b]{0.31\linewidth}
         \centering
         \includegraphics[width=\linewidth]{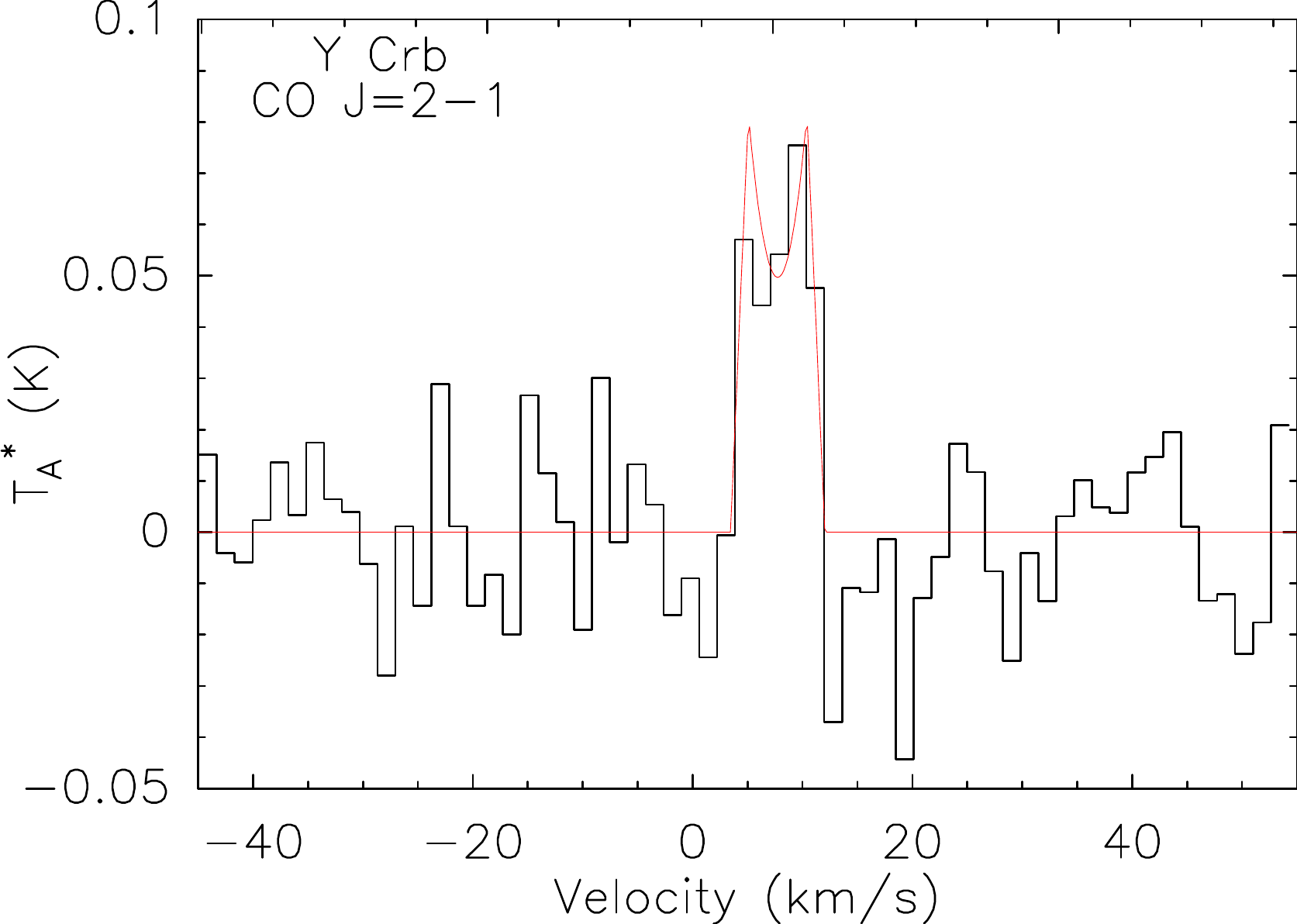}
     \end{subfigure}
     \begin{subfigure}[b]{0.31\linewidth}
         \centering
         \includegraphics[width=\linewidth]{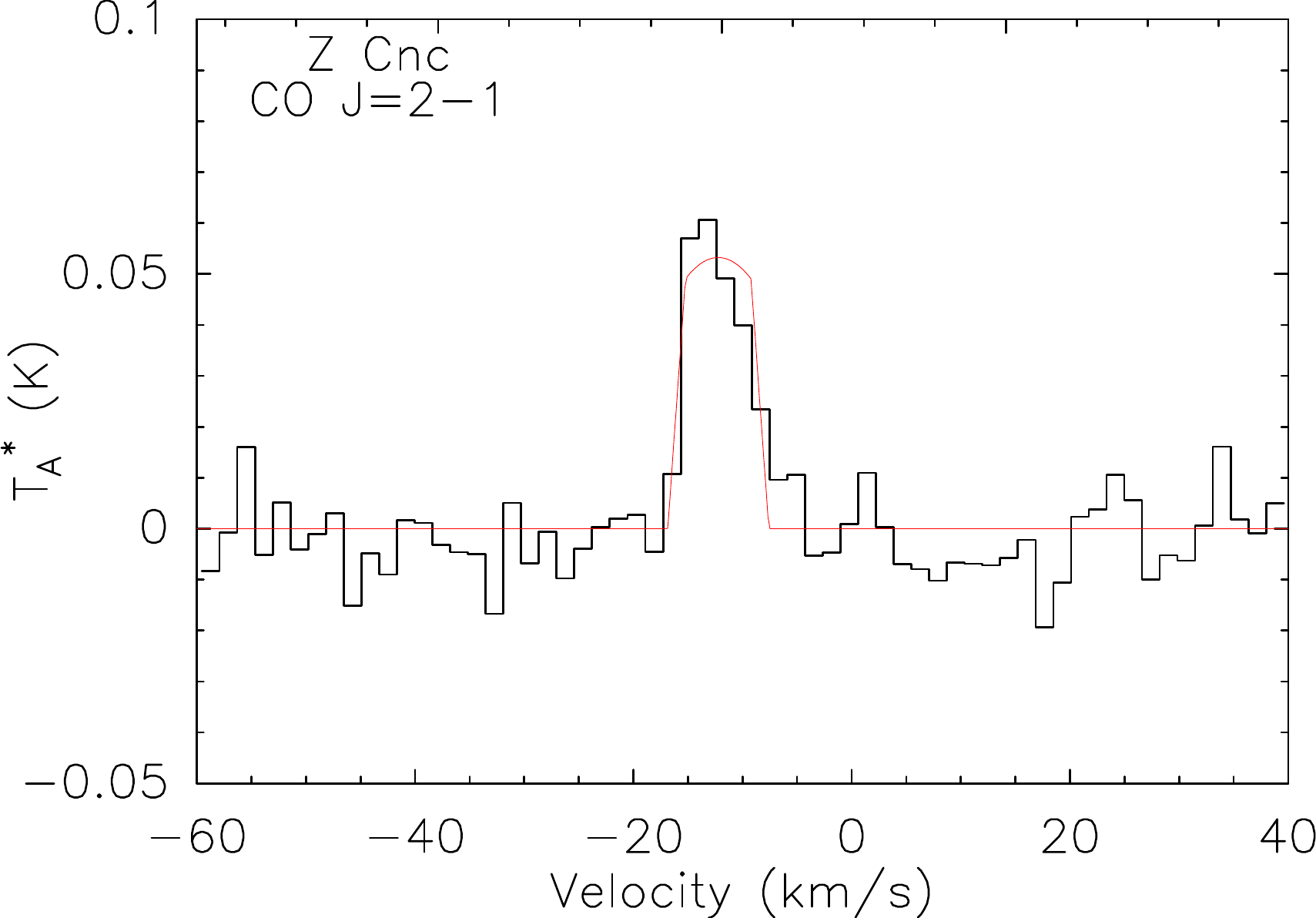}
     \end{subfigure}
     
     
     
        \caption{\iram\ spectra of the 15 sources detected in the \docedos\, transition (velocity resolution is $\delta$v=1.6\,\kms). Line profile fits are shown in red (see Sect.~\ref{resul}).}
        \label{fig:2-1_spectra}
\end{figure*}

\begin{figure*}[h!]
     \centering
     
     \begin{subfigure}[b]{0.31\linewidth}
         \centering
         \includegraphics[width=\linewidth]{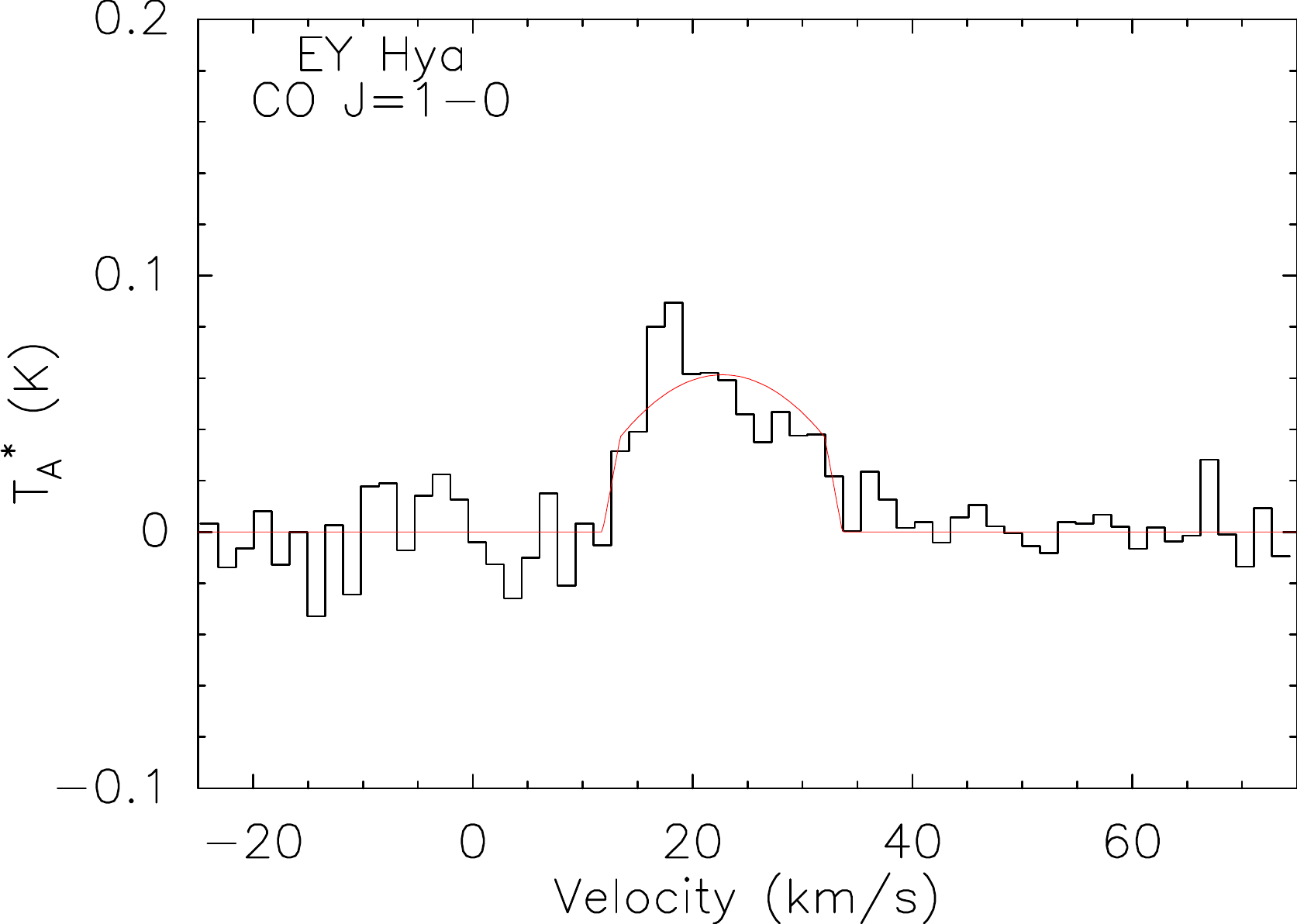}
     \end{subfigure}
     \begin{subfigure}[b]{0.31\linewidth}
         \centering
         \includegraphics[width=\linewidth]{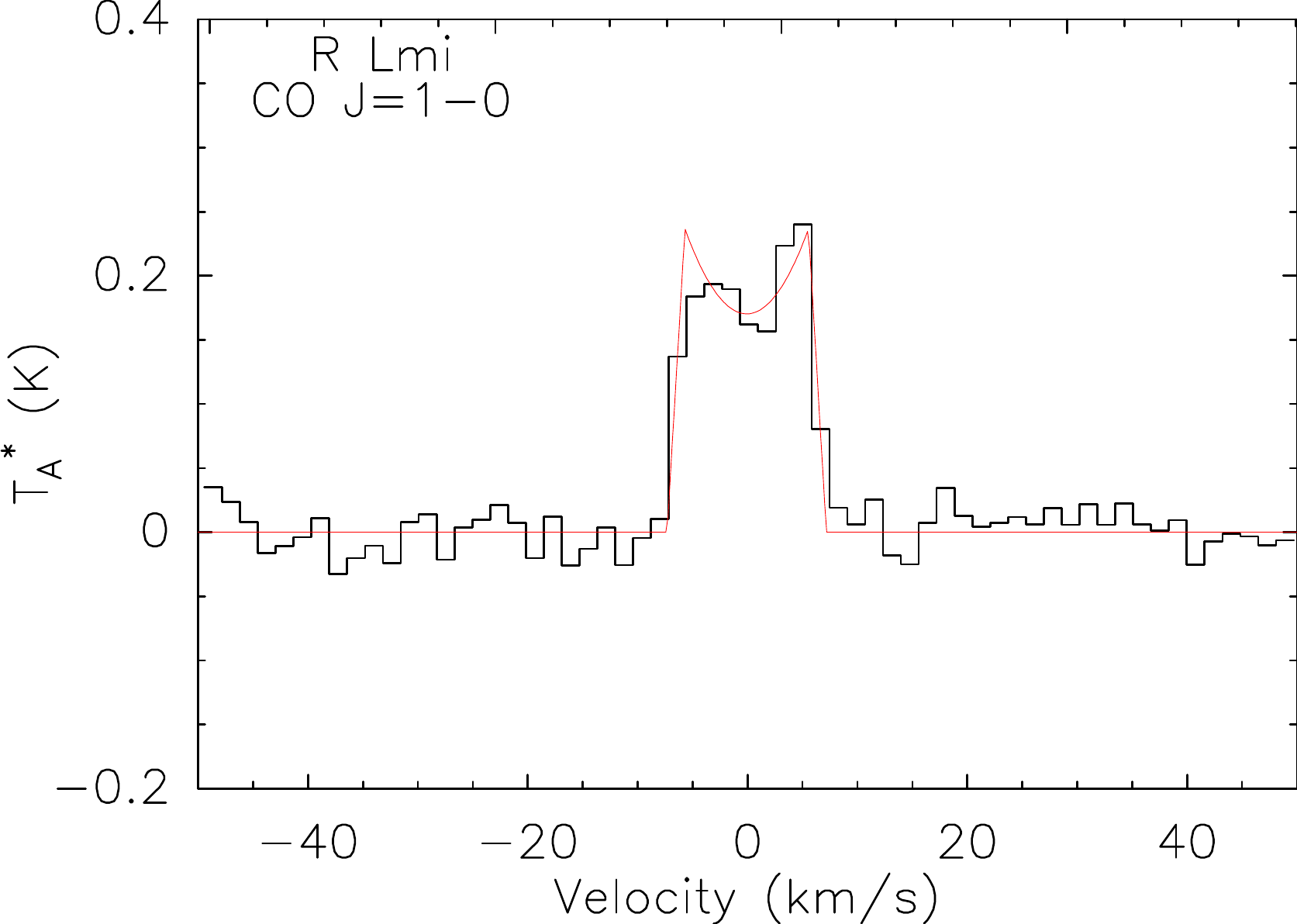}
     \end{subfigure}
      \begin{subfigure}[b]{0.31\linewidth}
         \centering
         \includegraphics[width=\linewidth]{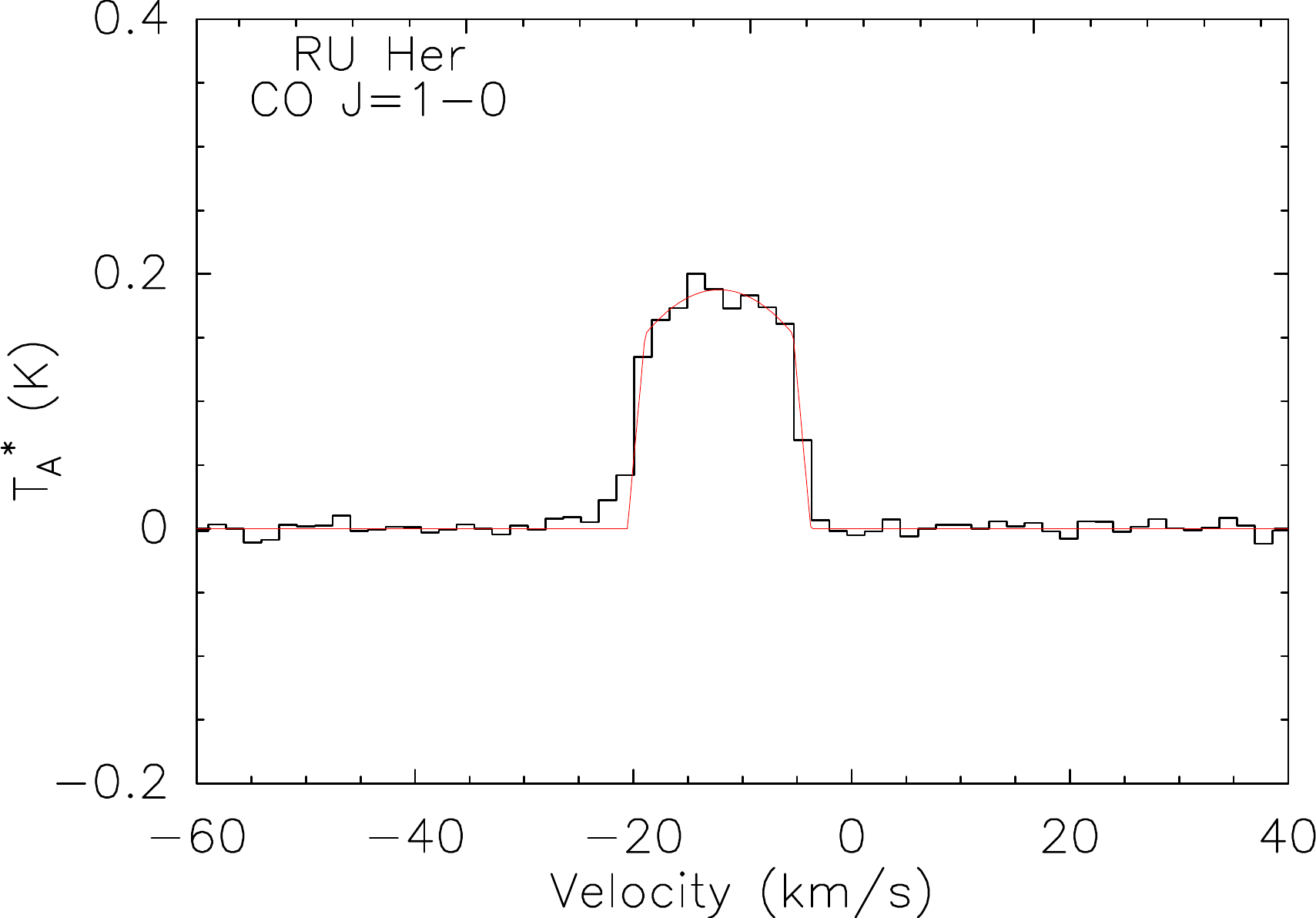}
     \end{subfigure}
     \begin{subfigure}[b]{0.31\linewidth}
         \centering
         \includegraphics[width=\linewidth]{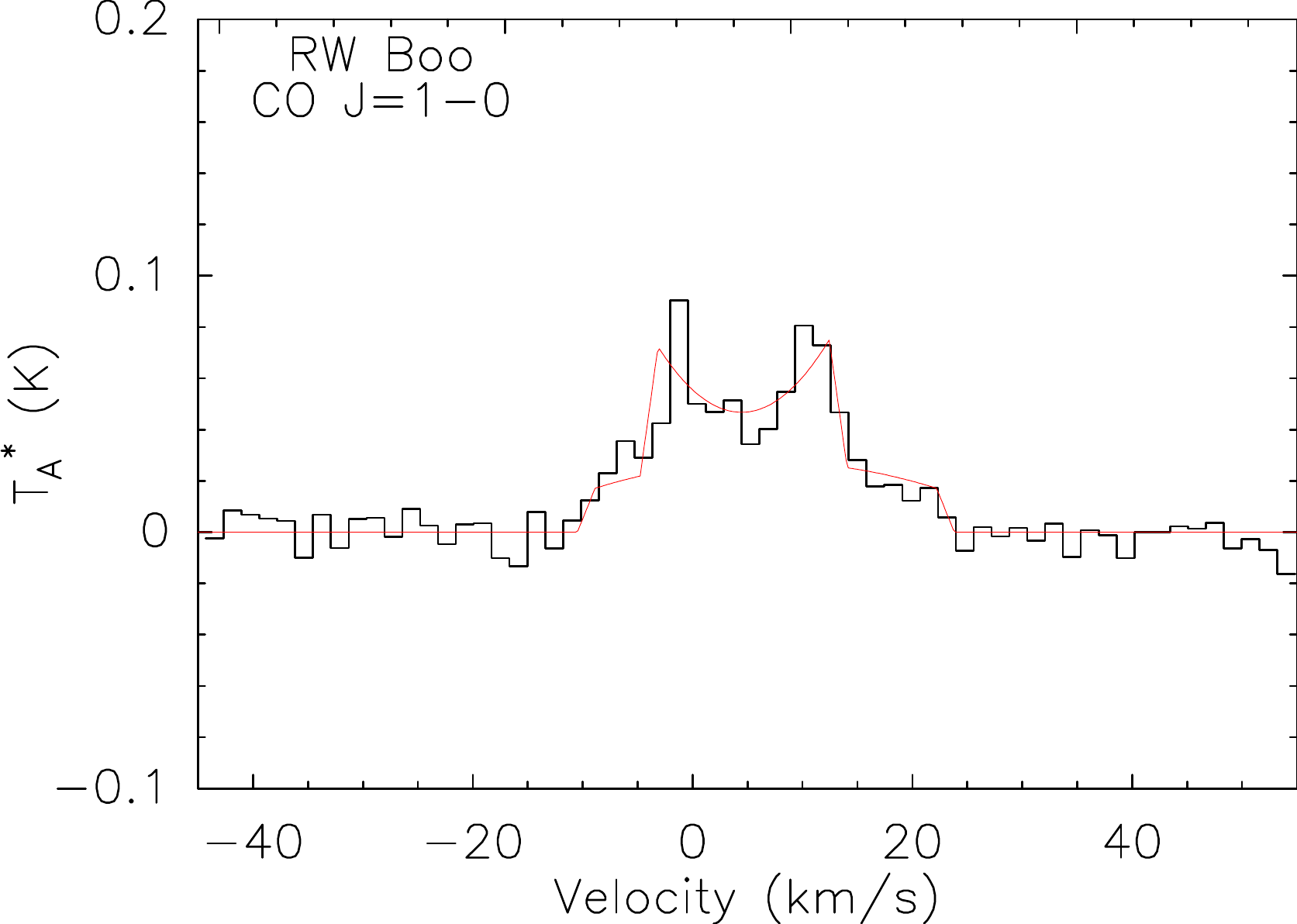}
     \end{subfigure}
     \begin{subfigure}[b]{0.31\linewidth}
         \centering
         \includegraphics[width=\linewidth]{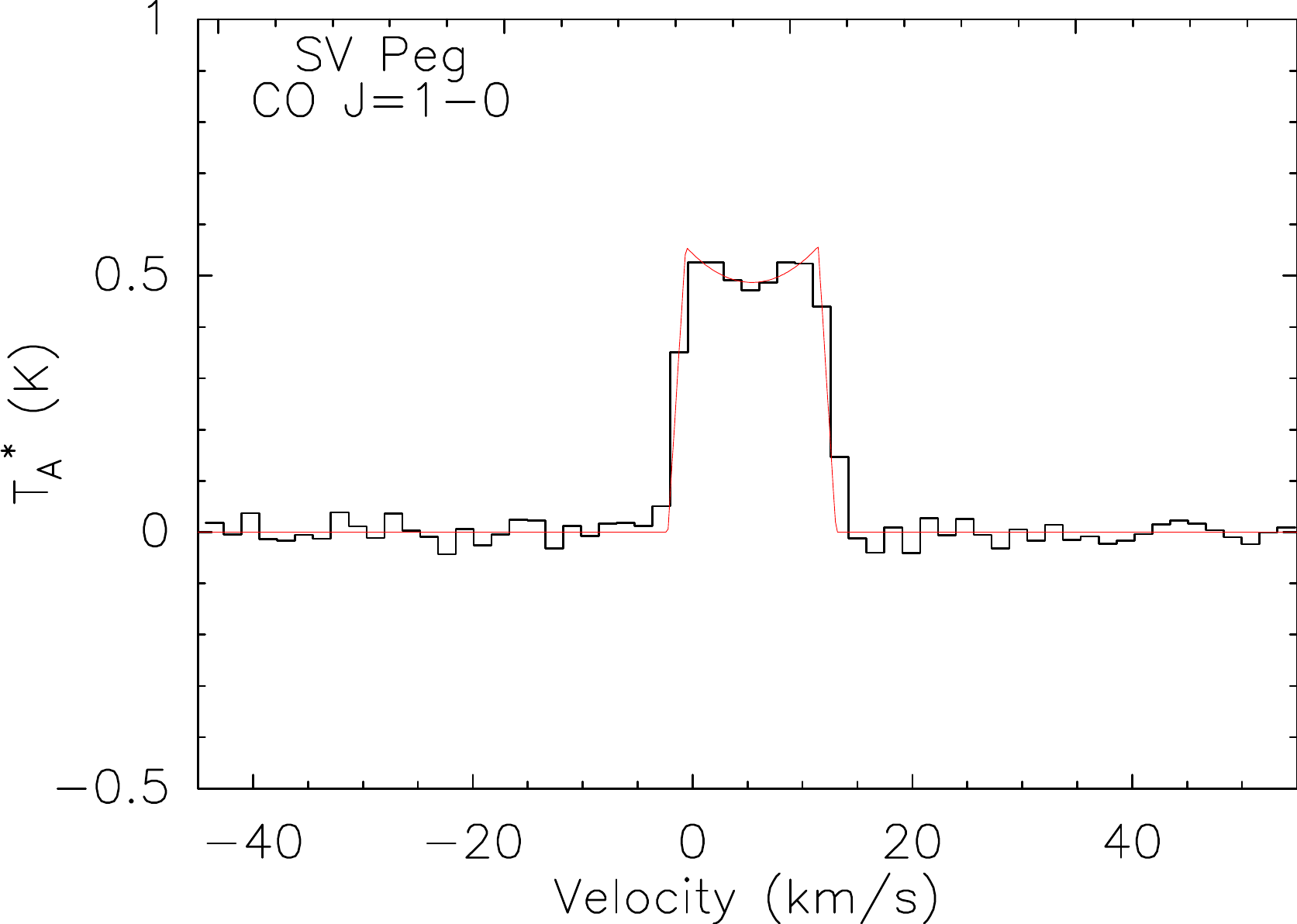}
     \end{subfigure}
     \begin{subfigure}[b]{0.31\linewidth}
         \centering
         \includegraphics[width=\linewidth]{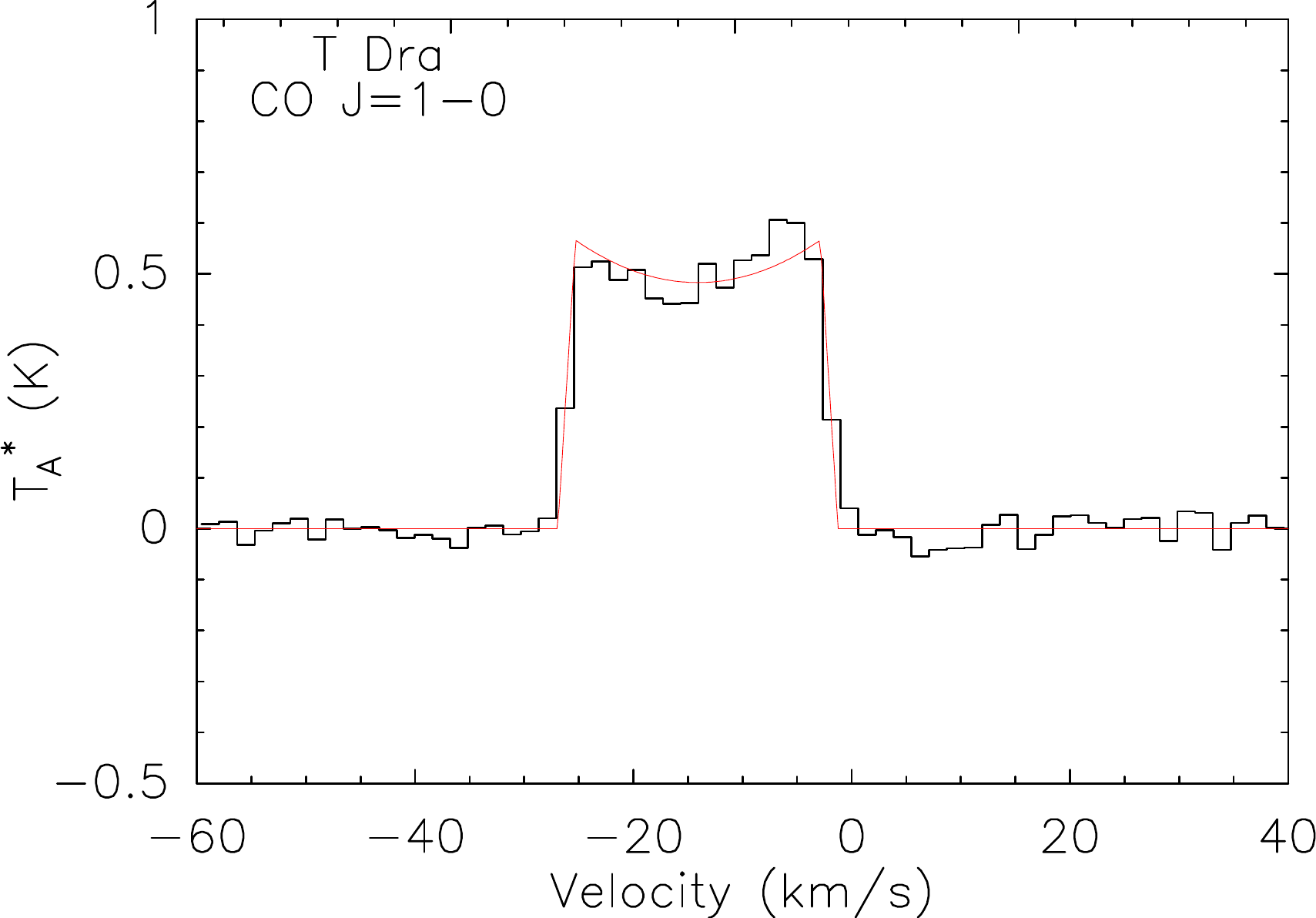}
     \end{subfigure}
     \begin{subfigure}[b]{0.31\linewidth}
         \centering
         \includegraphics[width=\linewidth]{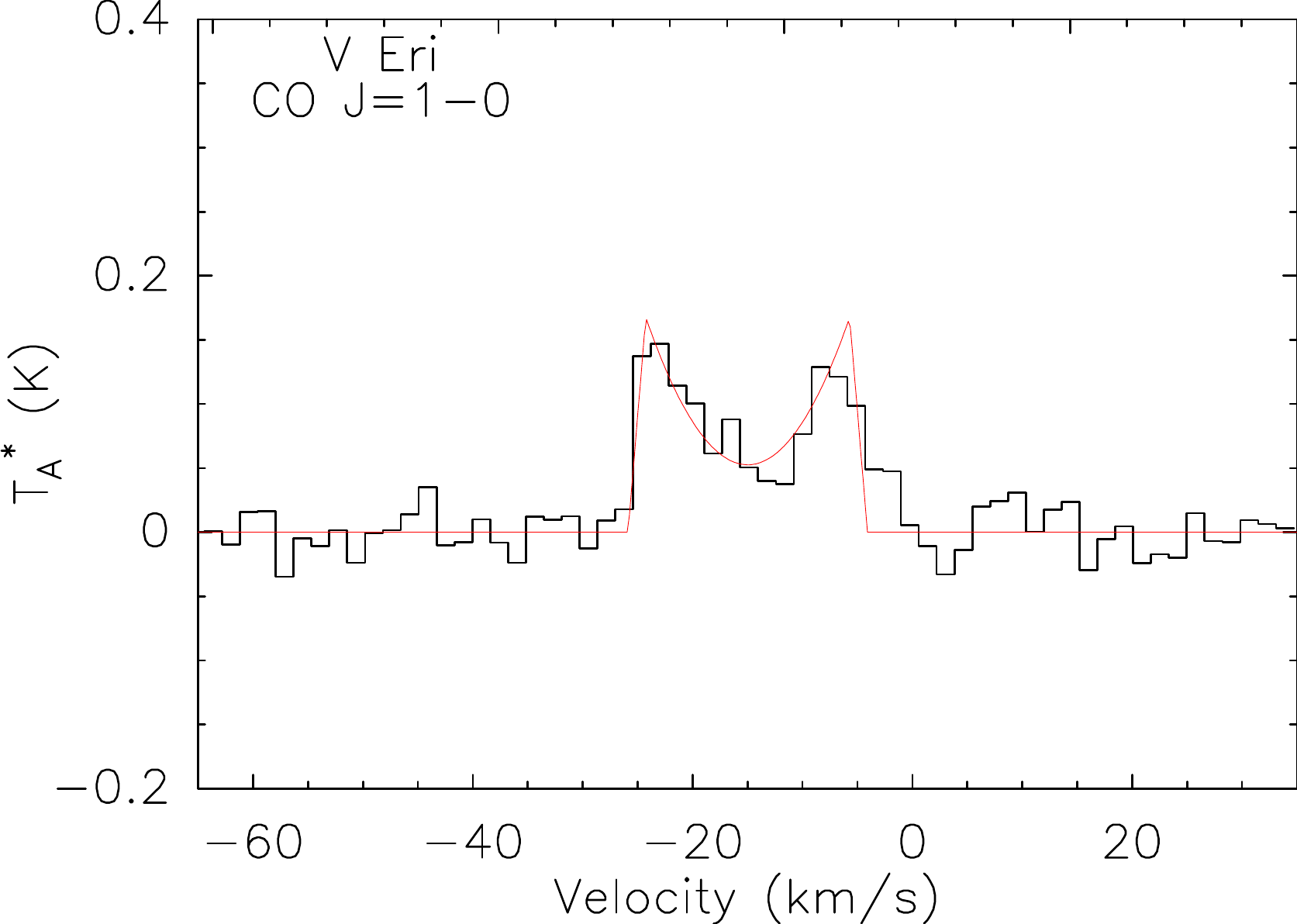}
     \end{subfigure}
     \begin{subfigure}[b]{0.31\linewidth}
         \centering
         \includegraphics[width=\linewidth]{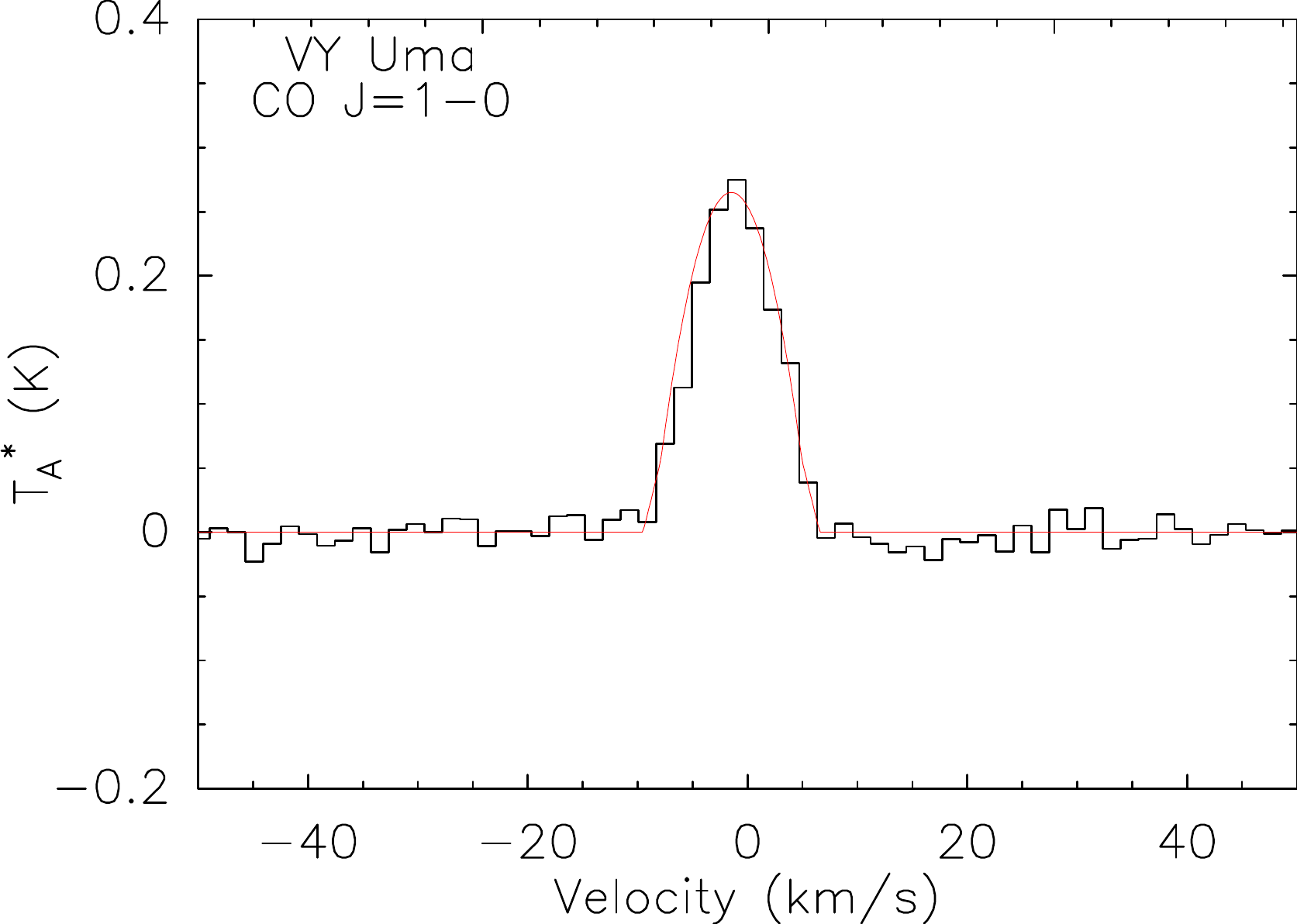}
     \end{subfigure}
     \begin{subfigure}[b]{0.31\linewidth}
         \centering
         \includegraphics[width=\linewidth]{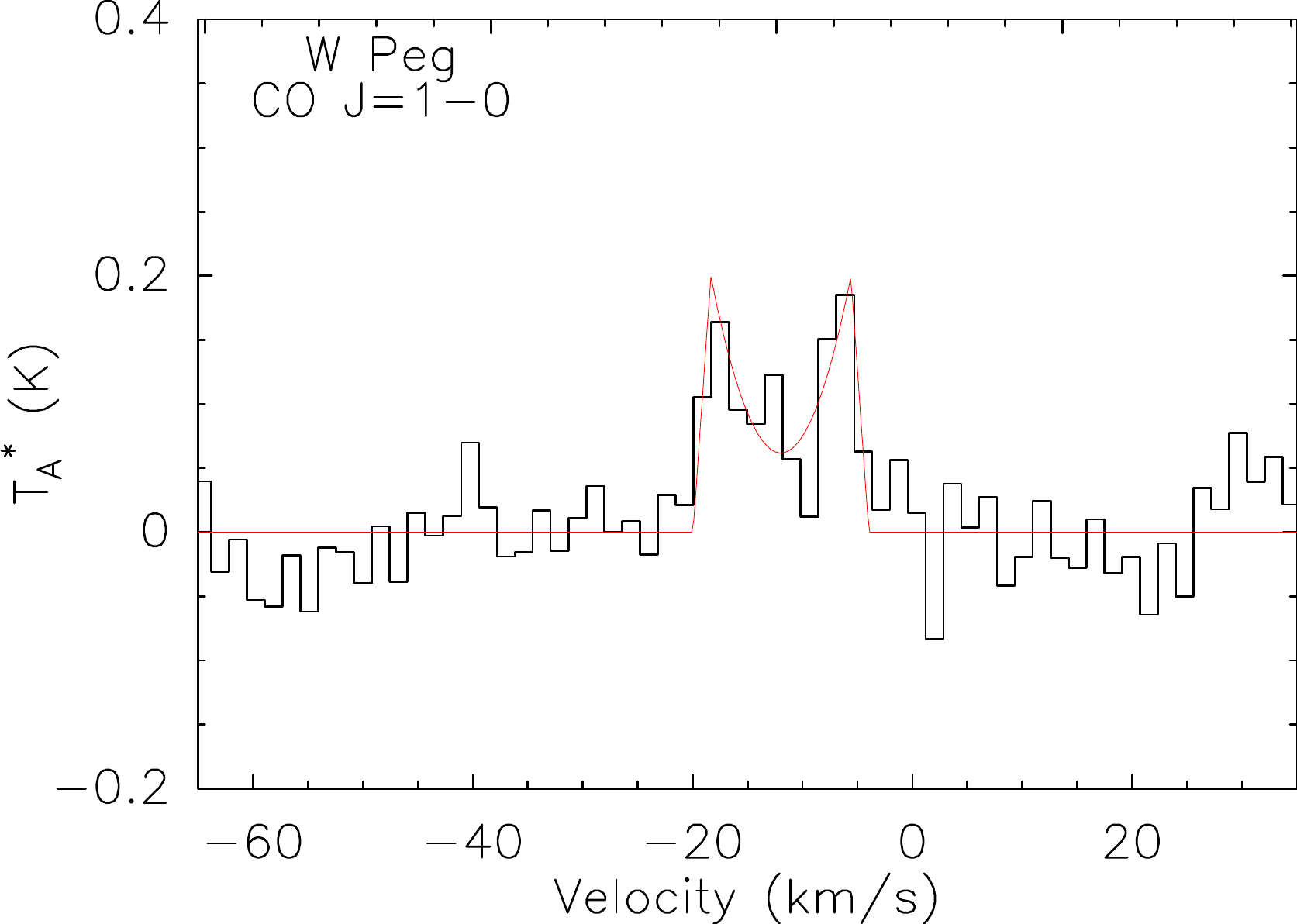}
     \end{subfigure}
     \begin{subfigure}[b]{0.31\linewidth}
         \centering
         \includegraphics[width=\linewidth]{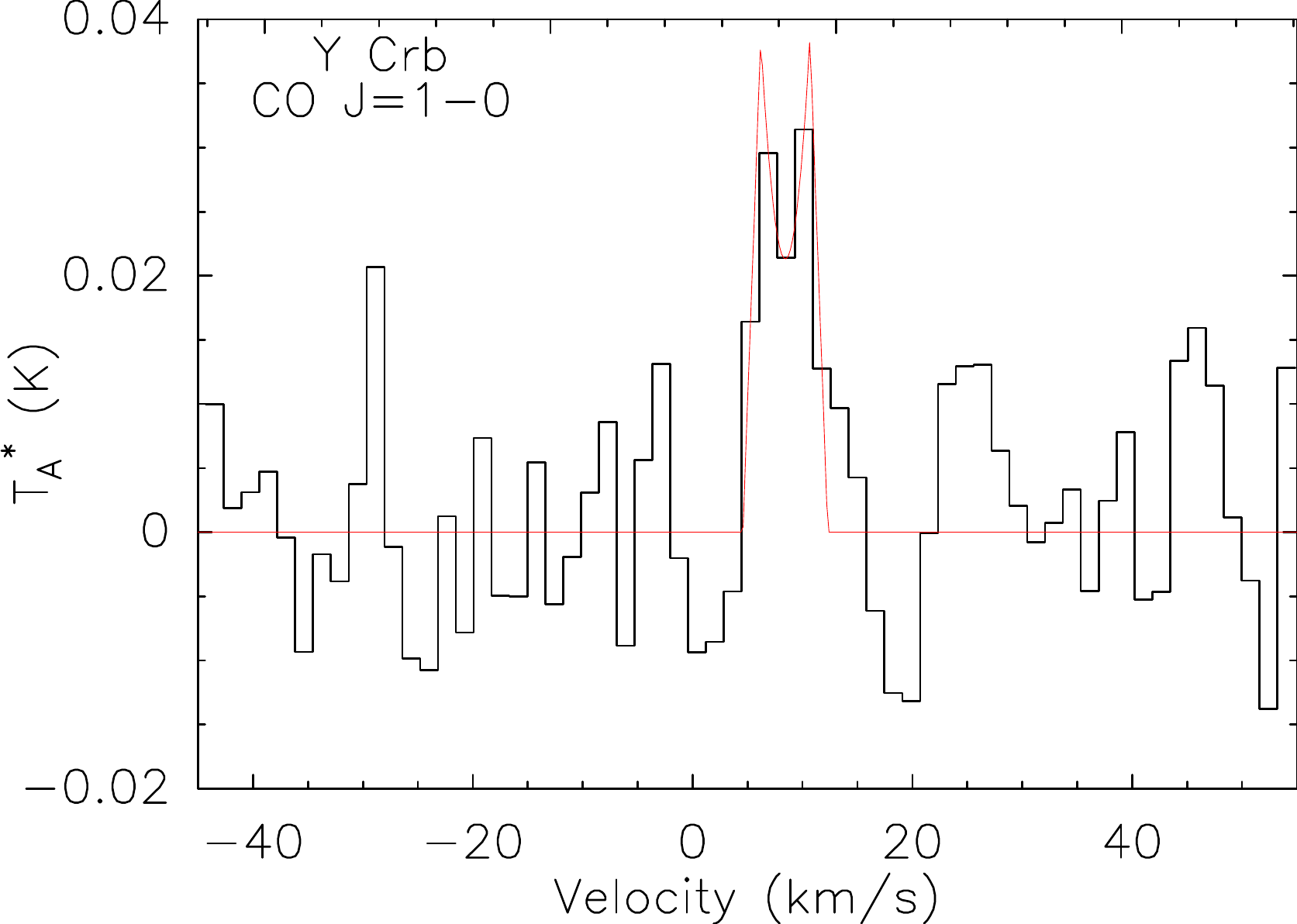}
     \end{subfigure}
        \caption{same as Fig.~\ref{fig:2-1_spectra} for the 10 sources detected in the \doceuno \, transition.}
        \label{fig:1-0_spectra}
\end{figure*}

We have searched for \doceuno\ \, and \docedos\ \, emission in a sample of 29 uvAGB binary candidates. Line emission was detected in 15 targets, 11 of them in both transitions. The observed spectra are presented in Figs. \ref{fig:2-1_spectra} and \ref{fig:1-0_spectra} for \docedos \, and \doceuno \, respectively. The main line measurements directly obtained from the observations are summarised in Table~\ref{tab:spectral_analysis}.

\subsection{Line profiles and expansion velocities}\label{sec:line_profiles}

The line profiles observed in our targets exhibit a range of shapes, including (pseudo) parabolic, double-horned, flat, triangular, Gaussian-like profiles, etc. Some objects also show profile asymmetries, such as horns with different intensities or quasi-triangular profiles with one side, in some cases the red-shifted one, being less intense. In the case of RW\,Boo, a double-component profile has been found as it had been previously reported by \cite{diaz-luis_2019}. The observed line profiles in our predominantly O-rich targets do not closely resemble the canonical profiles expected for the so-called standard CSE model, which assumes isotropic flow at a constant velocity. In the standard model, the line profile can vary between parabolic, flat, and double-horned shapes, depending on the line optical depth and the size of the envelope relative to the telescope beam \citep{Olofsson_1993}. However, this deviation from the idealised CSE model is a known property of O-rich CSEs, as discussed in previous studies by \cite{Margulis_1990, Knapp_1998}. These deviations can be attributed to geometric or velocity effects in the CSEs or multiple mass-loss components.

The \docedos\ and \doceuno\ line profile have enabled us to  estimate two key line parameters: the integrated flux and the average expansion velocities of the envelopes. These parameters were determined by fitting a shell-type profile using the software CLASS\footnote{\url{http://www.iram.fr/IRAMFR/GILDAS}} where the line shape is parametrised as 

\begin{equation}
    f(\nu)=\frac{A}{\Delta \nu}\frac{1+ 4 H [(\nu - \nu_{0})/\Delta \nu]^{2}}{1+H/3},
\end{equation}

\noindent
where $A$ is the integrated area, $\nu_{0}$ is the central frequency, $\Delta \nu$ is the full width at zero intensity, and $H$ is the horn-to-centre ratio of the line. The equivalent expansion velocity is

\begin{equation}
    \vexp=c \frac{\Delta \nu/2}{\nu_{0}},
\end{equation}

\noindent
where $c$ is the speed of light.

Additionally, we determined the full width at half maximum (FWHM) of the lines by fitting a Gaussian profile to each detected transition. For RW\,Boo, the \docedos\ and \doceuno\ lines exhibited a complex structure. In this case, the expansion velocities were estimated by combining two shell profiles: the primary component represented a two-horn profile, while the secondary component corresponded to a parabolic profile.

The line parameters derived as just described are listed in Table~\ref{tab:spectral_analysis}. 
The distribution of expansion velocities for our targets is depicted in Figure \ref{fig:histograms}. The majority of the sources exhibit moderate expansion velocities (\vexp\,$\sim$\,10\,\kms), which aligns with typical values observed in AGB stars. A few sources display narrower line profiles, indicating lower expansion velocities (\vexp\,$\lesssim$\,5\,\kms). These specific sources include R Uma, RZ Uma, Z Cnc (identified only in the \docedos\,transition), and Y Crb (identified in both lines).

\begin{table*}[h!]

\renewcommand{\arraystretch}{1.2}
\centering

\begin{adjustbox}{max width=\textwidth, height=0.95\textheight,keepaspectratio}

\begin{threeparttable}[b]

\caption{Spectral measurements for detected sources.}

\begin{tabular}{l c c c c c c c}
\hline\hline 
Source & line & $\int T_{A}^{*}dv$    & $\mathrm{T_{A}}$ & rms  &  $\mathrm{V_{sys}}$ & $\mathrm{V_{exp}}$  & FWHM \\ 
 & &  $(\mathrm{K\, km\, s^{-1}})$ &  $(\mathrm{mK})$ & $(\mathrm{mK})$ &  $(\mathrm{km\, s^{1}})$ &  $(\mathrm{km\, s^{1}})$ &  $(\mathrm{km \, s^{-1}})$ \\
\hline 

\noalign{\vspace{0.25cm}}

EY Hya & \doceuno & $1.12\pm0.10$ & $ 62$ & $11$ & $-24.6$ & $10.1$ & $14.7$  \\

EY Hya & \docedos & $4.74\pm0.07$  & $ 290$ & $7$& $-21.5$ & $10.7$  & $14.6$  \\

\noalign{\vspace{0.25cm}}

R Lmi & \doceuno & $2.42\pm0.18$ & $166$  & $18$& $-1.7$ & $6.4$  & $11.6$     \\

R Lmi & \docedos & $9.5\pm0.2$  & $740$ & $20$& $-2.0$ & $6.7$  & $11.2$       \\

\noalign{\vspace{0.25cm}}

R Uma & \doceuno  & $<0.22$ &   & $12$ &   &   &       \\

R Uma & \docedos & $0.40\pm0.09$ & $50$ & $9$& $-39.0$ & $4.7$  & $6.2$      \\

\noalign{\vspace{0.25cm}}

RR Eri & \doceuno  & $<0.72$ &   & $15$ &   &   &       \\

RR Eri & \docedos & $2.94\pm0.09$ & $171$ & $9$& $22.4$ & $12.0$  & $16.0$    \\

\noalign{\vspace{0.25cm}}

RT Cnc & \doceuno  & $<0.44$ &   & $9$ &   &   &       \\

RT Cnc & \docedos & $0.76 \pm0.08$  & $ 47$ & $8$& $-33.0$ & $7.2$  & $11.7$   \\

\noalign{\vspace{0.25cm}}

RU Her & \doceuno& $2.77\pm0.05$  & $187$  &  $5$& $12.2$ & $7.6$  & $12.8$ \\

RU Her & \docedos & $6.19 \pm0.06$  & $422$ & $6$& $12.0$ & $7.5$  & $12.7$    \\

\noalign{\vspace{0.25cm}}

RW Boo & \doceuno & $1.29\pm0.05$ & $ 38$ & $5$& $-5.1$ & $17.3$  & $20.3$ \\

 & & & $ 3$ & & & $7.7$ & \\

RW Boo & \docedos & $2.07 \pm 0.06$ & $ 54$ & $6$& $-4.4$ & $15.6$  & $19.8$   \\

& & & $ 29$ & & & $8.4$ &       \\

\noalign{\vspace{0.25cm}}

RZ Uma & \doceuno  & $<0.26$ &   & $11$ &   &   &       \\

RZ Uma & \docedos & $0.53 \pm0.10$  & $66$ &  $10$& $28.6$ & $4.7$  & $7.7$    \\

\noalign{\vspace{0.25cm}}

SV Peg & \doceuno & $7.48\pm0.19$  & $507$ &  $19$& $-5.3$ & $7.3$  & $12.5$  \\

SV Peg & \docedos & $13.9\pm0.4$ & $1120$ & $40$& $-5.4$ & $7.2$  & $11.3$      \\

\noalign{\vspace{0.25cm}}

T Dra & \doceuno & $12.1\pm 0.2$ & $490$ & $20$& $13.2$ & $12.0$  & $20.8$  \\

T Dra & \docedos & $14.18 \pm 0.10$ & $639$  & $10$& $13.9$ & $11.5$  & $19.4$ \\

\noalign{\vspace{0.25cm}}

V Eri & \doceuno & $2.29\pm0.17$ & $53$ & $17$& $14.1$ & $10.1$  & $22.9$  \\

V Eri & \docedos & $9.55\pm0.11$ & $393$ & $11$& $15.6$ & $10.4$  & $19.3$  \\

\noalign{\vspace{0.25cm}}

VY Uma & \doceuno & $2.52\pm0.10$ & $256$ &  $11$& $1.2$ & $6.7$  & $8.5$     \\

VY Uma & \docedos & $5.14\pm0.10$ & $579$ & $10$& $0.9$ & $6.5$  & $8.0$     \\

\noalign{\vspace{0.25cm}}

W Peg & \doceuno & $1.9\pm 0.3$   & $80$ & $30$& $15.7$ & $7.7$  & $16.3$    \\

W Peg & \docedos & $4.0\pm 0.7 $   & $370$ & $70$& $16.6$ & $7.1$  & $9.9$  \\

\noalign{\vspace{0.25cm}}

Y Crb & \doceuno & $0.15\pm0.08$  & $21$ & $8$& $-6.9$ & $3.1$  & $6.0$  \\

Y Crb & \docedos & $0.33\pm0.15$  & $58$ & $16$& $-4.3$ & $3.9$  & $6.0$ \\

\noalign{\vspace{0.25cm}}

Z Cnc & \doceuno  & $<20$ &   & $10$ &   &   &       \\

Z Cnc & \docedos & $0.42\pm0.08$ & $55$ & $8$& $9.3$ & $3.9$  & $6.7$  \\

\noalign{\vspace{0.25cm}}

\hline
\end{tabular}
\label{tab:spectral_analysis}
\begin{tablenotes}
\item \textbf{Notes.} Column (1): source, Col. (2): transition, Col. (3): area of the observed line profil, Col. (4): antenna temperature at the central velocity of the fit to the line profile, Col. (5): noise, Col. (6): Systemic velocity (line centroid) in the local standard of rest (LSR) frame, Col. (7): expansion velocity of the shell fit, Col. (8): Full Width at a Half Maximum of the Gaussian fit.
\end{tablenotes}

\end{threeparttable}
\end{adjustbox}

\renewcommand{\arraystretch}{1.0}

\end{table*}

\subsection{CO detection statistics} \label{detection_statistics}

Regarding the variability type, in terms of CO detection fractions, we observed the following trends: 83\% for Miras, 56\% for SRs, and 14\% for LB. These ratios are slightly lower compared to previous CO surveys of AGBs, where detection rates were around 90\% for Miras, 70\% for SRs, and 50\% for LB \citep[see e.g.][]{Margulis_1990,Kerschbaum_1999}. This effect, which is particularly notable for LB stars, could suggest that stars with UV excess have a lower likelihood of having CO-rich circumstellar envelopes compared to normal AGB stars. 
However, the relative order of detection ratios among the variability classes remains consistent.

The CO detection statistics is 50\% (13/26) among O-rich stars and 100\% (2/2) among C-rich stars. Since our sample is strongly biased towards O-rich stars, which represent 26 (90\%) of the targets, the CO detection statistics depending on chemistry cannot be considered meaningful, although it does go in the same direction as in many previous works finding a larger CO detection fraction among C-rich AGB stars than in other chemistry types \citep[e.g.][]{knapp1985}.

\begin{figure*}[h!]
    \centering
    \includegraphics[width=\linewidth]{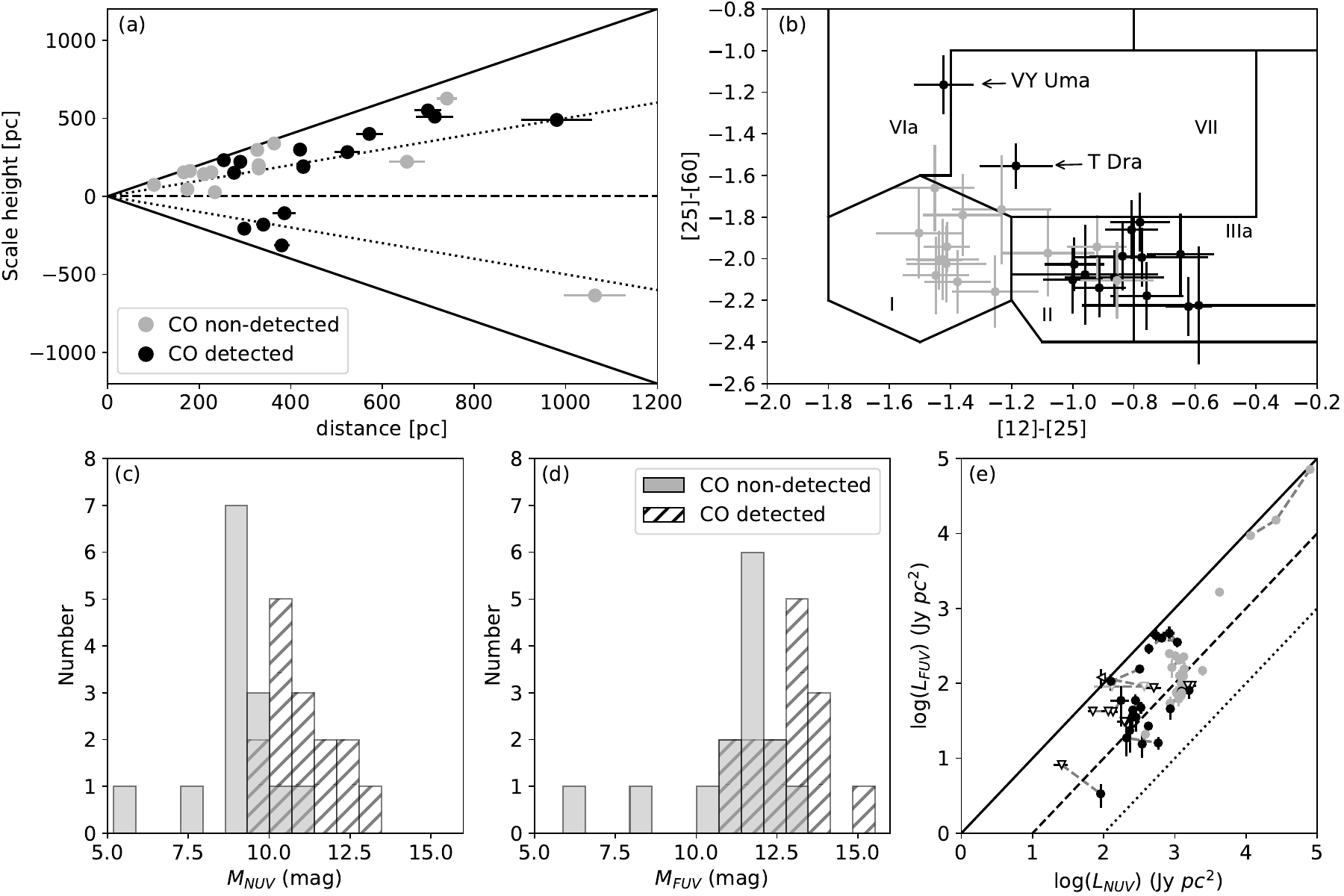}
    \caption{ Comparison between different parameters and the $^{12}$CO detectability of our targets.\protect\\
    (a) Scale height vs distance for our targets. The solid line represents a viewing angle of $\pm 90^{\circ}$, the dotted line represents a viewing angle of $\pm 30^{\circ}$, and the dashed line represents a viewing angle of $0^{\circ}$. The non-detected in $^{12}$CO are shown as red circles, sources detected in \docedos \, line are shown as green squares, and sources detected in both $^{12}$CO lines are shown as blue stars.\protect\\
    (b) IRAS [25]-[60] vs [12]-[25] two-colour diagram restricted to those regions in which our sources are located (i.e. regions I, II, IIIa, VIa, and VII). The colour scheme is the same as (a). \protect\\
    (c) Histogram with the distribution of the NUV magnitudes of our sample. The sources detected in $^{12}$CO are shown in white and sources non-detected in $^{12}$CO are shown in grey.\protect\\
    (d) Same as (c) for FUV magnitudes.\protect\\
    (e) Comparison between NUV and FUV luminosities.  The solid line represents the relation \rfuvnuv=1, the dashed line represents the relation \rfuvnuv=0.1, and the dotted line represents the relation \rfuvnuv=0.01. Solid circles correspond to well measured luminosities and empty triangles to upper limits in one of the luminosities; the colour scheme is the same as (a).}
    \label{fig:our_sample_distances}
\end{figure*}

Figure~\ref{fig:our_sample_distances} illustrates the detectability of $^{12}$CO in our sample, highlighting the comparison between different parameters to identify any discernible trends.
Figure~\ref{fig:our_sample_distances} (a) shows that CO detections are not strongly biased towards the nearest targets, but spread over a wide range of distances between 300 and 1000\,pc. In fact, there is a significant number of CO non-detections at distances lower than 300\,pc.

Figure~\ref{fig:our_sample_distances} (b) shows that the O-rich AGB stars of our sample located in regions I and VII were not detected, this is good agreement with the location of AGBs with thinnest CSEs in regions I or close to it \citep[see][]{van_der_veen_1988}, whereas most of the targets located in region II and all the targets located in region IIIa were detected. On the other hand, the C-rich AGB stars of our sample were detected (VY\,Uma and T\,Dra, which are located in regions VIa and VII respectively).

In Sect.~\ref{CO_IRAS60}, we specifically examine the correlation between CO and IRAS flux. As anticipated, we observe a proportional relationship between these fluxes, consistent with most of the sources with CO detections being characterised by high IRAS brightness ($\fsixty > 5$ Jy). with the exception of Y Crb, which has $\fsixty \sim 1.8$ Jy.

Since the main difference between our sources and most AGB stars is their UV excesses, we have analysed the distributions of NUV and FUV magnitudes in relation with the CO detection. Figs.~\ref{fig:our_sample_distances} (c) and (d) show the distributions of the NUV and FUV time-averaged magnitudes ($\rm M_{NUV}$ and $\rm M_{FUV}$ respectively) without reddening correction for the CO detected and non-detected sources. It is apparent that sources with CO detections have statistically higher UV magnitudes (lower UV fluxes) than those without CO detections. This difference is expected due to the fact that UV extinction is higher in envelopes with higher densities, we studied this anti-correlation with more detail in Sect.~\ref{trends}.

Figure~\ref{fig:our_sample_distances} (e) illustrates the comparison between FUV and NUV luminosities ($L_{\rm FUV}$ and $L_{\rm NUV}$, respectively) without reddening correction for CO detections and non-detections. The relationship between the two, NUV and FUV, luminosities appears to be proportional, consistent with previous studies \citep[see fig. 4 of][]{Ortiz_2019}. However, some sources observed at different epochs exhibit variations that deviate from this correlation. These variations result in significant changes in the \rfuvnuv \,ratio, with some sources showing an increase in one flux and a decrease in the other. Moreover, the sources with highest GALEX luminosities (i.e BD Cam and Y Gem) also show a high \rfuvnuv \,ratio (\rfuvnuv\,$>$\,0.2), whereas our entire sample cover a range \rfuvnuv\,$\simeq$\,0.06-1.0.
Without ignoring these considerations, the distribution does not reveal a clear trend between CO detections and the FUV/NUV ratio. However, it is evident that CO detections are predominantly found among sources with lower UV excess, as already deduced from Figs~\ref{fig:our_sample_distances} (c) and (d). 

\section{Analysis: CO population diagrams}\label{anal}

The main goal of this work is to characterise the molecular envelopes
of uvAGB stars (i.e. AGB binary candidates) as a class. In this section we
present our data analysis methodology to constrain the most relevant envelope parameters,
including the mass-loss rate of the AGB wind, and the results obtained.

The CO rotational lines provide a fundamental diagnostic of the
physical conditions of the molecular gas. As a first approximation,
we have applied the classical population (or rotational) diagram method, explained in \cite{goldsmith_1999}, to estimate the beam-average excitation temperature (\tex) and the $^{12}$CO column density ($N_{\rm CO}$) of the CSEs layers where the \doceuno \, and \docedos \, transitions observed are predominantly produced. This method has been successfully used in the analysis of the molecular emission from the envelopes of many evolved stars \citep[e.g.][]{ramos-medina_2018,Justtanont_2000}. 

The population diagram method relies on two main hypothesis: (i) optically thin emission and (ii) local thermodynamic equilibrium (LTE) conditions. As explained in \cite{goldsmith_1999}, under these two conditions, the column density of CO molecules in the upper state ($N_{\rm u}$) and its energy above the ground state ($E_{\rm u}$) are related by the following equation:

\begin{equation}
    ln \, \frac{N_{u}}{g_{u}} =  \ln \, \frac{3 \kb W}{8 \pi^{3} \nu S_{\rm ul} \mu^{2}} = ln \, \mathrm{N}  -ln \, C_{\tau} - ln \, 
    \left( \frac{\Delta \Omega_{a}}{\Delta \Omega_{s}}\right) - ln \, Z - \frac{E_{u}}{k_{B} \tex}.
    \label{eq:rotdiagram}
\end{equation}

\noindent
Here $g_{u}$ is the degeneracy of the upper level, Z is the partition function of the molecule, \kb\, the Boltzmann constant, 
$\nu$ is the rest frequency of the transition, S$_{\rm ul}$ is the line strength and $\mu$ is the permanent dipole moment of the molecule, $\mathrm{N}$ is the column density of the molecule (in our case ${N_{\rm CO}}$), $W=\int T_{A}^{*}dv$ is the integrated area of each spectral line, and $\Delta \Omega_{\rm a}$ and $\Delta \Omega_{\rm s}$ are the antenna and the source solid angle, respectively. The beam filling factor, defined as the ratio $(\Delta \Omega_{a}/\Delta \Omega_{s})$, has been estimated assuming a Gaussian distribution for both the antenna beam pattern and the source brightness distribution. Eq.~\ref{eq:rotdiagram} also incorporates the so-called opacity correction factor defined by \cite{goldsmith_1999} as $C_{\tau} =  \frac{\tau}{1-e^{-\tau}}$, where $\tau$ is the line peak optical depth. As discussed by these and other authors, this correction is valid as long as the lines are not very opaque ($\tau \lesssim 1$), which is the case of our sources (see below in this section).

According to Eq.~\ref{eq:rotdiagram}, a straight-line fit to the points in the population diagram provides \tex\ from the slope of the fit and $N$ from the y-axis intercept. The total CO column density has been converted to the total (H$_2$) mass in the CO-emitting volume for our targets in a simplified way as 

\begin{equation}
    \mathrm{M}=\mathrm{m_{H_{2}}} \Omega_{s}D^{2} \frac{N_{\rm CO}}{\xco}, 
\end{equation}

\noindent
where $\mathrm{m_{H_{2}}}=3.32\times 10^{-27}$ kg is the mass of the $\mathrm{H}_{2}$ molecule and $\xco=N_{\rm CO}/N_{\rm tot}$ is the $^{12}$CO-to-H$_2$ fractional abundance, assumed to be $\xco=2\times 10^{-4}$ for O-rich stars and $\xco=8 \times 10^{-4}$ for C-rich stars \citep{ramos-medina_2018,da_Silva_Santos_2019}. Assuming constant expansion velocity and using the values of the total emitting mass, we calculate the
mass-loss rates (\mloss) for each source in a simplified manner dividing the total mass by the crossing time of the CO-emitting layers, which is computed as the ratio between the characteristic radius of the CO envelope and the expansion velocity ($\mathrm{t_{exp}}=\rs/$\vexp).

The  extent of the \doceuno\ and \docedos\ emitting region in our targets is unknown a priori; however, it is a critical parameter to constrain the main envelope properties (\tex\ and $N$) using the population diagram technique described above. 
In this work, we have obtained a self-consistent estimate of the mass-loss rate and of the characteristic radius (\rs), considering that, as empirically demonstrated by \cite{Ramstedt_2020}, \rs\ is approximately 2-3 times smaller than the CO photodissociation radius (\rco) as obtained following the formulation by \cite{Mamon_1988}. We follow an iterative population diagram fitting process, starting by adopting a first value of \rs\, as input to derive a first guess of \mloss\ (computed as described above). This value of \mloss\ is then used to estimate \rco, which is given by 
\begin{equation}
    \rco= 7.3 \times 10^{16} \left( \frac{\Dot{M}}{10^{-6}} \right)^{0.58}\left( \frac{10}{\vexp} \right)^{0.4}\left( \frac{\xco}{4 \times 10^{-4}} \right)^{0.5}
,\end{equation}
\noindent
following the formulation by \cite{Mamon_1988} \citep[see also][]{Planesas_1990}. 
This new value of \rco\ is used to compute a new value of \rs, which is then used as a new input value to derive again \tex, $N$, and \mloss. This process is repeated until the input and output \rs\, converge to the same value.

The population diagrams of our sources, with the corresponding linear fits obtained once the iterative process we have just described has converged, are plotted in 
appendix~\ref{rotdiagrams}. The values of the different envelope parameters derived are given in Table~\ref{tab:parameters} and presented and discussed in the next section. For sources where only the \docedos \, transition is detected, the column density has been estimated assuming an excitation temperature of 10\,K, which is similar to the mean value of \tex\ derived for targets with detections in both \docedos \, and \doceuno. However, in the case of RR Eri, the \docedos \,/\doceuno \, line ratio was not consistent with \tex=10\,K but indicated a higher value of \tex=25\,K that has then been adopted.

From our analysis, we found optical depth values $\tau << 1$ for all sources except for T\,Dra, VY\,Uma and Y\,Crb, for which the CO peak-line optical depth takes values close to one for the \docedos\ line. In these cases, the corresponding opacity-correction factor applied to the envelope mass and, thus, to the mass-loss rate is still moderate, of $\sim1.6$.

We conducted a comparison between our estimated mass-loss rates (\mloss) and those previously reported in the literature (\mloss$_{\rm lit}$) for those sources for which this is possible (a total of 11 sources). Some of these prior estimations employed more detailed analysis, such as radiative transfer models. By examining this comparison, documented in Appendix~\ref{mloss_comp}, we found that the \mloss\ values presented in this paper exhibit good agreement with those from previous studies. Therefore, despite relying on an approximate method, the population diagram yields relatively accurate results. This can be attributed to the effetive thermalisation or near-thermalisation of CO molecules in the CSEs, along with the low to moderate CO line opacities in our targets \citep[for a more extensive discussion on this matter see][]{ramos-medina_2018}. We estimate that our \mloss\ values are accurate within a factor of approximately 3-4.

Considering the typical values of the radius of the CO-emitting layers found for our targets and the corresponding distances, we estimate angular diamaters ranging from approximately 1\farc2 to 10\arcsec, that is, smaller than the telescope beam ($\sim$22\arcsec\, and 11\arcsec\, at 3 and 1\,mm, respectively). Only T\,Dra and SV\,Peg, which possess larger sizes compared to the other sources, may exhibit partial resolution within the telescope beam. 

\begin{table*}[t]

\renewcommand{\arraystretch}{1.3}
\centering

\caption{Parameters estimated for the circumstellar envelopes.} 
\label{tab:parameters}

\begin{adjustbox}{max width=\textwidth}
\begin{threeparttable}[b]

\begin{tabular}{l c c c c c c c c c c c}
\hline\hline 
Source & $\mathrm{T_{rot}}$    & $\mathrm{N_{tot}}$ & \rsou & \rs& \rco & M &  $\Dot{M}$ & $\mathrm{v_{exp}}$ & $\mathrm{t_{exp}}$ & E(B-V) (CSM) & \textbf{$A_{V}$ (CSM)}\\  
 & $(\mathrm{K})$ & $(\mathrm{10^{16} cm^{-2}})$ & $(^\mathcal{{\prime \prime}})$ & $(\mathrm{10^{16} cm})$ & $(\mathrm{10^{16} cm})$ & $(\mathrm{10^{-4}     M_{\odot}})$ & $(\mathrm{10^{-7}M_{\odot} \, yr^{-1}})$ & $(\mathrm{km \, s^{-1}})$ & $(\mathrm{yr})$ & (mag) & \\  
\hline 

EY Hya & $13 \pm 6 $ & $2.9 \pm 1.0$ & $2.3$ & $1.5$ & $3.7$ & $2.7$ & $5.8$ & $10.4 \pm 0.3$ & $500$ & 0.026 & 0.0084\\               

T Dra & $5.8 \pm 1.2$ & $11 \pm 2$ &  $4.8$  & $7.1$ &  $18$ & $55$ & $29$ & $11.8 \pm 0.3$  & $1900$ & 0.025 & 0.0081\\

R Lmi & $15 \pm 7$ & $3.5 \pm 1.3$  & $3.4$ & $1.5$ & $3.7$ & $3.0$ & $4.2$ & $6.6 \pm 0.1$ & $700$ & 0.031 & 0.010\\

R Uma & $10^{*}$ & $3.0 \pm 0.7$ & $0.9$ & $0.7 $ & $2.1$ & $0.6$ & $1.3$ & $4.7 \pm 0.4$ & $500$ & 0.027 & 0.0087\\

RR Eri & $25^{*} $ & $ 2.4 \pm 0.1 $ & $1.9$ & $1.1$ & $2.9$ & $1.2$ & $4.1$ & $12.0 \pm 0.3$ & $300$ & 0.022 & 0.0071\\

RT Cnc & $ 10^{*} $ & $1.5 \pm 0.2 $ & $1.3$ & $0.5 $ & $1.3$ & $0.15$ & $0.7$ & $7.2 \pm 0.2$ & $200$ & 0.013 & 0.0042\\

RU Her & $8 \pm 2$ & $5.7 \pm 1.5$  & $2.8$ & $3.0 $ & $7.6$ & $20$ & $16$ & $7.5 \pm 0.1$ & $1300$ & 0.051 & 0.017\\

RW Boo$^{\dagger}$ & $6.0 \pm 1.2$ & $2.2 \pm 0.7$ & $2.7$ & $1.1$ & $2.6$ & $1.0$ & $3.5$ & $12.3 \pm 1.2$ & $300$ & 0.020& 0.0065\\

RW Boo$^{1}$ & $7.7 \pm 1.2$ & $1.6 \pm 0.7$ & $1.4$ & $0.5$ & $1.3$ & $0.2$ & $0.8$ & $8.0 \pm 0.3$ & $200$ & 0.020& 0.0065\\

RW Boo$^{2}$ & $5.3 \pm 1.2$ & $1.9 \pm 0.7$ & $2.6$ & $1.0$ & $2.5$ & $0.8$ & $3.9$ & $16.6 \pm 0.9$ & $200$ & 0.020& 0.0065\\


RZ Uma & $10^{*}$ & $ 2.5 \pm 0.5$  & $1.0 $ & $0.8 $ & $2.0$ &  $0.6$ & $1.2$ & $4.7 \pm 0.3$ & $500$ & 0.022 & 0.0071\\

SV Peg & $8 \pm 2$ & $5.6 \pm 1.5$ & $5.0$  & $2.9 $ & $7.4$ & $20$ & $15$ & $7.3 \pm 0.1$ & $1300$ & 0.050 & 0.016\\

V Eri & $14 \pm 7$ & $3.0 \pm 1.1$ & $3.4$ & $1.5$ & $3.8$ & $2.7$ & $5.9$ & $10.2 \pm 0.2$ & $500$ & 0.027 & 0.0087\\

VY Uma & $7.5 \pm 1.9$ & $5.1 \pm 1.1$ & $3.0$  & $1.9$ & $4.8$ & $6.9$ & $2.0$ & $6.6 \pm 0.1$ & $900$  & 0.011 & 0.0035\\

W Peg  & $7 \pm 2$ & $3.4 \pm 1.2$  & $2.8$ & $1.5 $ & $3.7$ & $2.9$ & $4.5$ & $7.4 \pm 0.3$ & $600$ & 0.030 & 0.0097\\

Y Crb & $7 \pm 3$ & $3.0 \pm 1.7$ & $0.9$   & $0.9$ & $2.4$ & $1.1$ & $1.3$ & $3.5 \pm 0.4$  & $800$ & 0.027 & 0.0087\\

Z Cnc & $10^{*}$ & $2.0 \pm 0.5$  & $0.9$ & $0.6$ & $1.5$ & $0.3$ & $0.6$ & $3.9 \pm 0.2$ & $500$ & 0.018 & 0.0058\\

\hline
\end{tabular} 
\begin{tablenotes}
\item \textbf{Notes.} Column (1): name of the source, col (2): excitation temperature, col (3): CO column density, col (4): source radius in arcseconds, col (6): source radius in cm, col (6): photodisociation radius in cm, col (7): mass of the circumstellar envelope, col (8): mass-loss rate, col (9): average expansion velocity, col (10): expansion time, Col (11): reddening produced by the CSM (see Sect.~\ref{ext_corr}). (*) Fixed temperature. ($\dagger$) estimated with an average RW\,Boo \vexp. (1) estimated for the RW\,Boo low \vexp component. (2) estimated for the RW\,Boo high \vexp \,component (see table~\ref{tab:spectral_analysis}), components (1) and (2) are shown for illustrative purposes and were not taken into account in the analysis. 
\end{tablenotes}

\end{threeparttable}
\end{adjustbox}

\renewcommand{\arraystretch}{1.0}

\end{table*}

\begin{figure*}[h!]
    \centering
    \includegraphics[width=\linewidth]{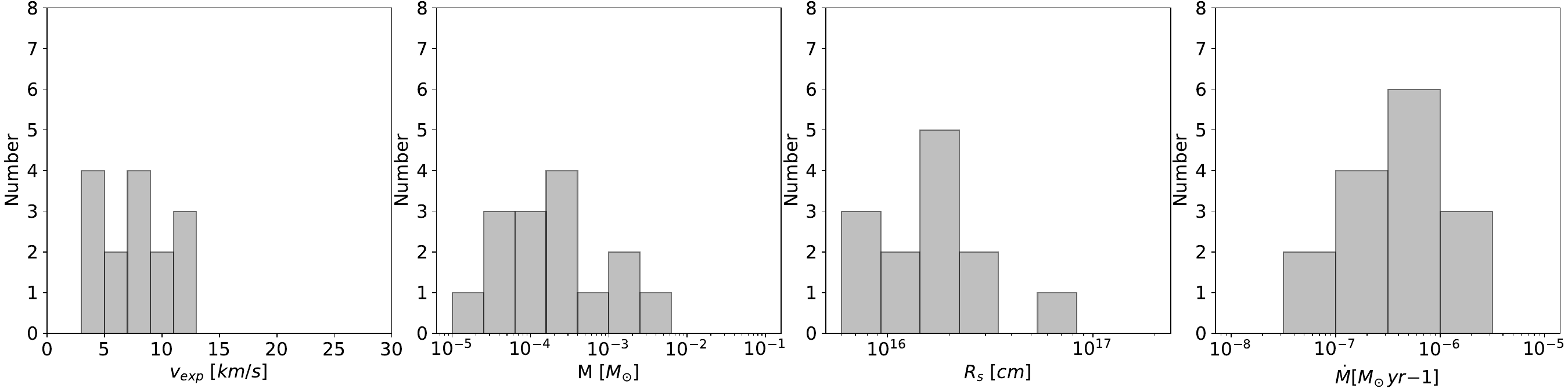}
    \caption{Distribution of the CO envelope expansion velocity (\vexp), mass ($M$), characteristic radius (\rs), and mass-loss rate (\mloss) of the uvAGB with CO detections in our sample.}
    \label{fig:histograms}
\end{figure*}

\section{Properties of the circumstellar envelopes}  
\label{trends}

In this section, we will discuss the primary characteristics of the molecular envelopes surrounding our sample of uvAGB stars, which have been inferred from our detections of CO line emissions (Table~\ref{tab:parameters}). Furthermore, we will compare the obtained values with those derived from various AGB samples mentioned in the existing literature. In the case of RW Boo, we present the solution for the average \vexp \, as well as two additional solutions corresponding to each of the two \vexp \, components found in the spectral lines -- these latter two solutions are presented for illustrative purposes, but were not taken into account in the posterior analysis.

Figure \ref{fig:histograms} presents the distributions of the key envelope parameters obtained in Sect.~\ref{anal}, specifically the expansion velocity (\vexp), mass ($M$), characteristic radius of the CO-emitting volume ($R_{\mathrm{s}}$), and mass-loss rate (\mloss).

\subsection{Expansion velocity}
The distribution of expansion velocities in our sample exhibits a relatively even spread, encompassing values ranging from approximately 3 to 13 km/s (Fig. \ref{fig:histograms}, left). These values fall within the range of expansion velocities observed in AGB envelopes, although they tend to occupy the lower part of the population, which typically reaches up to $\sim$30 km/s according to previous studies \citep{Hofner_2018}. This outcome can be attributed to the composition of our sample, which primarily comprises O-rich AGB stars (90\%) alongside a substantial proportion of irregular and semi-regular AGB stars (79\%). It has been shown that O-rich have lower expansion velocities than C-rich AGBs \citep[e.g.][]{Margulis_1990, Hofner_2018}. In addition, it is widely recognised that irregulars and semi-regulars exhibit systematically lower terminal velocities compared to Mira-type variables \citep[see][]{Olofsson_2002}.

The lowest expansion velocities observed in this study are those of the O-rich semi-regulars Y\,Crb and Z\,Cnc (\vexp=3.5 and 3.9\,\kms\,respectively). According to \cite{Winters_2003} the mass-loss mechanism in low outflow-velocity (\vexp$<$5\,\kms) AGB stars is fundamentally dominated by pulsations without the dust radiation pressure playing a dominant role, as opposed to the majority of AGBs. Another possibility is that the narrow profiles observed in some of our sources may be associated with stable, rotating structures in a circumbinary configuration, similar to what has been observed in L$_2$\,Pup \citep[see][]{Kervella_2019}.

At the high end of our velocity distribution, with \vexp\,$\sim$12\,\kms, we find the C-rich Mira T\,Dra and the O-rich semi-regulars RR\,Eri and RW\,Boo. For RW\,Boo a double-shell profile was found (see Sect.~\ref{sec:line_profiles}) indicating the presence of a slow (\vexp$\sim$8.0\,\kms) and a fast (\vexp$\sim$16.6\,\kms) wind components, in this analysis we have considered the parameters estimated with an average expansion velocity (\vexp$=$12.3\,\kms). Finally, within the sensitivity limit of our observations, we do not identify in any of the uvAGB stars in our sample the presence of massive, fast (up to a few hundred \,\kms) molecular outflows like those commonly present in the subsequent evolutionary stages of pre-PN and PN \citep[e.g.][]{Bujarrabal_2001, sanchez-contreras_2012}.

\subsection{Size and mass}

The characteristic radius of the envelope layers where the \doceuno\ and \docedos\ emission observed is predominantly produced, which has been self-consistently estimated together with the mass-loss rate from our population diagram analysis (Sect.~\ref{anal}), takes values ranging from 6\ex{15}\,cm to 2\ex{17}\,cm with a mean around 2\ex{16}\,cm. This is in good agreement with values obtained from previous works using similar empirical relationships or model-based indirect estimations \citep[e.g.][]{Groenewegen_2017} but also from direct measurements from interferometric CO mapping \citep[e.g.][]{Ramstedt_2020} for other samples of AGB CSEs with similar mass-loss rates. 

The derived sizes of the envelopes and their corresponding expansion velocities indicate relatively short crossing or kinematical times, ranging from $\sim$200 to $\sim$2000 years (Table~\ref{tab:parameters}). Consequently, CO millimetre-wavelength observations are not sensitive to the mass-loss process that occurred more than a few thousand years ago. This finding aligns with the relatively low values of the envelope mass discovered in our study, ranging from $10^{-5}$ to $5 \times 10^{-3}$\,\msun, with a peak around $3 \times 10^{-4}$\,\msun. Our analysis reveals a slightly higher mass distribution compared to the results reported by \cite{ramos-medina_2018} and \cite{da_Silva_Santos_2019} based on high-$J$ CO emission observations using the {\it Herschel} telescope. This discrepancy is consistent with the fact that far-infrared observations probe a smaller volume, closer to the central regions, compared to the millimetre-wavelength observations.

\subsection{Mass-loss rate}\label{mloss_rate}

The distribution of \mloss\ of our sample covers a range from 6\ex{-8} to 3\ex{-6}\,\my\, with a mean value of 6\ex{-7}\,\my. The mass-loss rates obtained in our study fall within the range of values reported in previous investigations of large samples of AGB stars \citep[see][]{Hofner_2018}. However, it is worth noting that our sample lacks objects exhibiting extremely high or extremely low mass-loss rates, specifically those reaching up to few\ex{-4}\,\my\ or falling below a few\ex{-8}\my. The uvAGB stars in our sample with the lowest and highest mass-loss rates are Z\,Cnc and T\,Dra, respectively.

\begin{figure}[hb!]
    \centering
    \includegraphics[width=0.95\linewidth]{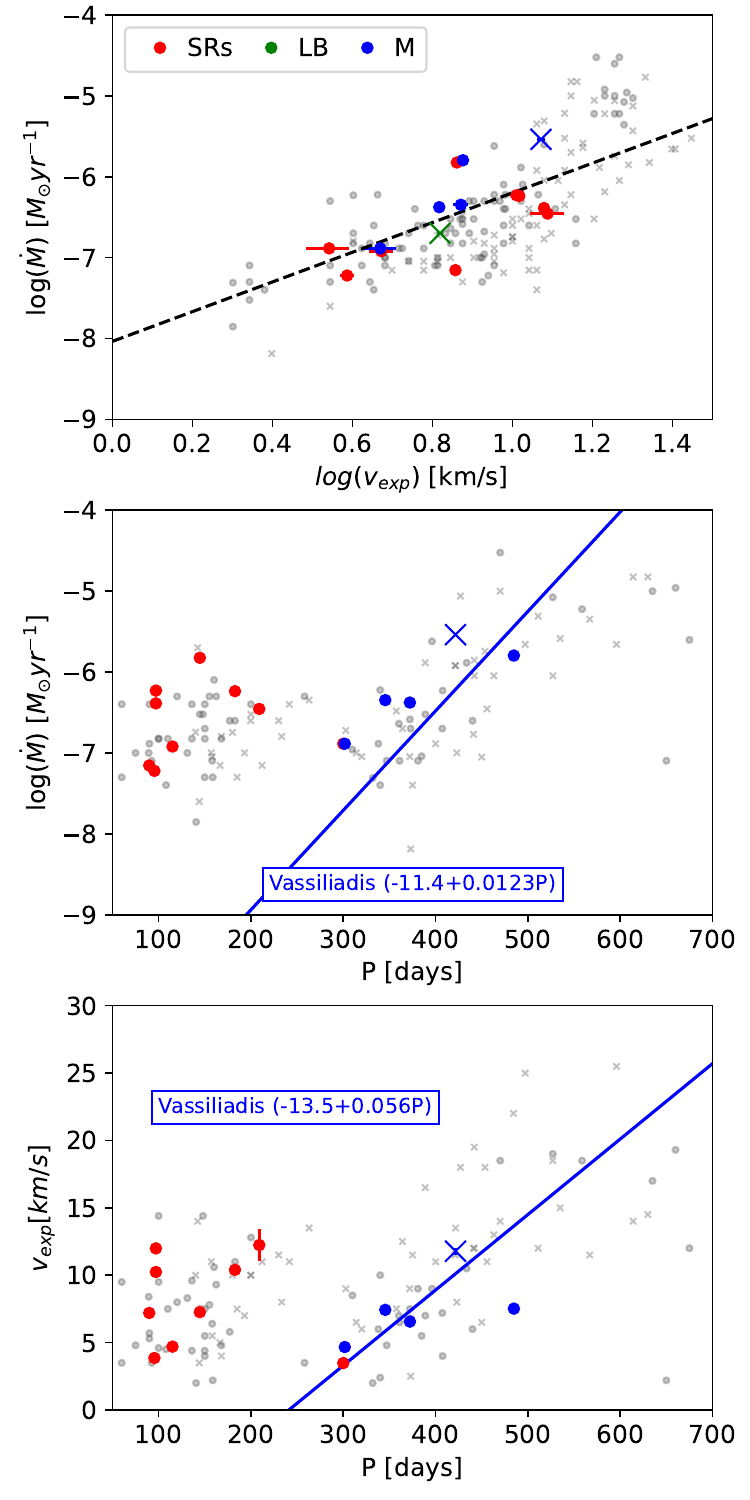}
    \caption{Comparison between some fundamental parameters of the AGBs envelopes (\mloss, \vexp \, and \ppuls) and with well-known correlations found in the literature. Upper: Relationship between \mloss \, and \vexp. Centre: Relationship between \mloss \, and \ppuls. Lower: Relationship between \vexp \, and \ppuls. The colours of the markers represent the variability type. Red: semi-irregulars (SRs), green: irregulars (LB), and blue:  Miras (M). The dashed line represents the linear fit performed over our sample and the solid lines represent the linear relationships for \mloss \, and \vexp \, with \ppuls \, found for Mira AGBs from \cite{Vassiliadis_1993}. Grey markers represent AGB stars with \mloss\ and \vexp\ values recompiled by \cite{Van_de_Sande_2021}. The shape of the markers represents the chemistry type in the two samples: circles for O-rich AGB stars and crosses for  C-rich AGB stars.}
    \label{fig:parameters}
\end{figure}

For AGB stars there are well-known correlations between some of the fundamental parameters of their envelopes, in particular, between \mloss\ and \vexp\ and between these and the stellar pulsation period ($P$) for regular and semi-irregular variables. In Fig.\ref{fig:parameters} we explore those same relationships to see if they are followed in the same way for our sample of uvAGBs. In this comparison, we use the sample of AGB stars recompiled by \cite{Van_de_Sande_2021}. Pulsation periods for all targets (AGBs and uvAGBs) are obtained from General catalogue variable stars (GCVS).

The mass-loss mechanism primarily manifests itself through two key parameters: the mass-loss rate (\mloss) and the expansion velocity (\vexp). Extensive studies on large samples of AGB stars have consistently demonstrated a correlation between these parameters of the type \mloss\, $\propto$\,\vexp$^\alpha$ with values of $\alpha$=1.4-3.3 depending on chemistry and variability type of the stars \citep[see e.g.][]{Young_1995, Knapp_1998, Olofsson_2002}.
In our specific study of uvAGBs, we obtained a very similar result (Fig.~\ref{fig:parameters}, top) showing a weak relationship $\mloss \simeq 10^{-8}\vexp^{1.8}$. However, it is important to note that our sample size is relatively smaller compared to previous studies, which may partially explain the relatively low Pearson's correlation coefficient found, $r$=0.65. In line with previous findings by \cite{Olofsson_2002}, a weaker correlation between \mloss\, and \vexp\, was observed for O-rich AGB stars when compared to their C-rich counterparts. The spread in the \mloss\, versus \vexp\, trend found in previous studies of AGB stars is in any case large, which probably reflects dust-to-gas ratio variations among other factors \citep{ Netzer_1993,Habing_1994}.

In the case of regular variable stars, such as Mira-type stars, it is widely recognised that a correlation exists between both \mloss\ and \vexp\ and the stellar pulsation period, $P$ \citep[e.g.][]{Vassiliadis_1993}. The uvAGB stars in our sample adhere closely to this relationship, as depicted in Fig.~\ref{fig:parameters} (middle and lower panels). The semi-regular uvAGB stars, however, tend to cluster in the low-period region (\ppuls$<$200 days), exhibiting mass-loss rates (and expansion velocities) in the range \mloss$\sim$10$^{-7}$-10$^{-6}$\my\ (\vexp$\sim$5-15\,\kms), which is  greater than would be expected based on their relatively short periods if they were Mira variables. Notably, no discernible correlation with the period is observed for these semi-regular uvAGB stars, which aligns with findings from larger semi-regular AGB samples \citep{Van_de_Sande_2021}. In these cases, it suggests that there is another influential factor, beyond pulsation, playing a significant role in governing the mass-loss process of these objects.

\subsection{Mass outflow momentum and \mloss-\lbol\ relationship} \label{mass outflow}

In our study of uvAGBs, we have examined
the ratio $\beta$=\mloss\,\vexp\,$c$/\lbol, which compares the momentum rate in the mass outflow (\mloss\,\vexp) to the momentum rate in the stellar radiation (\lbol/$c$). This parameter, also known as the overpressure or radiation pressure efficiency, is directly proportional to the dust opacity of the wind if radiation pressure on dust grains is the main wind driving force \citep{Knapp_1986, Lefevre_1989}.

\begin{figure}[hb!]
    \centering
    \includegraphics[width=\linewidth]{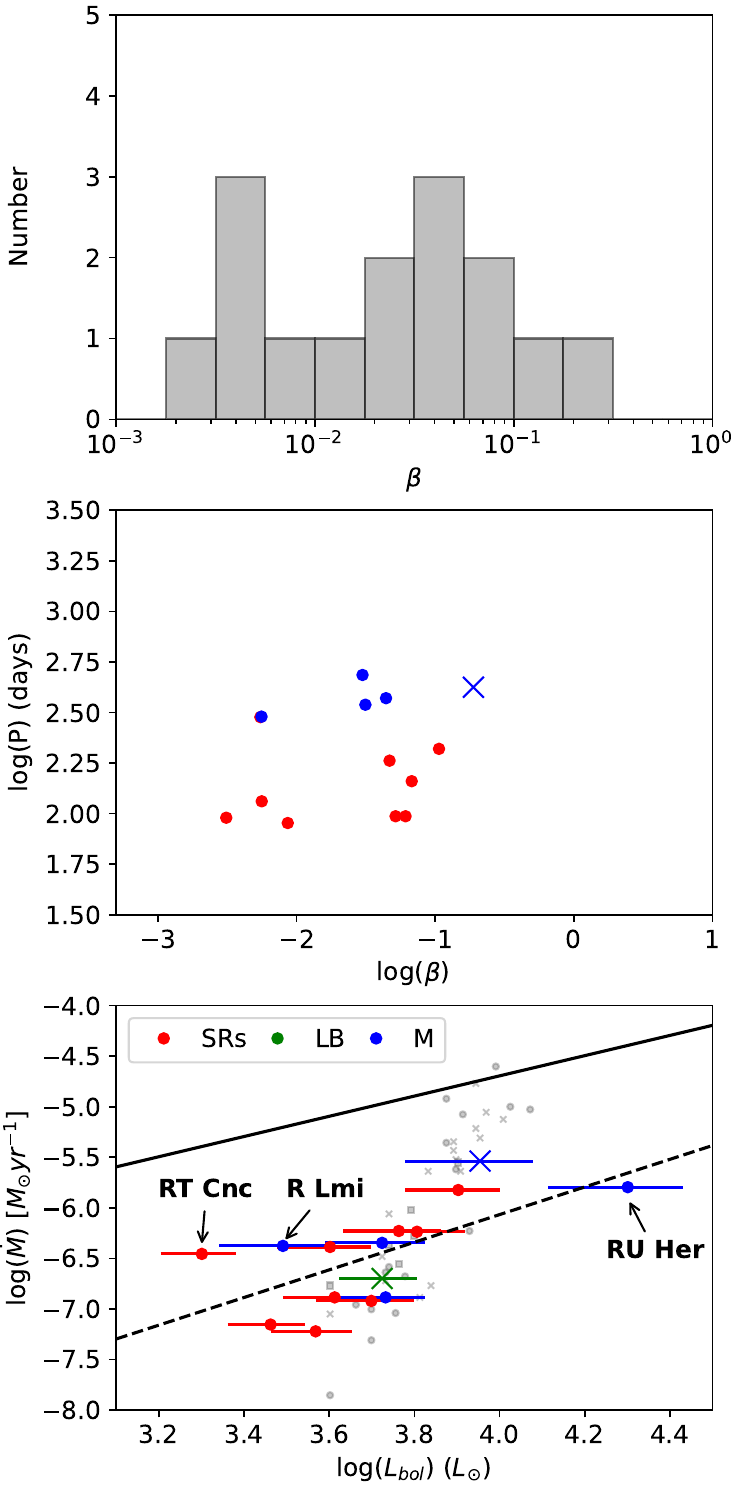}
    \caption{Comparison between envelope parameters related with the mass-loss mechanism. Upper: Distribution of the $\beta$ parameter, defined as the ratio of the outflow to the stellar radiation momentum ($\beta$=\mloss\,\vexp\,$c$/\lbol, Sect.~\ref{mass outflow}). Centre: Relationship between stellar pulsation period ($P$) and $\beta$. Lower: Relationship between \mloss\ and \lbol\,; the solid line represents the single-scattering limit in the case \vexp=10\kms\, \cite[see][]{groenewegen_2018} and the dashed line the best linear fit to our dataset. The colours and shapes of the markers are the same as in Fig.\,\ref{fig:parameters}. Grey markers represent AGB stars with \mloss\ and \lbol\ values recompiled from \cite{Danilovich_2015} (circles: O-rich AGB stars, squares: S stars, and crosses: C-rich AGB stars).
    }
    \label{fig:histogram_beta}
\end{figure}

Analysing the distribution of beta values for our sample (Fig.~\ref{fig:histogram_beta}, upper panel),
we consistently find values well below 1 for all the sources, aligning well with the values observed in the majority of AGB stars and with the values expected if radiation pressure is the principal operative mechanism.  As pointed out by \cite{Knapp_1986}, the $\beta$ values much lower than 1 (found in our sample but also in some AGB stars) indeed suggests that these objects are not losing mass as efficiently as it is possible. It is worth mentioning also that while some AGB stars may exhibit $\beta$ values slightly higher than 1, which can be attributed to multiple scattering \citep{Lefevre_1989}, none of the objects in our sample demonstrate such behaviour. The sources with the largest values of $\beta$$\sim$0.1-0.2 are T\,Dra and RW\,Boo. In the case of the progeny of AGB stars (i.e. post-AGBs or pre-PNe), it is frequently observed that the $\beta$ values are notably high, which is considered as an indication that a distinct mass-loss mechanism, one that differs from radiation pressure acting on dust, is occurring in these stages \citep{Bujarrabal_2001}.

Stellar pulsation is the other mechanism determining the mass-loss rate. We find a rough proportionality between $\beta$ and the pulsation period $P$ for our sample (Fig.~\ref{fig:histogram_beta}, central panel), in agreement with previous results for AGB stars \citep{Knapp_1986}. The observed trend reflects an underlying correlation between \mloss\, and $P$: the mass-loss rate increases with longer periods and consequently, the dust opacity also increases. We find that Mira variables and semi-regulars occupy two distinct regions in the $P$-$\beta$ diagram. The SRs are located in the lower part, reflecting that despite their shorter periods, they exhibit similar mass-loss rates compared to Mira variables as shown in Fig.~\ref{fig:parameters} (middle panel).

In the last panel of Figure \ref{fig:histogram_beta}, we conducted a comparison between \mloss\ and \lbol\ for our sample. All our targets fall below the single-scattering limit, represented by the solid line, as already discussed and indicated by the low values of $\beta$ observed. We observed a correlation that can be described by the power-law relation $\mloss \sim 3 \times 10^{-12} \lbol^{1.4}$, represented by the dashed line in the figure. A power-law relationship is in line with expectations and is commonly observed in AGB stars. It indicates that as the luminosity increases, either due to larger stellar masses or a more advanced stage of evolution on the AGB, the mass-loss rate tends to increase as well \citep{Hofner_2018, groenewegen_2018}.
We conducted a comparison between our values of \mloss\ and \lbol\ with those reported by \cite{Danilovich_2015}, who estimated mass-loss rates from radiative transfer modelling of CO line emission in a sample of 53 AGB stars. 
Upon initial inspection, there is an apparent discrepancy in the exponent of the power-law relationship between the two samples. However, when considering the majority of sources in both studies, there is a general agreement and similarity in the distribution of points. The only notable exceptions are three objects (RT Cnc, RU Her, and, perhaps R\,LMi) located at the extreme high and low ends of the luminosity range. If we exclude these objects from the analysis, the slopes or power-law exponents of the correlations in both studies match more closely. 
This indicates that, overall, there is consistency between the mass-loss rate and luminosity relationships derived from our study and that of \cite{Danilovich_2015}, with the exception of a few outliers.

\section{CO emission and CO-derived envelope parameters correlations with IR/UV properties}\label{luminosities}

In this section we investigate correlations between the CO emission intensity and the infrared and ultraviolet continuum emission, as well as the bolometric luminosity. We also explore trends between the primary envelope parameters derived from CO and the distinctive ultraviolet emission properties exhibited by our targets.

\subsection{CO versus\ IRAS 60\micron\ emission}\label{CO_IRAS60}

The correlation between CO line intensity and the IRAS 60\micron\ emission has been well established for low-to-intermediate mass evolved stars, including AGB and post-AGB stars, as documented in numerous previous studies \citep[][among others]{Nyman_1992, Bujarrabal_1992, Olofsson_1993}. This correlation is attributed to the fact that both the CO emission and the IRAS 60\micron\ emission are indicative of the amount of material (in the form of gas and dust, respectively) present in their envelopes.

\begin{figure}[h!]
    \centering
    \includegraphics[width=\linewidth]{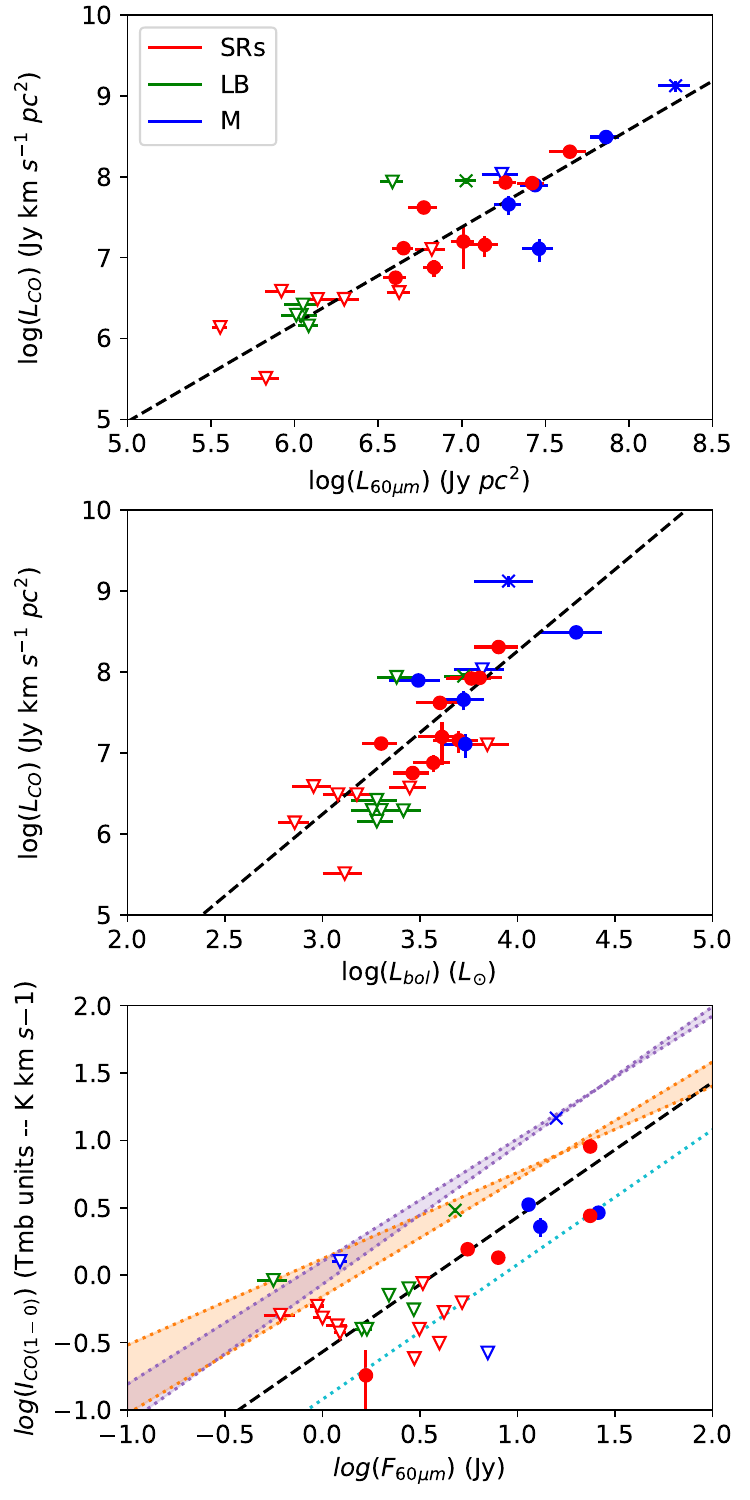}
    \caption{Comparisons of the CO integrated intensity, \irassesenta\ emission, and bolometric luminosity (Tables~\ref{tab:fluxes} and \ref{tab:sources}). In the upper and middle panels $L_{\rm CO}$ is compared with $L_{\rm 60 \micron}$ and \lbol. In the lower panel the \doceuno\ main-beam brightness temperature (\tmb) is plotted against the \irassesenta\ flux after properly scaling \tmb\ when observed with a telescope other than the IRAM-30m antenna (see \ref{CO_IRAS60}). The dashed line is the best linear fit to our data; the dotted lines are linear correlations from \cite{Nyman_1992} and \cite{Bujarrabal_1992} for O-rich AGBs (orange), C-rich AGBs (purple), and pre-PNe (cyan). The colours and shapes of the markers are the same as in Fig.\,\ref{fig:parameters}.}
    \label{fig:CO_comparisons}
\end{figure}

In our study of uvAGB stars, we also observe a strong correlation between the velocity-integrated luminosity of the \docedos\ transition ($L_{\rm CO}$) and the IRAS 60\micron\ luminosity ($L_{\rm60 \micron}$) as shown in Fig.~\ref{fig:CO_comparisons} (upper panel). This finding supports the notion that the gas and dust mass-loss rates are proportional to each other.
Given that the infrared radiation is a significant or dominant component of the energy emitted by the dusty envelopes around AGB stars, it is reasonable to expect a correlation between the CO luminosity and the bolometric luminosity, which is in fact observed in our sample of uvAGBs (Fig.~\ref{fig:CO_comparisons}, central panel). 

A linear fit to the $L_{\rm CO}$ versus $L_{\rm60 \micron}$ and \lbol\ data points yields 

\begin{equation}
    L_{\rm CO} \approx 0.1 \times 10^{1.2} L_{\rm60 \micron}
,\end{equation}

\begin{equation}
    L_{\rm CO} \approx 0.2 \times 10^{2.0} \lbol
.\end{equation}

We note that most of the low-luminosity ($\lbol<$3000\,\ls) sources are CO non-detections (triangles in Fig.~\ref{fig:CO_comparisons}). Some of these low-luminosity targets could be RGB or early-AGB stars, with mass-loss rates still well below the maximum values reached at the tip of the AGB phase \citep[see e.g.][]{Hofner_2018}. However, the large uncertainties in their luminosities (see Sect.~\ref{sample}) does not allow us to robustly assert that these are not AGB stars. Nevertheless, they do not affect the posterior analysis as the CO non-detections were not used in the fitting; the derived upper limits on their CO fluxes are quite consistent with various relationships obtained for the detected targets discussed in this section (see e.g. the relationships between CO and \irassesenta /bolometric luminosities shown in Fig.~\ref{fig:CO_comparisons}).

In the literature, there are different approaches used by several authors to compare the relationship between the CO intensity and the IRAS 60\micron\ emission. In principle, we focused on using the luminosity as a reference magnitude and specifically examined the \docedos\ transition due to its higher number of detections compared to the \doceuno transition. However, a common method employed in exploring the CO-\fsixty\ emission relationship is by comparing the main-beam brightness temperature (\tmb\ in K) of the \doceuno\ transition with the IRAS\,60\micron\ flux (in Jy). 
In Fig.~\ref{fig:CO_comparisons}, lower panel, we show the distribution of the uvAGBs in this study using this same representation together with the relationships derived for different samples of AGB stars in the past.

Specifically, we refer to the relationships presented in \cite{Nyman_1992} for AGBs in different regions of the two colour IRAS diagram (the same regions are shown in Fig~\ref{fig:IRAS_colour_diagram_NUV+FUV}), including O-rich sources (located into regions II and III) and C-rich sources (located into regions VIa and VII). To account for the difference in beam size between the SEST\,15m telescope, used as a reference in the studies of \cite{Nyman_1992}, and the IRAM\,30m antenna, used in this work, we multiplied their relationships by a factor of 4 for a proper comparison with ours. Furthermore, we present the relationship described by \cite{Bujarrabal_1992} for PPNe, which shows that PPNe are positioned noticeably below the relationships observed for AGBs. \cite{Bujarrabal_1992} interpreted this as an indication that PPNe generally have a lower gas-to-dust mass ratio. This could be attributed to significant CO photodissociation in more advanced stages of evolution beyond the AGB phase, in which the envelopes become more diluted, and the central stars become hotter.

We found a relationship $I_{\rm CO} \sim 0.27 \fsixty$ for our sample of 10 targets with 1--0 detections (2 C-rich and 8 O-rich). The only two C-rich uvAGBs in our sample lie very close to the expectations based on the CO-\fsixty\ relationship reported by \cite{Nyman_1992} for this same chemical type. The O-rich uvAGBs in our sample, however, seem to fall below the relationship for O-rich AGBs \citep{Nyman_1992} and much closer to the relationship for PPNe \citep{Bujarrabal_1992}. Since our uvAGBs are characterised by an excess of emission in the UV, this result does not seem overly surprising and could indicate, as in the case of PPNs, that a non-negligible fraction of the molecular gas is dissociated by the central source. We further discuss this in Section\,\ref{dis_COvsUV} in connection with the CO-to-UV relationship.

\subsection{CO versus NUV and FUV emission}\label{UV_lumi}

\begin{figure*}[h!]
    \centering
    \includegraphics[width=\linewidth]{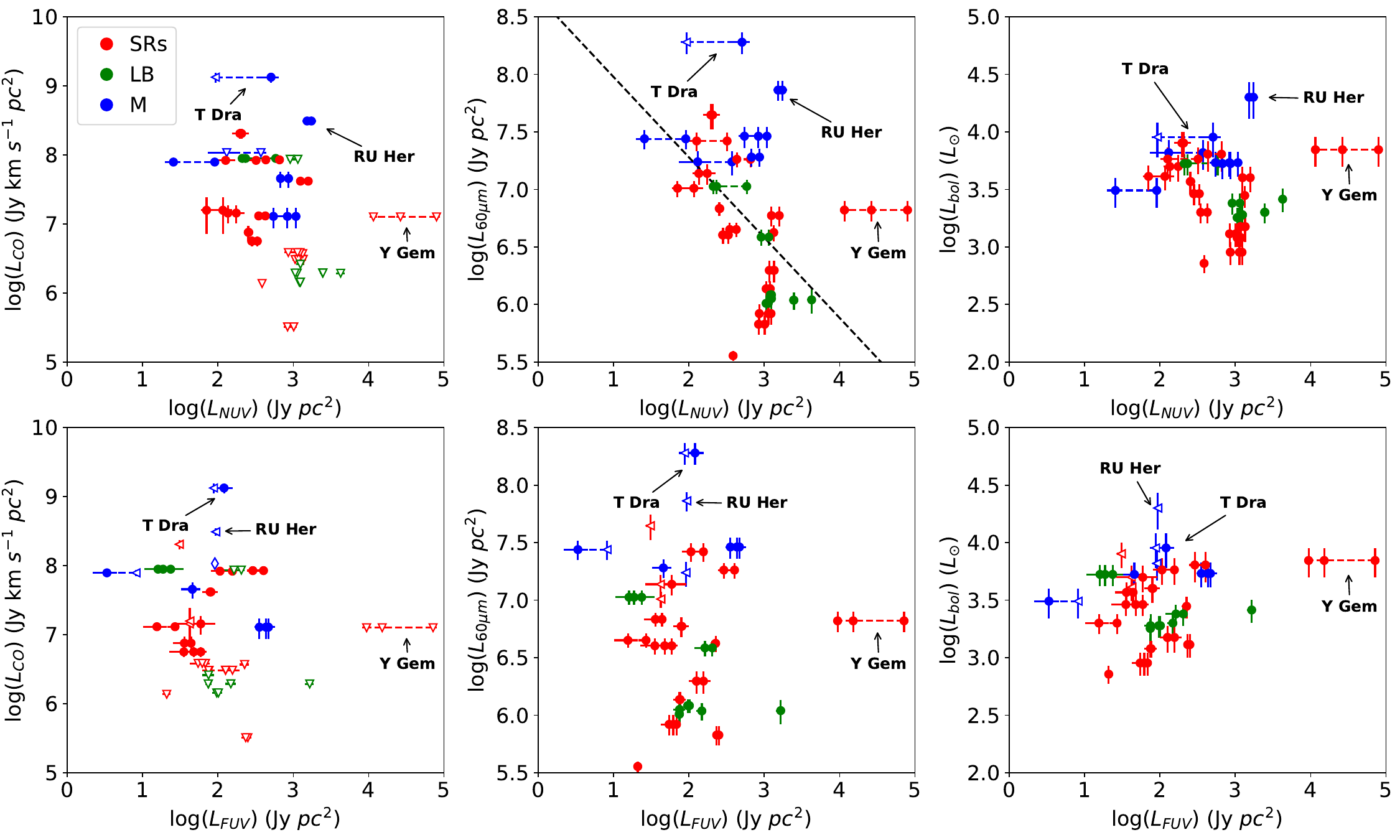}
    \caption{Comparisons between the \docedos\ velocity-integrated luminosity ($L_{\rm CO}$, left), IRAS\,60\micron\ luminosity ($L_{\rm 60\micron}$, middle), and bolometric luminosity (\lbol) with the luminosity in the GALEX NUV and FUV bands (top and bottom panels, respectively). The colour-coding used for variability types is shown in the top left panel. Triangles are upper limits. The dashed line represents the best linear fit to the \fsixty-to-$L_{\rm NUV}$ data points (Pearson's correlation coefficient r=-0.48). The trends in the rest of the variables represented have not been confirmed (\S\,\ref{UV_lumi}).}
    \label{fig:UV_comparison}
\end{figure*}

A potential trend between the CO emission and the distinctive UV properties of uvAGB stars has not been explored before. For our sample, $L_{\rm CO}$ is compared with the extinction-corrected luminosities in the NUV and FUV bands ($L_{\rm NUV}$ and $L_{\rm FUV}$; see Fig.~\ref{fig:UV_comparison}, left panels). As previously discussed in Sect.~\ref{detection_statistics}, the CO non-detections (indicated by triangles) tend to cluster towards the higher range of NUV and FUV luminosities, suggesting a potential inverse relationship between CO and UV intensity. Establishing a definitive trend is challenging, however,  since the CO-NUV and -FUV diagrams exhibit a considerable spread, the number of CO detections is relatively small or moderate (15 sources) and (unlike CO) the UV emission is notably variable. 

To address the limitation of a relatively small number of CO detections, and considering the established proportionality between CO and IRAS 60\micron\ emission (discussed in Sect.~\ref{CO_IRAS60}), we also examine the relationship between UV and IRAS 60\micron\ luminosity (Fig.~\ref{fig:UV_comparison}, middle panels). In these diagrams, the presence of an anti-correlation appears to emerge more clearly but only for the NUV band, for which we derive a Pearson's correlation coefficient of r=$-$0.48. 
For the FUV band, however, the anti-correlation is not confirmed (r=$-$0.01) as the CO intensities exhibit a significant spread (on the order of dex$\sim$2-3) for very similar values of $L_{\rm FUV}$.

In the right panels of Figure~\ref{fig:UV_comparison}, we examine the relationship between the total bolometric luminosity and the NUV and FUV luminosity for our target stars. We tentatively observe a weak anti-correlation (r=$-$0.30) in the NUV band, which is likely influenced by the underlying $L_{\rm 60 \micron}$-to-$L_{\rm NUV}$ relationship. However, in the FUV band, no clear trend is evident. This observation aligns well with the results reported by \cite{Montez_2017} in a comprehensive study of GALEX uvAGBs, where no significant correlation was found between FUV band emission and other bands ($UBVRIJHK$), in contrast to the stronger correlations observed in the NUV flux.

In all the diagrams presented and discussed above, we assigned different labels to our uvAGB stars based on their variability class to examine potential trends. However, no significant relationships were observed, except for the expected observation that Mira-type variables generally exhibit stronger CO emission compared to semi-regulars and irregulars, which is consistent with broader studies of AGB stars.

Finally, it is worth noting that the sources T\,Dra, RU\,Her and Y\,Gem stand out as outliers in all these diagrams, deviating from the general pattern observed in the sample.  
T\,Dra exhibits exceptionally intense CO and IRAS\,60\micron\ emission compared to the rest of the sample, yet its NUV and FUV emission falls within the average range. Similarly, RU Her shows the second more intense CO and IRAS\,60\micron\ emissions as well as the largest \lbol, it also shows a large NUV flux despite the fact that it was not detected in the FUV band. In contrast, Y\,Gem is the strongest UV emitter in the sample, displaying also significant UV variations, and at the same time it shows relatively high values of $L_{\rm 60 \micron}$ \citep[Even though, a weak \docedos\ emission line was reported by][]{Sahai_2011}. 
Remarkably, T\,Dra and Y\,Gem share an additional characteristic, they are both X-ray emitters. This X-ray emission perhaps suggests a more extreme nature for these uvAGB stars (potentially connected with strong binary interactions), warranting further investigation into the underlying mechanisms driving their distinct behaviour.

\subsection{Envelope parameters versus\ NUV and FUV emission} \label{effects}

\begin{figure*}[h!]
    \centering
    \includegraphics[width=\linewidth]{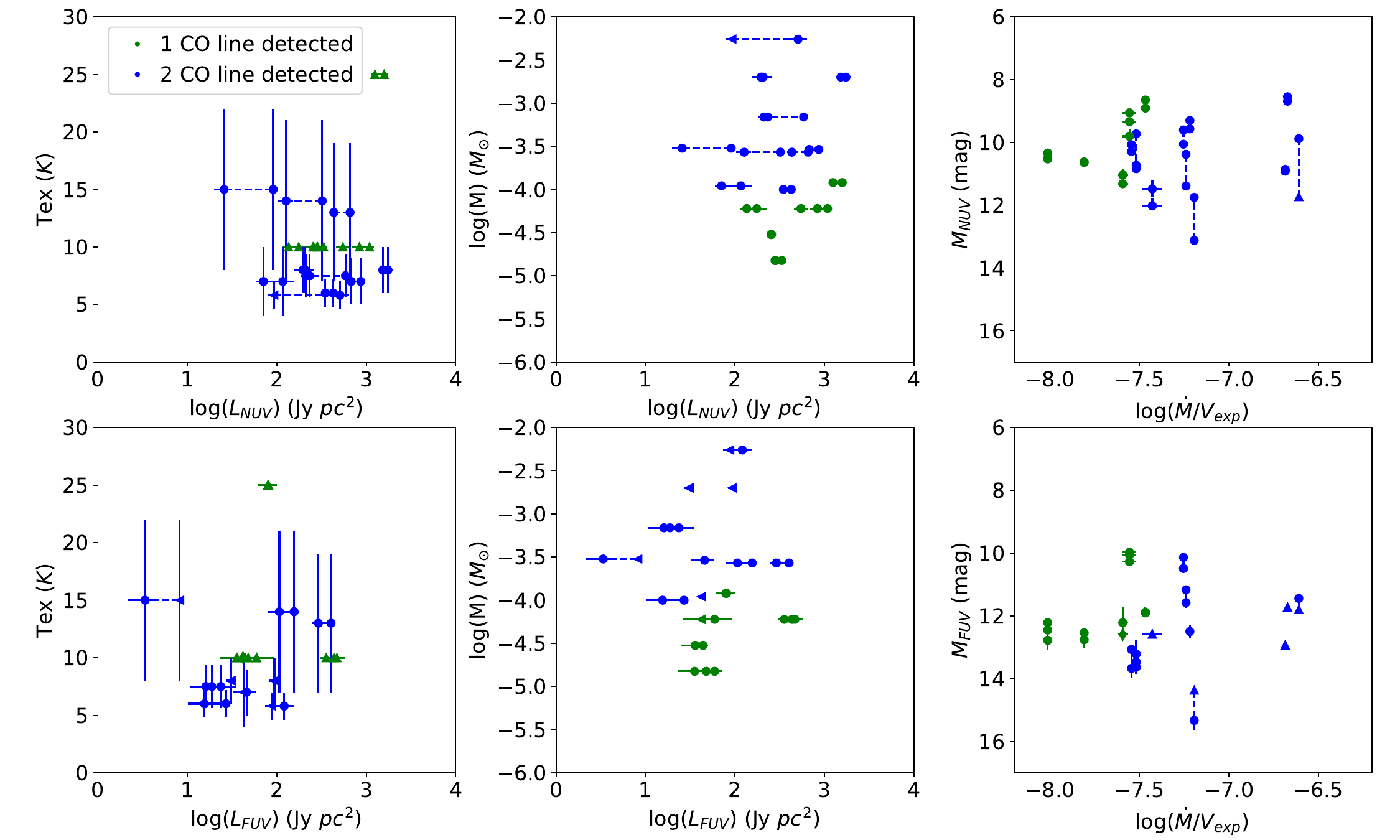}
    \caption{Comparison of main envelope parameters (\tex, M, and \mloss/\vexp) and UV properties of our sample of uvAGBs (NUV and FUV; top and bottom panels, respectively).}
    \label{fig:UV_vs_param}
\end{figure*}

In the investigation of potential trends between the two main envelope parameters derived from the CO population diagram analysis, namely the excitation temperature (\tex) and the mass in the CO-emitting volume (M), no clear correlation or trend was observed in any of the cases with any of the UV bands, as shown in the left and middle panels of Figure~\ref{fig:UV_vs_param}. However, it is important to consider two important factors that could make it challenging to identify a correlation even if it exists. Firstly, the small number of sources, particularly those with two detected lines, which provide more reliable values of \tex\ and M. In this case, only 10 sources met the criteria, limiting the statistical power of the analysis. Secondly, the variability in the NUV and FUV bands introduces additional dispersion on the x-axis of the diagrams. This variability can obscure any underlying correlation between the envelope parameters and the UV bands.

There is an additional factor that introduces complexity in discerning a clear trend of the UV emission with the envelope temperature, which is the lower limit nature of \tex\ derived from the CO population diagram analysis. This is because some regions in the outer layers of the envelope, where the low-$J$ CO transition emission originates, may reach densities close to, but slightly below, the critical densities of the CO 1--0 and 2--1 transitions (i.e. $\lesssim$(1-5)\ex{4}\,\cm3). Consequently, the derived \tex\ values may underestimate the true \tkin\ of the gas.

Possible mild deviations from LTE are not anticipated to strongly impact the mass estimation. This has been discussed in detail by \cite{ramos-medina_2018} and \cite{da_Silva_Santos_2019}. The reason for this is that even though \tex\ may deviate from the true \tkin, it precisely represents the level population distribution. As a result, the computation of the total number of emitting molecules (and hence, the mass) remains robust since it involves summing up the populations of all levels.

In our study, we also investigated the relationship between extinction-corrected GALEX magnitudes ($M_{\rm NUV}$ and $M_{\rm FUV}$) and \mloss/\vexp\ as a density proxy. 
We did not observe a confirmation of the linear anti-correlation depicted in Figure 10 of \cite{Montez_2017} in the NUV band. However, it is important to note some differences between our study and that by \cite{Montez_2017}. First, we included GALEX UV photometry for different epochs when available. Second, we dereddened the GALEX magnitudes using distance-corrected values of the total extinction (interstellar and circumstellar, see tables~\ref{tab:sources} and \ref{tab:parameters}), instead of relying on the default extinction estimate included in the GALEX catalogue, which accounts for the interstellar extinction across the whole Galaxy in the direction of a given target. Third, in the study by \cite{Montez_2017}, the mass-loss rates and absolute magnitudes were estimated using distances from the {\it Hipparcos} catalogue, whereas in our current paper, we utilised distances from the {\it Gaia} catalogue. It is important to mention that there is a discrepancy between the {\it Gaia} and {\it Hipparcos} distances for $\sim$30\% of the sources in the Montez et al.\ sample, with a factor of 2 difference, and this percentage increases to $\sim$65\% when considering a factor of 1.3. A recent study by \cite{Scicluna_2022} has demonstrated that {\it Hipparcos} distances are particularly unreliable for distances greater than $\sim$200\,pc. In addition, the uncertainties in the distances in {\it Gaia} and {\it Hipparcos} catalogues might be underestimated (see Sect.~\ref{sample}). Finally, our sample size is smaller and has a narrower range of $M_{\rm NUV}$ values compared to \cite{Montez_2017}, who compiled data from a larger set of AGB stars. In the FUV band, neither \cite{Montez_2017} nor our study found a clear trend. Considering these differences, our results are consistent with those of \cite{Montez_2017}, suggesting compatibility without indicating an anti-correlation between \mloss/\vexp\ and $M_{\rm NUV}$.

In previous studies of AGB stars, several authors have observed an anti-correlation between the envelope mass and effective temperature \citep[e.g.][]{ramos-medina_2018,da_Silva_Santos_2019}. However, in our investigation of uvAGBs, we did not find conclusive evidence supporting this relationship. This lack of evidence may be attributed to the relatively small number of sources in our sample and the relatively narrow ranges observed for $M$ and \tex.

\section{Discussion}\label{dis}

\subsection{Extinction effects and enhanced CO photodissociation} \label{dis_COvsUV}

In Sect.~\ref{UV_lumi}, we find that there is an anti-correlation between the CO and IRAS\,60\micron\ emission and the UV luminosity in the uvAGBs in our study. Such anti-correlation is expected to exists to some extent primarily due to the fact that UV emission can be more easily detected when the extinction caused by the envelope is low. This anti-correlation is also shown in Fig.~\ref{fig:IRAS_colour_diagram_NUV+FUV}, where uvAGBS, specially fuvAGBs, are located in the regions correspondent to AGBs with thin envelopes. On the other hand, objects with higher mass-loss rates, and consequently stronger CO and infrared emission, have higher extinction, making UV detection more challenging or even impossible even if an UV excess exists. This natural relationship between mass loss, extinction, and UV emission contributes to the observed anti-correlation between CO/IRAS\,60\micron\ and UV in our sample.
  
On the other hand, the presence of excess UV radiation from the central source, regardless of its origin, is anticipated to induce enhanced photodissociation of CO in the inner envelope regions. This increased photodissociation would result in weaker CO emission overall. Another consequence from this is that the relationship between CO and IRAS 60\micron\ emission would exhibit different slopes for AGB stars with and without UV excess, which is indeed a confirmed result from this study as described in Sect.~\ref{CO_IRAS60} and shown in  Fig.~\ref{fig:CO_comparisons}.

We hypothesise that the relatively weaker relationship observed between CO (or IRAS 60\micron) emission and $L_{\rm FUV}$, compared to $L_{\rm NUV}$, can be attributed, at least in part, to the lower overall extinction near the FUV wavelength of approximately 1500\,\AA. In contrast, the NUV bandpass is situated in the vicinity of the UV extinction bump around 2175\,\AA, where the dust extinction curve shows a relative maximum. This suggests that the NUV band is more sensitive to extinction-related effects than the FUV band. Additionally, it is also very likely that different emission mechanisms contribute in different proportions in the NUV band and in the FUV band. In particular, the FUV band may be dominated by emission associated with the presence of a binary, e.g. accretion as suggested by \cite{Sahai_2008}, whereas the NUV band may have a greater contribution from intrinsic emission, e.g. chromospheric emission.

The lack of correlation between FUV and any of the envelope or stellar properties explored so far in uvAGBs lend support to the notion that the UV excess observed in our stars, with a vast majority exhibiting relatively high FUV-to-NUV flux ratios \rfuvnuv$\gtrsim$0.06, has an extrinsic origin, meaning that  it is not directly linked to the intrinsic fundamental properties of AGB stars, such as their overall luminosity, but to binary-induced physical processes. 

\subsection{Proportion of uvAGBs in large samples of AGB stars}

As part of this study, we conducted a comparison between the CO-derived envelope properties and emission characteristics of our sample of uvAGBs and those from AGB stars from larger samples found in the literature. Our goal was to identify any unique or distinguishing features that could differentiate these two groups.
This comparison is not straightforward, since the larger samples of AGB stars surely also include UV-excess AGB stars that had not been identified as such. These unidentified sources add an additional layer of complexity to the analysis, making it challenging to isolate and differentiate the specific properties of uvAGBs.

In this section, we aim to provide constraints on the proportion of uvAGB stars within the larger population of AGB stars. The catalogue of GALEX UV emission from AGB stars by \cite{Montez_2017}, consisting of 316 AGB stars observed by GALEX, serves as a basis for our analysis. Among these stars, 179 (57\%) were detected in the NUV band, while only 38 (12\%) were detected in the FUV band. However, it is important to acknowledge that these numbers represent an upper limit to the true proportion of uvAGB stars. This is due to the fact that the original sample of $\sim$500 well-studied, nearby AGB stars, from which the catalogue was compiled, was initially assembled to search for X-ray emission. As a result, the sample included pre-selected UV- and/or X-ray emitters, biasing the proportion of uvAGB stars in the catalogue.

As part of our analysis, we conducted an independent search for uvAGB stars by cross-matching the \cite{Suh_2021} catalogue of AGB stars with the GALEX archive (see Sect.~\ref{sample}). By applying matching criteria similar to those used by \cite{Montez_2017}, we identified a total of 9838 and 1464 AGB stars that have been observed with GALEX in NUV and FUV bands respectively. Among these, 1019 (10.4\%) were detected in the NUV band, and 66 (4.5\%) were detected in the FUV band.  These results indicate that AGB stars are not frequently detected at UV wavelengths, particularly in the FUV range. However, it is important to consider that the \cite{Suh_2021} catalogue includes a substantial number of AGB stars located at large distances (>1000 pc). This introduces an observational bias against detecting uvAGBs, as a significant portion of them would remain undetected due to the substantial interstellar extinction at those distances. For this reason, the low proportion of uvAGBs has to be considered as a lower limit. 

The discrepancies in the detection rates of nuvAGBs and fuvAGBs introduce significant uncertainties in estimating the ratios of these UV-emitting AGB stars in the overall population. Additionally, the distribution of galactic AGB stars is skewed towards the galactic centre, resulting in limited sky coverage by GALEX and other UV telescopes in those regions. Consequently, a considerable number of UV-emitting AGB stars located in the galactic centre remain unidentified. This spatial bias contributes to a gap in the statistics of UV emissions in AGB stars, preventing a comprehensive understanding of the full population of UV-emitting AGBs.

In summary, the proportion of UV-emitting AGB stars (uvAGBs) falls within the range of 10-60\% for NUV (nuvAGBs) and 4-12\% for FUV (fuvAGBs). These estimates indicate that the fraction of uvAGBs within the AGB population, specially nuvAGBs, is not negligible, which can complicate the identification of genuine uvAGB properties from the broader AGB population. 

\section{Summary and conclusions}\label{summ}

We present observations with the \iram\ of the \doceuno\ and \docedos\ line emission in a sample of 29 AGB stars with $UV$ excess. Except for RU Her, SV Peg, TU And, and Y Crb, all our targets show an excess  in the NUV and in the FUV. The presence of FUV excess in AGB stars is commonly unambiguously attributed to the influence of a binary system, either through the presence of a hot companion or an accretion disk (see Sect.~\ref{intro}). Notably, several stars in our sample exhibit X-ray emission as evidence for binarity (see Sect.~\ref{sample}), and all the stars in the  sample show proper motion anomalies \citep[see][]{Kervella_2019}. Consequently, the uvAGBs in our sample are considered potential AGB binary candidates. Our sample primarily consists of O-rich stars, representing  90\% of the sample. There is also a substantial over-representation of semi-irregular  variable stars, constituting 55\% of the total population. This proportion deviates from that commonly observed in larger samples of AGB stars documented in the literature, where Mira-type variables tend to be more prevalent (see Sect.~\ref{sample}).

We  detect CO emission in a total of 15 targets,   10 of which exhibit emission in both the \docedos\,and \doceuno\,transitions. We observe a trend for the CO non-detections to be associated with the stars with the highest UV excesses. 

The widths of the observed CO line profiles are indicative of expansion velocities in the range \vexp=3-13\,\kms, within the range for AGB stars. Line profiles vary between flat, double-horned, triangular, Gaussian, and parabolic, not following too closely the canonical profile shapes expected from the  standard CSE model (this is not exclusive of uvAGBs, but a common property of O-rich AGBs). At least in one case, RW\,Boo, there is clear evidence of a composite line profile indicative of two kinematic components expanding at 8.0 and 16.6\,\kms, respectively.

For CO detections, we performed a population diagram analysis to derive the mean excitation temperature (\tex) and mass ($M$) of the CO-emitting volume (see Sect.~\ref{anal}). We also estimated in a self-consistent manner the characteristic radius of the CO-emitting layers (\rs) and the CO photo-dissociation radius (\rco) together with the mass-loss rate (\mloss). 

Excitation temperatures \tex$\sim$5-15\,K are prevalent in our targets, except for RR Eri which requires a slightly higher value of \tex$\gtrsim$25\,K. These are probably lower limits to the gas kinetic temperatures in the outer (1-5)\ex{16}\,cm) envelope regions traced by low-$J$ CO transitions. We   determined envelope masses 0.3-55\,$\times 10^{-4}$\msun, which correspond to the amount of mass lost by these objects within the past $\sim$200-2000 years, considering the deduced envelope size and measured expansion velocities.

We find values of the mass-loss rates between $6\times 10^{-8}$ and $3\times 10^{-6}$\,\my\ (i.e. within the range of values reported in previous studies of AGB stars), but there are no objects with extremely high or extremely low mass-loss rates.  
We  explored several relationships between various envelope parameters that are well established for AGB stars. Specifically, we focused on \mloss, \vexp, and the stellar pulsation period (see Sect.~\ref{mloss_rate}). We find that the uvAGBs in our studies closely adhere to these relationships (Fig.~\ref{fig:parameters}). We also estimate the   radiation pressure efficiency $\beta$=\mloss\vexp\,$c$/\lbol\  and found values below 1 for all our targets, consistent with dust-driven winds. However, the efficiency of mass loss is not as high as theoretically possible as it is also observed in many AGB stars.

We investigated correlations between the CO emission intensity and both the infrared emission at 60\micron\ and the distinctive ultraviolet emission of our uvAGBs. 
We corroborate the existence of proportionality between the CO intensity and the \irassesenta\  emission in our sample. However, we observe a slope that is lower compared to previous studies conducted on O-rich and C-rich AGB stars, and closer to the slope observed in post-AGB objects or pPNe. This suggests that uvAGBs may have a lower gas-to-dust mass ratio compared to other AGB stars. This could be attributed to a higher fraction of atomic gas relative to molecular gas (particularly CO), which may be influenced by the presence of excess high-energy radiation emissions such as NUV, FUV, and in some cases X-rays. As for the CO-to-UV emission comparison, we find a tentative anti-correlation of CO with the extinction-corrected NUV emission; however, no clear trend is observed with the FUV excess. The FUV excess shows in general very scattered values and lack of correlation with any of the investigated envelope parameters or with the emission at other wavelengths. This would be in good agreement with the idea that FUV emission does not have an intrinsic origin, and it would be an unequivocal sign of binarity. 

 This is the first (and so far only) dedicated CO-based study of uvAGB stars as a class. We find that our sample of uvAGB stars does not exhibit notable discrepancies when compared to the broader category of AGB stars, except for the different CO-to-IRAS\,60\micron\ trend, which is more similar to that found for pre-PNe. In principle, our findings fit well with the dust-driven wind scenario, and there is no need to invoke alternative mass-loss mechanisms to explain the characteristics of the envelopes around uvAGBs. This conclusion is based on results obtained from single-dish low-$J$ CO line emission observations.

Assuming that the uvAGB stars in our sample are bona fide binaries, our findings indicate that the effects of companions on the outer regions of AGB winds, as traced by the $J$=2--1 and $J$=1--0 transitions, are subtle, and will require higher sensitivity and higher-dynamic range spatially resolved mapping to be identified (as demonstrated by the challenging identification of arcs and spirals within the faint haloes around certain AGB and post-AGB stars, see Sect.~\ref{intro}). The presence of companions can have more noticeable effects in the inner regions ($\lesssim$(100-500) au) of the primary AGB star's wind, which can be studied using higher excitation lines, such as higher-$J$ CO transitions. 

It should also be noted  that the AGB samples with which we have compared ours are not only composed of single AGBs, but certainly include (in an unknown proportion) binary AGB stars of any type, including some uvAGBs. This makes it difficult to adequately isolate   the unique or distinctive characteristics of the molecular envelopes of uvAGB (binary candidate) stars from this comparison. 

Finally, we find that there is a larger proportion of irregular and semi-regular variables among AGB stars with FUV+NUV emission than among AGB stars with only NUV emission (see Sect.~\ref{sample}). This raises the possibility that SR and LB variability in AGB stars may be an indicator of binarity and associated accretion activity that enhances FUV emission. However, a study of a larger sample is needed to explore this tentative result more thoroughly, taking into account the observational bias that larger samples will likely cover a larger distance range, making it more difficult to detect and classify variability and to detect UV emission.

\begin{acknowledgements}
      This research has been funded by grant No. PID2019-105203GB-C22 by the Spanish Ministry of Science and Innovation/State Agency of Research MCIN/AEI/ 10.13039/501100011033 and by “ERDF A way of making Europe”. JAH is supported by INTA grant PRE\_MDM\_05. We thank the \iram\ staff for their support during the observations. We thank an anonymous referee for his/her review which has helped us improve the paper. 
      
      This research has made use of the SIMBAD database, operated at CDS, Strasbourg, France. This work presents results from the European Space Agency (ESA) space mission Gaia. Gaia data are being processed by the Gaia Data Processing and Analysis Consortium (DPAC). Funding for the DPAC is provided by national institutions, in particular the institutions participating in the Gaia MultiLateral Agreement (MLA). The Gaia mission website is \url{https://www.cosmos.esa.int/gaia}. The Gaia archive website is \url{https://archives.esac.esa.int/gaia}. Some of the data presented in this paper were obtained from the Multimission Archive at the Space Telescope Science Institute (MAST). STScI is operated by the Association of Universities for Research in Astronomy, Inc., under NASA contract NAS5-26555. Support for MAST for non-HST data is provided by the NASA Office of Space Science via grant NAG5-7584 and by other grants and contracts. 

      R.S.’s contribution to the research described here was carried out at the Jet Propulsion Laboratory, California Institute of Technology, under a contract with NASA, and funded in part by NASA via ADAP awards, and multiple HST GO awards from the Space Telescope Science Institute.
\end{acknowledgements}

\bibliographystyle{aa} 
\bibliography{main}

\begin{thebibliography}{83}
\expandafter\ifx\csname natexlab\endcsname\relax\def\natexlab#1{#1}\fi

\bibitem[{IRA(1988)}]{IRAS_catalogue}
 1988, {Infrared Astronomical Satellite (IRAS) Catalogs and Atlases.Volume 1: Explanatory Supplement.}, Vol.~1

\bibitem[{{Am{\^o}res} {et~al.}(2021){Am{\^o}res}, {Jesus}, {Moitinho}, {Arsenijevic}, {Levenhagen}, {Marshall}, {Kerber}, {K{\"u}nzel}, \& {Moura}}]{amores_2021}
{Am{\^o}res}, E.~B., {Jesus}, R.~M., {Moitinho}, A., {et~al.} 2021, \mnras, 508, 1788

\bibitem[{{Andriantsaralaza} {et~al.}(2022){Andriantsaralaza}, {Ramstedt}, {Vlemmings}, \& {De Beck}}]{Andriantsaralaza_2022}
{Andriantsaralaza}, M., {Ramstedt}, S., {Vlemmings}, W.~H.~T., \& {De Beck}, E. 2022, \aap, 667, A74

\bibitem[{{Balick} \& {Frank}(2002)}]{Balick_2002}
{Balick}, B. \& {Frank}, A. 2002, \araa, 40, 439

\bibitem[{{Bujarrabal} {et~al.}(1992){Bujarrabal}, {Alcolea}, \& {Planesas}}]{Bujarrabal_1992}
{Bujarrabal}, V., {Alcolea}, J., \& {Planesas}, P. 1992, \aap, 257, 701

\bibitem[{{Bujarrabal} {et~al.}(2001){Bujarrabal}, {Castro-Carrizo}, {Alcolea}, \& {S{\'a}nchez Contreras}}]{Bujarrabal_2001}
{Bujarrabal}, V., {Castro-Carrizo}, A., {Alcolea}, J., \& {S{\'a}nchez Contreras}, C. 2001, \aap, 377, 868

\bibitem[{{Bujarrabal} {et~al.}(1989){Bujarrabal}, {Gomez-Gonzalez}, \& {Planesas}}]{Bujarrabal_1989}
{Bujarrabal}, V., {Gomez-Gonzalez}, J., \& {Planesas}, P. 1989, \aap, 219, 256

\bibitem[{{Bujarrabal} {et~al.}(1986){Bujarrabal}, {Planesas}, {Gomez-Gonzalez}, {Martin-Pintado}, \& {del Romero}}]{Bujarrabal_1986}
{Bujarrabal}, V., {Planesas}, P., {Gomez-Gonzalez}, J., {Martin-Pintado}, J., \& {del Romero}, A. 1986, \aap, 162, 157

\bibitem[{{Carter} {et~al.}(2012){Carter}, {Lazareff}, {Maier}, {Chenu}, {Fontana}, {Bortolotti}, {Boucher}, {Navarrini}, {Blanchet}, {Greve}, {John}, {Kramer}, {Morel}, {Navarro}, {Pe{\~n}alver}, {Schuster}, \& {Thum}}]{Carter_2012}
{Carter}, M., {Lazareff}, B., {Maier}, D., {et~al.} 2012, \aap, 538, A89

\bibitem[{{Castro-Carrizo} {et~al.}(2010){Castro-Carrizo}, {Quintana-Lacaci}, {Neri}, {Bujarrabal}, {Sch{\"o}ier}, {Winters}, {Olofsson}, {Lindqvist}, {Alcolea}, {Lucas}, \& {Grewing}}]{Castro-Carrizo_2010}
{Castro-Carrizo}, A., {Quintana-Lacaci}, G., {Neri}, R., {et~al.} 2010, \aap, 523, A59

\bibitem[{{Conti} {et~al.}(2011){Conti}, {Bianchi}, \& {Shiao}}]{Conti_2011}
{Conti}, A., {Bianchi}, L., \& {Shiao}, B. 2011, \apss, 335, 329

\bibitem[{{da Silva Santos} {et~al.}(2019){da Silva Santos}, {Ramos-Medina}, {S{\'a}nchez Contreras}, \& {Garc{\'\i}a-Lario}}]{da_Silva_Santos_2019}
{da Silva Santos}, J.~M., {Ramos-Medina}, J., {S{\'a}nchez Contreras}, C., \& {Garc{\'\i}a-Lario}, P. 2019, \aap, 622, A123

\bibitem[{{Danilovich} {et~al.}(2015){Danilovich}, {Teyssier}, {Justtanont}, {Olofsson}, {Cerrigone}, {Bujarrabal}, {Alcolea}, {Cernicharo}, {Castro-Carrizo}, {Garc{\'\i}a-Lario}, \& {Marston}}]{Danilovich_2015}
{Danilovich}, T., {Teyssier}, D., {Justtanont}, K., {et~al.} 2015, \aap, 581, A60

\bibitem[{{De Marco}(2009)}]{De_Marco_2009}
{De Marco}, O. 2009, \pasp, 121, 316

\bibitem[{{Decin} {et~al.}(2020){Decin}, {Montarg{\`e}s}, {Richards}, {Gottlieb}, {Homan}, {McDonald}, {El Mellah}, {Danilovich}, {Wallstr{\"o}m}, {Zijlstra}, {Baudry}, {Bolte}, {Cannon}, {De Beck}, {De Ceuster}, {de Koter}, {De Ridder}, {Etoka}, {Gobrecht}, {Gray}, {Herpin}, {Jeste}, {Lagadec}, {Kervella}, {Khouri}, {Menten}, {Millar}, {M{\"u}ller}, {Plane}, {Sahai}, {Sana}, {Van de Sande}, {Waters}, {Wong}, \& {Yates}}]{Decin_2020}
{Decin}, L., {Montarg{\`e}s}, M., {Richards}, A.~M.~S., {et~al.} 2020, Science, 369, 1497

\bibitem[{{D{\'\i}az-Luis} {et~al.}(2019){D{\'\i}az-Luis}, {Alcolea}, {Bujarrabal}, {Santander-Garc{\'\i}a}, {Castro-Carrizo}, {G{\'o}mez-Garrido}, \& {Desmurs}}]{diaz-luis_2019}
{D{\'\i}az-Luis}, J.~J., {Alcolea}, J., {Bujarrabal}, V., {et~al.} 2019, \aap, 629, A94

\bibitem[{{Duquennoy} \& {Mayor}(1991)}]{Duquennoy_1991}
{Duquennoy}, A. \& {Mayor}, M. 1991, \aap, 248, 485

\bibitem[{{El-Badry} {et~al.}(2021){El-Badry}, {Rix}, \& {Heintz}}]{El-Badry_2021}
{El-Badry}, K., {Rix}, H.-W., \& {Heintz}, T.~M. 2021, \mnras, 506, 2269

\bibitem[{{Gaia Collaboration} {et~al.}(2021){Gaia Collaboration}, {Brown}, {Vallenari}, {Prusti}, {de Bruijne}, {Babusiaux}, {Biermann}, {Creevey}, {Evans}, {Eyer}, {Hutton}, {Jansen}, {Jordi}, {Klioner}, {Lammers}, {Lindegren}, {Luri}, {Mignard}, {Panem}, {Pourbaix}, {Randich}, {Sartoretti}, {Soubiran}, {Walton}, {Arenou}, {Bailer-Jones}, {Bastian}, {Cropper}, {Drimmel}, {Katz}, {Lattanzi}, {van Leeuwen}, {Bakker}, {Cacciari}, {Casta{\~n}eda}, {De Angeli}, {Ducourant}, {Fabricius}, {Fouesneau}, {Fr{\'e}mat}, {Guerra}, {Guerrier}, {Guiraud}, {Jean-Antoine Piccolo}, {Masana}, {Messineo}, {Mowlavi}, {Nicolas}, {Nienartowicz}, {Pailler}, {Panuzzo}, {Riclet}, {Roux}, {Seabroke}, {Sordo}, {Tanga}, {Th{\'e}venin}, {Gracia-Abril}, {Portell}, {Teyssier}, {Altmann}, {Andrae}, {Bellas-Velidis}, {Benson}, {Berthier}, {Blomme}, {Brugaletta}, {Burgess}, {Busso}, {Carry}, {Cellino}, {Cheek}, {Clementini}, {Damerdji}, {Davidson}, {Delchambre}, {Dell'Oro}, {Fern{\'a}ndez-Hern{\'a}ndez}, {Galluccio}, {Garc{\'\i}a-Lario},
  {Garcia-Reinaldos}, {Gonz{\'a}lez-N{\'u}{\~n}ez}, {Gosset}, {Haigron}, {Halbwachs}, {Hambly}, {Harrison}, {Hatzidimitriou}, {Heiter}, {Hern{\'a}ndez}, {Hestroffer}, {Hodgkin}, {Holl}, {Jan{\ss}en}, {Jevardat de Fombelle}, {Jordan}, {Krone-Martins}, {Lanzafame}, {L{\"o}ffler}, {Lorca}, {Manteiga}, {Marchal}, {Marrese}, {Moitinho}, {Mora}, {Muinonen}, {Osborne}, {Pancino}, {Pauwels}, {Petit}, {Recio-Blanco}, {Richards}, {Riello}, {Rimoldini}, {Robin}, {Roegiers}, {Rybizki}, {Sarro}, {Siopis}, {Smith}, {Sozzetti}, {Ulla}, {Utrilla}, {van Leeuwen}, {van Reeven}, {Abbas}, {Abreu Aramburu}, {Accart}, {Aerts}, {Aguado}, {Ajaj}, {Altavilla}, {{\'A}lvarez}, {{\'A}lvarez Cid-Fuentes}, {Alves}, {Anderson}, {Anglada Varela}, {Antoja}, {Audard}, {Baines}, {Baker}, {Balaguer-N{\'u}{\~n}ez}, {Balbinot}, {Balog}, {Barache}, {Barbato}, {Barros}, {Barstow}, {Bartolom{\'e}}, {Bassilana}, {Bauchet}, {Baudesson-Stella}, {Becciani}, {Bellazzini}, {Bernet}, {Bertone}, {Bianchi}, {Blanco-Cuaresma}, {Boch}, {Bombrun}, {Bossini},
  {Bouquillon}, {Bragaglia}, {Bramante}, {Breedt}, {Bressan}, {Brouillet}, {Bucciarelli}, {Burlacu}, {Busonero}, {Butkevich}, {Buzzi}, {Caffau}, {Cancelliere}, {C{\'a}novas}, {Cantat-Gaudin}, {Carballo}, {Carlucci}, {Carnerero}, {Carrasco}, {Casamiquela}, {Castellani}, {Castro-Ginard}, {Castro Sampol}, {Chaoul}, {Charlot}, {Chemin}, {Chiavassa}, {Cioni}, {Comoretto}, {Cooper}, {Cornez}, {Cowell}, {Crifo}, {Crosta}, {Crowley}, {Dafonte}, {Dapergolas}, {David}, {David}, {de Laverny}, {De Luise}, {De March}, {De Ridder}, {de Souza}, {de Teodoro}, {de Torres}, {del Peloso}, {del Pozo}, {Delbo}, {Delgado}, {Delgado}, {Delisle}, {Di Matteo}, {Diakite}, {Diener}, {Distefano}, {Dolding}, {Eappachen}, {Edvardsson}, {Enke}, {Esquej}, {Fabre}, {Fabrizio}, {Faigler}, {Fedorets}, {Fernique}, {Fienga}, {Figueras}, {Fouron}, {Fragkoudi}, {Fraile}, {Franke}, {Gai}, {Garabato}, {Garcia-Gutierrez}, {Garc{\'\i}a-Torres}, {Garofalo}, {Gavras}, {Gerlach}, {Geyer}, {Giacobbe}, {Gilmore}, {Girona}, {Giuffrida}, {Gomel}, {Gomez},
  {Gonzalez-Santamaria}, {Gonz{\'a}lez-Vidal}, {Granvik}, {Guti{\'e}rrez-S{\'a}nchez}, {Guy}, {Hauser}, {Haywood}, {Helmi}, {Hidalgo}, {Hilger}, {H{\l}adczuk}, {Hobbs}, {Holland}, {Huckle}, {Jasniewicz}, {Jonker}, {Juaristi Campillo}, {Julbe}, {Karbevska}, {Kervella}, {Khanna}, {Kochoska}, {Kontizas}, {Kordopatis}, {Korn}, {Kostrzewa-Rutkowska}, {Kruszy{\'n}ska}, {Lambert}, {Lanza}, {Lasne}, {Le Campion}, {Le Fustec}, {Lebreton}, {Lebzelter}, {Leccia}, {Leclerc}, {Lecoeur-Taibi}, {Liao}, {Licata}, {Lindstr{\o}m}, {Lister}, {Livanou}, {Lobel}, {Madrero Pardo}, {Managau}, {Mann}, {Marchant}, {Marconi}, {Marcos Santos}, {Marinoni}, {Marocco}, {Marshall}, {Martin Polo}, {Mart{\'\i}n-Fleitas}, {Masip}, {Massari}, {Mastrobuono-Battisti}, {Mazeh}, {McMillan}, {Messina}, {Michalik}, {Millar}, {Mints}, {Molina}, {Molinaro}, {Moln{\'a}r}, {Montegriffo}, {Mor}, {Morbidelli}, {Morel}, {Morris}, {Mulone}, {Munoz}, {Muraveva}, {Murphy}, {Musella}, {Noval}, {Ord{\'e}novic}, {Orr{\`u}}, {Osinde}, {Pagani}, {Pagano},
  {Palaversa}, {Palicio}, {Panahi}, {Pawlak}, {Pe{\~n}alosa Esteller}, {Penttil{\"a}}, {Piersimoni}, {Pineau}, {Plachy}, {Plum}, {Poggio}, {Poretti}, {Poujoulet}, {Pr{\v{s}}a}, {Pulone}, {Racero}, {Ragaini}, {Rainer}, {Raiteri}, {Rambaux}, {Ramos}, {Ramos-Lerate}, {Re Fiorentin}, {Regibo}, {Reyl{\'e}}, {Ripepi}, {Riva}, {Rixon}, {Robichon}, {Robin}, {Roelens}, {Rohrbasser}, {Romero-G{\'o}mez}, {Rowell}, {Royer}, {Rybicki}, {Sadowski}, {Sagrist{\`a} Sell{\'e}s}, {Sahlmann}, {Salgado}, {Salguero}, {Samaras}, {Sanchez Gimenez}, {Sanna}, {Santove{\~n}a}, {Sarasso}, {Schultheis}, {Sciacca}, {Segol}, {Segovia}, {S{\'e}gransan}, {Semeux}, {Shahaf}, {Siddiqui}, {Siebert}, {Siltala}, {Slezak}, {Smart}, {Solano}, {Solitro}, {Souami}, {Souchay}, {Spagna}, {Spoto}, {Steele}, {Steidelm{\"u}ller}, {Stephenson}, {S{\"u}veges}, {Szabados}, {Szegedi-Elek}, {Taris}, {Tauran}, {Taylor}, {Teixeira}, {Thuillot}, {Tonello}, {Torra}, {Torra}, {Turon}, {Unger}, {Vaillant}, {van Dillen}, {Vanel}, {Vecchiato}, {Viala}, {Vicente},
  {Voutsinas}, {Weiler}, {Wevers}, {Wyrzykowski}, {Yoldas}, {Yvard}, {Zhao}, {Zorec}, {Zucker}, {Zurbach}, \& {Zwitter}}]{GAIA}
{Gaia Collaboration}, {Brown}, A.~G.~A., {Vallenari}, A., {et~al.} 2021, \aap, 649, A1

\bibitem[{{Goldsmith} \& {Langer}(1999)}]{goldsmith_1999}
{Goldsmith}, P.~F. \& {Langer}, W.~D. 1999, \apj, 517, 209

\bibitem[{{Green} {et~al.}(2019){Green}, {Schlafly}, {Zucker}, {Speagle}, \& {Finkbeiner}}]{Green_2019}
{Green}, G.~M., {Schlafly}, E., {Zucker}, C., {Speagle}, J.~S., \& {Finkbeiner}, D. 2019, \apj, 887, 93

\bibitem[{{Groenewegen}(2017)}]{Groenewegen_2017}
{Groenewegen}, M.~A.~T. 2017, \aap, 606, A67

\bibitem[{{Groenewegen} {et~al.}(2002{\natexlab{a}}){Groenewegen}, {Sevenster}, {Spoon}, \& {P{\'e}rez}}]{Groenewegen_2002a}
{Groenewegen}, M.~A.~T., {Sevenster}, M., {Spoon}, H.~W.~W., \& {P{\'e}rez}, I. 2002{\natexlab{a}}, \aap, 390, 501

\bibitem[{{Groenewegen} {et~al.}(2002{\natexlab{b}}){Groenewegen}, {Sevenster}, {Spoon}, \& {P{\'e}rez}}]{Groenewegen_2002b}
{Groenewegen}, M.~A.~T., {Sevenster}, M., {Spoon}, H.~W.~W., \& {P{\'e}rez}, I. 2002{\natexlab{b}}, \aap, 390, 511

\bibitem[{{Groenewegen} \& {Sloan}(2018)}]{groenewegen_2018}
{Groenewegen}, M.~A.~T. \& {Sloan}, G.~C. 2018, \aap, 609, A114

\bibitem[{{Habing} {et~al.}(1994){Habing}, {Tignon}, \& {Tielens}}]{Habing_1994}
{Habing}, H.~J., {Tignon}, J., \& {Tielens}, A.~G.~G.~M. 1994, \aap, 286, 523

\bibitem[{{Hillwig}(2018)}]{Hillwig_2018}
{Hillwig}, T. 2018, Galaxies, 6, 85

\bibitem[{{H{\"o}fner} \& {Olofsson}(2018)}]{Hofner_2018}
{H{\"o}fner}, S. \& {Olofsson}, H. 2018, \aapr, 26, 1

\bibitem[{{Hunsch} {et~al.}(1998){Hunsch}, {Schmitt}, {Schroder}, \& {Zickgraf}}]{Hunsch_1998}
{Hunsch}, M., {Schmitt}, J. H.~M.~M., {Schroder}, K.-P., \& {Zickgraf}, F.-J. 1998, \aap, 330, 225

\bibitem[{{Justtanont} {et~al.}(2000){Justtanont}, {Barlow}, {Tielens}, {Hollenbach}, {Latter}, {Liu}, {Sylvester}, {Cox}, \& {Rieu}}]{Justtanont_2000}
{Justtanont}, K., {Barlow}, M.~J., {Tielens}, A.~G.~G.~M., {et~al.} 2000, \aap, 360, 1117

\bibitem[{{Kahane} \& {Jura}(1994)}]{Kahane_1994}
{Kahane}, C. \& {Jura}, M. 1994, \aap, 290, 183

\bibitem[{{Kastner}(1992)}]{Kastner_1992}
{Kastner}, J.~H. 1992, \apj, 401, 337

\bibitem[{{Kerschbaum} \& {Olofsson}(1999)}]{Kerschbaum_1999}
{Kerschbaum}, F. \& {Olofsson}, H. 1999, \aaps, 138, 299

\bibitem[{{Kerschbaum} {et~al.}(1996){Kerschbaum}, {Olofsson}, \& {Hron}}]{Kerschbaum_1996}
{Kerschbaum}, F., {Olofsson}, H., \& {Hron}, J. 1996, \aap, 311, 273

\bibitem[{{Kervella} {et~al.}(2019){Kervella}, {Arenou}, {Mignard}, \& {Th{\'e}venin}}]{Kervella_2019}
{Kervella}, P., {Arenou}, F., {Mignard}, F., \& {Th{\'e}venin}, F. 2019, \aap, 623, A72

\bibitem[{{Kim} \& {Taam}(2012)}]{Kim_2012}
{Kim}, H. \& {Taam}, R.~E. 2012, \apj, 744, 136

\bibitem[{{Knapp}(1986)}]{Knapp_1986}
{Knapp}, G.~R. 1986, \apj, 311, 731

\bibitem[{{Knapp} \& {Morris}(1985)}]{knapp1985}
{Knapp}, G.~R. \& {Morris}, M. 1985, \apj, 292, 640

\bibitem[{{Knapp} {et~al.}(1998){Knapp}, {Young}, {Lee}, \& {Jorissen}}]{Knapp_1998}
{Knapp}, G.~R., {Young}, K., {Lee}, E., \& {Jorissen}, A. 1998, \apjs, 117, 209

\bibitem[{{Lefevre}(1989)}]{Lefevre_1989}
{Lefevre}, J. 1989, \aap, 219, 265

\bibitem[{{Lima} {et~al.}(2022){Lima}, {Luna}, \& {Nu{\~n}ez}}]{Lima_2022}
{Lima}, I.~J., {Luna}, G.~J.~M., \& {Nu{\~n}ez}, N.~E. 2022, The Astronomer's Telegram, 15332, 1

\bibitem[{{Loup} {et~al.}(1993){Loup}, {Forveille}, {Omont}, \& {Paul}}]{Loup_1993}
{Loup}, C., {Forveille}, T., {Omont}, A., \& {Paul}, J.~F. 1993, \aaps, 99, 291

\bibitem[{{Mamon} {et~al.}(1988){Mamon}, {Glassgold}, \& {Huggins}}]{Mamon_1988}
{Mamon}, G.~A., {Glassgold}, A.~E., \& {Huggins}, P.~J. 1988, \apj, 328, 797

\bibitem[{{Margulis} {et~al.}(1990){Margulis}, {van Blerkom}, {Snell}, \& {Kleinmann}}]{Margulis_1990}
{Margulis}, M., {van Blerkom}, D.~J., {Snell}, R.~L., \& {Kleinmann}, S.~G. 1990, \apj, 361, 673

\bibitem[{{Martin} {et~al.}(2005){Martin}, {Fanson}, {Schiminovich}, {Morrissey}, {Friedman}, {Barlow}, {Conrow}, {Grange}, {Jelinsky}, {Milliard}, {Siegmund}, {Bianchi}, {Byun}, {Donas}, {Forster}, {Heckman}, {Lee}, {Madore}, {Malina}, {Neff}, {Rich}, {Small}, {Surber}, {Szalay}, {Welsh}, \& {Wyder}}]{GALEX}
{Martin}, D.~C., {Fanson}, J., {Schiminovich}, D., {et~al.} 2005, \apjl, 619, L1

\bibitem[{{Montez} {et~al.}(2017){Montez}, {Ramstedt}, {Kastner}, {Vlemmings}, \& {Sanchez}}]{Montez_2017}
{Montez}, Rodolfo, J., {Ramstedt}, S., {Kastner}, J.~H., {Vlemmings}, W., \& {Sanchez}, E. 2017, \apj, 841, 33

\bibitem[{{Neri} {et~al.}(1998){Neri}, {Kahane}, {Lucas}, {Bujarrabal}, \& {Loup}}]{Neri_1998}
{Neri}, R., {Kahane}, C., {Lucas}, R., {Bujarrabal}, V., \& {Loup}, C. 1998, \aaps, 130, 1

\bibitem[{{Netzer} \& {Elitzur}(1993)}]{Netzer_1993}
{Netzer}, N. \& {Elitzur}, M. 1993, \apj, 410, 701

\bibitem[{{Neugebauer} {et~al.}(1984){Neugebauer}, {Habing}, {van Duinen}, {Aumann}, {Baud}, {Beichman}, {Beintema}, {Boggess}, {Clegg}, {de Jong}, {Emerson}, {Gautier}, {Gillett}, {Harris}, {Hauser}, {Houck}, {Jennings}, {Low}, {Marsden}, {Miley}, {Olnon}, {Pottasch}, {Raimond}, {Rowan-Robinson}, {Soifer}, {Walker}, {Wesselius}, \& {Young}}]{Neugebauer_1984}
{Neugebauer}, G., {Habing}, H.~J., {van Duinen}, R., {et~al.} 1984, \apjl, 278, L1

\bibitem[{{Nyman} {et~al.}(1992){Nyman}, {Booth}, {Carlstrom}, {Habing}, {Heske}, {Sahai}, {Stark}, {van der Veen}, \& {Winnberg}}]{Nyman_1992}
{Nyman}, L.~A., {Booth}, R.~S., {Carlstrom}, U., {et~al.} 1992, \aaps, 93, 121

\bibitem[{{Ochsenbein} {et~al.}(2000){Ochsenbein}, {Bauer}, \& {Marcout}}]{vizier}
{Ochsenbein}, F., {Bauer}, P., \& {Marcout}, J. 2000, \aaps, 143, 23

\bibitem[{{Olofsson} {et~al.}(1987){Olofsson}, {Eriksson}, \& {Gustafsson}}]{Olofsson_1987}
{Olofsson}, H., {Eriksson}, K., \& {Gustafsson}, B. 1987, \aap, 183, L13

\bibitem[{{Olofsson} {et~al.}(1993){Olofsson}, {Eriksson}, {Gustafsson}, \& {Carlstrom}}]{Olofsson_1993}
{Olofsson}, H., {Eriksson}, K., {Gustafsson}, B., \& {Carlstrom}, U. 1993, \apjs, 87, 267

\bibitem[{{Olofsson} {et~al.}(2002){Olofsson}, {Gonz{\'a}lez Delgado}, {Kerschbaum}, \& {Sch{\"o}ier}}]{Olofsson_2002}
{Olofsson}, H., {Gonz{\'a}lez Delgado}, D., {Kerschbaum}, F., \& {Sch{\"o}ier}, F.~L. 2002, \aap, 391, 1053

\bibitem[{{Ortiz} \& {Guerrero}(2016)}]{Ortiz_2016}
{Ortiz}, R. \& {Guerrero}, M.~A. 2016, \mnras, 461, 3036

\bibitem[{{Ortiz} \& {Guerrero}(2021)}]{Ortiz_2021}
{Ortiz}, R. \& {Guerrero}, M.~A. 2021, \apj, 912, 93

\bibitem[{{Ortiz} {et~al.}(2019){Ortiz}, {Guerrero}, \& {Costa}}]{Ortiz_2019}
{Ortiz}, R., {Guerrero}, M.~A., \& {Costa}, R. D.~D. 2019, \mnras, 482, 4697

\bibitem[{{Planesas} {et~al.}(1990){Planesas}, {Bachiller}, {Martin-Pintado}, \& {Bujarrabal}}]{Planesas_1990}
{Planesas}, P., {Bachiller}, R., {Martin-Pintado}, J., \& {Bujarrabal}, V. 1990, \apj, 351, 263

\bibitem[{{Ramos-Medina} {et~al.}(2018){Ramos-Medina}, {S{\'a}nchez Contreras}, {Garc{\'\i}a-Lario}, \& {da Silva Santos}}]{ramos-medina_2018}
{Ramos-Medina}, J., {S{\'a}nchez Contreras}, C., {Garc{\'\i}a-Lario}, P., \& {da Silva Santos}, J.~M. 2018, \aap, 618, A171

\bibitem[{{Ramstedt} {et~al.}(2012){Ramstedt}, {Montez}, {Kastner}, \& {Vlemmings}}]{Ramstedt_2012}
{Ramstedt}, S., {Montez}, R., {Kastner}, J., \& {Vlemmings}, W.~H.~T. 2012, \aap, 543, A147

\bibitem[{{Ramstedt} {et~al.}(2020){Ramstedt}, {Vlemmings}, {Doan}, {Danilovich}, {Lindqvist}, {Saberi}, {Olofsson}, {De Beck}, {Groenewegen}, {H{\"o}fner}, {Kastner}, {Kerschbaum}, {Khouri}, {Maercker}, {Montez}, {Quintana-Lacaci}, {Sahai}, {Tafoya}, \& {Zijlstra}}]{Ramstedt_2020}
{Ramstedt}, S., {Vlemmings}, W.~H.~T., {Doan}, L., {et~al.} 2020, \aap, 640, A133

\bibitem[{{Sahai} {et~al.}(2008){Sahai}, {Findeisen}, {Gil de Paz}, \& {S{\'a}nchez Contreras}}]{Sahai_2008}
{Sahai}, R., {Findeisen}, K., {Gil de Paz}, A., \& {S{\'a}nchez Contreras}, C. 2008, \apj, 689, 1274

\bibitem[{{Sahai} {et~al.}(2007){Sahai}, {Morris}, {S{\'a}nchez Contreras}, \& {Claussen}}]{Sahai_2007}
{Sahai}, R., {Morris}, M., {S{\'a}nchez Contreras}, C., \& {Claussen}, M. 2007, \aj, 134, 2200

\bibitem[{{Sahai} {et~al.}(2011){Sahai}, {Neill}, {Gil de Paz}, \& {S{\'a}nchez Contreras}}]{Sahai_2011}
{Sahai}, R., {Neill}, J.~D., {Gil de Paz}, A., \& {S{\'a}nchez Contreras}, C. 2011, \apjl, 740, L39

\bibitem[{{Sahai} {et~al.}(2018){Sahai}, {S{\'a}nchez Contreras}, {Mangan}, {Sanz-Forcada}, {Muthumariappan}, \& {Claussen}}]{Sahai_2018}
{Sahai}, R., {S{\'a}nchez Contreras}, C., {Mangan}, A.~S., {et~al.} 2018, \apj, 860, 105

\bibitem[{{Sahai} {et~al.}(2022){Sahai}, {Sanz-Forcada}, {Guerrero}, {Ortiz}, \& {Contreras}}]{Sahai_2022}
{Sahai}, R., {Sanz-Forcada}, J., {Guerrero}, M., {Ortiz}, R., \& {Contreras}, C.~S. 2022, Galaxies, 10, 62

\bibitem[{{Sahai} {et~al.}(2016){Sahai}, {Sanz-Forcada}, \& {S{\'a}nchez Contreras}}]{Sahai_2016}
{Sahai}, R., {Sanz-Forcada}, J., \& {S{\'a}nchez Contreras}, C. 2016, in Journal of Physics Conference Series, Vol. 728, Journal of Physics Conference Series, 042003

\bibitem[{{Sahai} {et~al.}(2015){Sahai}, {Sanz-Forcada}, {S{\'a}nchez Contreras}, \& {Stute}}]{Sahai_2015}
{Sahai}, R., {Sanz-Forcada}, J., {S{\'a}nchez Contreras}, C., \& {Stute}, M. 2015, \apj, 810, 77

\bibitem[{{Samus'} {et~al.}(2017){Samus'}, {Kazarovets}, {Durlevich}, {Kireeva}, \& {Pastukhova}}]{GCVS}
{Samus'}, N.~N., {Kazarovets}, E.~V., {Durlevich}, O.~V., {Kireeva}, N.~N., \& {Pastukhova}, E.~N. 2017, Astronomy Reports, 61, 80

\bibitem[{{S{\'a}nchez Contreras} \& {Sahai}(2012)}]{sanchez-contreras_2012}
{S{\'a}nchez Contreras}, C. \& {Sahai}, R. 2012, \apjs, 203, 16

\bibitem[{{Schlegel} {et~al.}(1998){Schlegel}, {Finkbeiner}, \& {Davis}}]{Schlegel_1998}
{Schlegel}, D.~J., {Finkbeiner}, D.~P., \& {Davis}, M. 1998, \apj, 500, 525

\bibitem[{{Sch{\"o}ier} \& {Olofsson}(2001)}]{schoier_2001}
{Sch{\"o}ier}, F.~L. \& {Olofsson}, H. 2001, \aap, 368, 969

\bibitem[{{Scicluna} {et~al.}(2022){Scicluna}, {Kemper}, {McDonald}, {Srinivasan}, {Trejo}, {Wallstr{\"o}m}, {Wouterloot}, {Cami}, {Greaves}, {He}, {Hoai}, {Kim}, {Jones}, {Shinnaga}, {Clark}, {Dharmawardena}, {Holland}, {Imai}, {van Loon}, {Menten}, {Wesson}, {Chawner}, {Feng}, {Goldman}, {Liu}, {MacIsaac}, {Tang}, {Zeegers}, {Amada}, {Antoniou}, {Bemis}, {Boyer}, {Chapman}, {Chen}, {Cho}, {Cui}, {Dell'Agli}, {Friberg}, {Fukaya}, {Gomez}, {Gong}, {Hadjara}, {Haswell}, {Hirano}, {Hony}, {Izumiura}, {Jeste}, {Jiang}, {Kaminski}, {Keaveney}, {Kim}, {Kraemer}, {Kuan}, {Lagadec}, {Lee}, {Li}, {Liu}, {Liu}, {de Looze}, {Lykou}, {Maraston}, {Marshall}, {Matsuura}, {Min}, {Otsuka}, {Oyadomari}, {Parsons}, {Patel}, {Peeters}, {Pham}, {Qiu}, {Randall}, {Rau}, {Redman}, {Richards}, {Serjeant}, {Shi}, {Sloan}, {Smith}, {Suh}, {Toal{\'a}}, {Uttenthaler}, {Ventura}, {Wang}, {Yamamura}, {Yang}, {Yun}, {Zhang}, {Zhang}, {Zhao}, {Zhu}, \& {Zijlstra}}]{Scicluna_2022}
{Scicluna}, P., {Kemper}, F., {McDonald}, I., {et~al.} 2022, \mnras, 512, 1091

\bibitem[{{Soker}(1994)}]{Soker_1994}
{Soker}, N. 1994, \mnras, 270, 774

\bibitem[{{Suh}(2021)}]{Suh_2021}
{Suh}, K.-W. 2021, \apjs, 256, 43

\bibitem[{{Taylor}(2005)}]{topcat}
{Taylor}, M.~B. 2005, in Astronomical Society of the Pacific Conference Series, Vol. 347, Astronomical Data Analysis Software and Systems XIV, ed. P.~{Shopbell}, M.~{Britton}, \& R.~{Ebert}, 29

\bibitem[{{Van de Sande} {et~al.}(2021){Van de Sande}, {Walsh}, \& {Millar}}]{Van_de_Sande_2021}
{Van de Sande}, M., {Walsh}, C., \& {Millar}, T.~J. 2021, \mnras, 501, 491

\bibitem[{{van der Veen} \& {Habing}(1988)}]{van_der_veen_1988}
{van der Veen}, W.~E.~C.~J. \& {Habing}, H.~J. 1988, \aap, 194, 125

\bibitem[{{Vassiliadis} \& {Wood}(1993)}]{Vassiliadis_1993}
{Vassiliadis}, E. \& {Wood}, P.~R. 1993, \apj, 413, 641

\bibitem[{{Wenger} {et~al.}(2000){Wenger}, {Ochsenbein}, {Egret}, {Dubois}, {Bonnarel}, {Borde}, {Genova}, {Jasniewicz}, {Lalo{\"e}}, {Lesteven}, \& {Monier}}]{SIMBAD}
{Wenger}, M., {Ochsenbein}, F., {Egret}, D., {et~al.} 2000, \aaps, 143, 9

\bibitem[{{Winters} {et~al.}(2003){Winters}, {Le Bertre}, {Jeong}, {Nyman}, \& {Epchtein}}]{Winters_2003}
{Winters}, J.~M., {Le Bertre}, T., {Jeong}, K.~S., {Nyman}, L.~{\r{A}}., \& {Epchtein}, N. 2003, \aap, 409, 715

\bibitem[{{Young}(1995)}]{Young_1995}
{Young}, K. 1995, \apj, 445, 872

\bibitem[{{Zuckerman} \& {Dyck}(1986)}]{Zuckerman_1986}
{Zuckerman}, B. \& {Dyck}, H.~M. 1986, \apj, 304, 394

\end{thebibliography}




\begin{appendix}

\onecolumn

\section{Detection limits for non-detected sources}

\begin{table*}[h!]

\renewcommand{\arraystretch}{1.2}
\centering

\begin{adjustbox}{}

\begin{threeparttable}[b]

\caption{Spectral measurements for non-detected sources.}

\begin{tabular}{l c c c c c c c c c c}
\hline\hline 
Source & line &  rms & & Source & line &  rms & & Source & line &  rms  \\ 
 & &   $(\mathrm{mK})$ & &  & &   $(\mathrm{mK})$ & &  & &   $(\mathrm{mK})$  \\
\hline 

\noalign{\vspace{0.25cm}}

AT\,Dra & \doceuno & $16$  & & DF\,Leo & \doceuno & $9$  & & ST\,Uma & \doceuno & $9$  \\

AT\,Dra & \docedos & $13$  & & DF\,Leo & \docedos & $11$ & & ST\,Uma & \docedos & $14$  \\

\noalign{\vspace{0.25cm}}

BC\,Cmi & \doceuno & $14$  & & FH\,Vir & \doceuno & $10$ & & TU\,And & \doceuno & $30$  \\

BC\,Cmi & \docedos & $20$  & & FH\,Vir & \docedos & $9$  & & TU\,And & \docedos & $40$  \\

\noalign{\vspace{0.25cm}}

BD\,Cam & \doceuno & $9$  & & IN\,Hya & \doceuno & $11$  & & UY\,Leo & \doceuno & $16$  \\

BD\,Cam & \docedos & $13$ & & IN\,Hya & \docedos & $12$  & & UY\,Leo & \docedos & $50$  \\

\noalign{\vspace{0.25cm}}

BY\,Boo & \doceuno & $13$  & & OME\,Vir & \doceuno & $14$  & & Y\,Gem & \doceuno & $9$  \\

BY\,Boo & \docedos & $30$  & & OME\,Vir & \docedos & $20$  & & Y\,Gem & \docedos & $12$  \\

\noalign{\vspace{0.25cm}}

CG\,Uma & \doceuno & $9$  & & RR\,Umi & \doceuno & $15$  \\

CG\,Uma & \docedos & $14$ & & RR\,Umi & \docedos & $12$  \\
\hline
\end{tabular}
\label{tab:spectral_analysis}

\end{threeparttable}
\end{adjustbox}

\renewcommand{\arraystretch}{1.0}

\end{table*}

\section{Ancyllary data} \label{sources fluxes}

Infrared fluxes (at 12\,$\mu$m, 25\,$\mu$m, 60\,$\mu$m and 100 \,$\mu$m) of our targets are from the Infrared Astronomical Satellite \citep[IRAS,][]{Neugebauer_1984} and obtained from the IRAS Point Source Catalog v2.1 \citep[PSC,][]{IRAS_catalogue}. Ultraviolet, NUV and FUV, GALEX multi-epoch fluxes are retrieved from the GALEX MAST archive \cite{Conti_2011}. 
IRAS and GALEX photometry for our targets is presented in Table~\ref{tab:fluxes}.

The GALEX MAST archive provides both magnitudes in AB system and fluxes in ${\rm \mu Jy}$, the conversion between them is: \footnote{for more information see {\tt \url{https://asd.gsfc.nasa.gov/archive/galex/}}}

\begin{equation}
  log_{10}f_{\nu}= -0.4 m_{{\rm AB}} + 9.56
\end{equation}

In addition, the extinction correction can be performed over the observed fluxes following the equation:

\begin{equation}
    f_{\nu}^{\rm corr}=f_{\nu}\times 10^{0.4 A_{V}} = f_{\nu}\times 10^{0.4 R_{V}E(B-V)}
\end{equation}

Finally, the luminosity of each source can be estimated for each frequency following the equation: 

\begin{equation}
    L_{\nu}=f_{\nu} \times 4 \pi D^{2}
\end{equation}

Added by TeX Support
\begin{longtable}{l c c c c c c c}
\caption{\label{tab:fluxes} Fluxes used in this study.} \\
\hline\hline
Source & IRAS $[\mathrm{12 \mu m}]$ & IRAS $[\mathrm{25 \mu m}]$ & IRAS $[\mathrm{60 \mu m}]$ & IRAS $[\mathrm{100 \mu m}]$ & Galex epoch & GALEX NUV & GALEX FUV  \\  
       & $(\mathrm{Jy})$ & $(\mathrm{Jy})$ & $(\mathrm{Jy})$ &  $(\mathrm{Jy})$ & $(\mathrm{Jy})$ &  $(\mathrm{\mu Jy})$ & $(\mathrm{\mu Jy})$    \\ 
\hline
\endfirsthead
\caption{continued.}\\
\hline\hline
Source & IRAS $[\mathrm{12 \mu m}]$ & IRAS $[\mathrm{25 \mu m}]$ & IRAS $[\mathrm{60 \mu m}]$ & IRAS $[\mathrm{100 \mu m}]$ & Galex epoch & GALEX NUV & GALEX FUV  \\  
       & $(\mathrm{Jy})$ & $(\mathrm{Jy})$ & $(\mathrm{Jy})$ &  $(\mathrm{Jy})$ & $(\mathrm{Jy})$ &  $(\mathrm{\mu Jy})$ & $(\mathrm{\mu Jy})$    \\
\hline

\endhead
\hline
\endfoot

AT\,Dra & $48.1\pm1.4$ & $13.1\pm0.5$ & $2.19\pm0.15 $ & $0.9\pm0.3 $ & 2011-07-25 & $2145\pm 8$ & \\
 & & & & & 2004-06-26 & $2217 \pm17$ & $172 \pm9 $ \\
 & & & & & 2004-06-26 & $2208 \pm16 $ & $181 \pm9$ \\
BC\,Cmi & $23.4\pm0.9 $ & $6.6\pm0.4$ & $0.94\pm0.08 $ & $\leq 1.0 $ & 2006-01-01 & $1013 \pm16$ & $55 \pm5 $ \\
BD\,Cam & $41\pm 2$ & $10.8\pm0.5$ & $1.59\pm0.19$ & $\leq 2.03$ & 2005-12-28 & $5970 \pm 40$ & $2330\pm30$ \\
BY\,Boo & $69 \pm 3 $ & $18.60\pm0.11 $ & $2.9\pm0.4 $ & $\leq 1.16$ & 2004-06-05  & $3240\pm 30$ & $215 \pm9 $ \\
 & & & & & 2008-07-03 & $3040 \pm 9$ & $213 \pm 5$ \\
CG\,Uma & $40 \pm 2 $ & $10.9 \pm0.8 $ & $1.69 \pm0.17 $ & $\leq 1.00$ & 2004-01-18 & $1491 \pm 23$ & $108 \pm 9$ \\
 & & & & & 2004-01-31 & $1505 \pm4 $ & \\
DF\,Leo & $10.7 \pm0.4$ & $2.8 \pm0.2 $ & $0.61 \pm0.07 $ & $\leq 1.0$ & 2007-02-23 & $673 \pm4 $ & $39.6 \pm1.7$ \\
 & & & & & 2004-02-23 & $471 \pm10 $ & $32 \pm4 $\\
 & & & & & 2009-01-30 & $623 \pm11$ & $36 \pm 4$\\
EY\,Hya & $102 \pm16$ & $50\pm 2$ & $8.0\pm0.7$ & $2.8\pm0.2 $ & 2007-02-19 & $133.6 \pm1.7$ & $95.0 \pm1.7 $ \\
 & & & & & 2006-02-20 & $88 \pm5 $ & $69 \pm 7$ \\
FH\,Vir & $18.8\pm 1.1 $ & $6.0\pm0.5 $ & $1.19\pm0.18 $ & $\leq 1.0$ & 2006-05-01 & $417 \pm 3$ & $44.8 \pm1.9$ \\
 & & & & & 2004-04-14 & $426 \pm13 $ & $56\pm9 $ \\
 & & & & & 2004-04-14 & $366 \pm12$ & $45 \pm6$ \\
IN\,Hya & $18.2\pm0.9$ & $5.2\pm0.4$ & $1.00\pm0.11$ & $\leq 1.0$ & 2006-02-23 & $580 \pm14$ & $44 \pm6 $ \\
 & & & & & 2004-02-09 & $656 \pm4 $ & $43.4\pm1.9 $ \\
OME\,Vir & $62\pm 3 $ & $15.6\pm 1.2 $ & $2.8\pm0.3 $ & $\leq 1.2$ & 2004-0327 & $2950 \pm30$ & $179 \pm12 $ \\
R\,Lmi & $430 \pm 50$ & $176 \pm 19 $ & $26 \pm 3 $ & $7.9 \pm 0.9$ & 2006-02-28 & $17 \pm2 $ & $\leq 5.75$ \\
 & & & & & 2005-04-11 & $59.3 \pm1.1 $ & $2.4\pm0.7$ \\
R\,Uma & $94 \pm19$ & $55\pm 8$ & $7.0\pm0.8 $ & $2.8\pm0.2 $ & 2003-1208 & $59 \pm4 $ & $55 \pm 7$ \\
 & & & & & 2006-01-06 & $117 \pm 2$ & $44.9 \pm1.9$  \\
 & & & & & 2003-12-08 & $90 \pm8 $ & $59 \pm7 $  \\
RR\,Eri & $44 \pm 2$ & $20.4\pm 1.0 $ & $3.3 \pm0.3$ & $0.82\pm0.10 $ & 2007-10-20 & $485\pm10 $ & $28 \pm 4$ \\
 & & & & & 2003-11-26 & $382 \pm3 $ & $27.1 \pm1.3 $\\
RR\,Umi & $124\pm 4$ & $33.0\pm 1.3 $ & $5.2\pm0.5 $ & $1.12\pm0.15$ & 2005-03-07 & $6490 \pm 40$ & $1930\pm30$ \\
 & & & & & 2007-04-06 & $7830\pm60$ & $1810 \pm50$ \\
RT\,Cnc & $73\pm 3$ & $29.1\pm 1.5 $ & $4.20\pm0.42$ & $1.40\pm0.11$ & 2006-02-20 & $206 \pm 7$ & $33 \pm 4$ \\
 & & & & & 2009-01-03 & $173 \pm7$ & $40 \pm 6$ \\
 & & & & & 2006-02-20 & $175 \pm11$ & $24 \pm7 $ \\
RU\,Her & $173\pm 19$ & $78 \pm 3$ & $11.4\pm0.9$ & $4.5\pm0.4$ & 2004-05-29 & $85 \pm3 $ & $\leq 5.75$ \\
 & & & & & 2005-07-27 & $74.8\pm1.0 $ & $\leq 5.75$\\
RW\,Boo & $61\pm3$ & $29.8\pm1.2$ & $5.6\pm0.5$ & $2.8\pm0.3 $ & 2007-04-08 & $369 \pm13 $ & $17\pm 5$ \\
 & & & & & 2006-05-05 & $451 \pm2 $ & $29.5\pm1.6$\\
RZ\,Uma & $46.5\pm1.9$ & $22.1\pm0.9$ & $4.0\pm0.4$ & $1.92\pm0.17$ & 2006-12-26 & $20 \pm 4$ & $8 \pm 4$ \\
 & & & & & 2005-11-21 & $15.6 \pm0.9 $ & $\leq 5.75$ \\
ST\,Uma & $43.9\pm1.8$ & $18.8\pm0.9$ & $3.1\pm0.3$ & $1.11\pm0.12$ & 2004-03-08 & $994 \pm14$ & $168 \pm13$ \\
SV\,Peg & $265\pm 11$ & $146\pm9$ & $24\pm3$ & $9.9\pm1.0$ & 2006-09-20 &  $30 \pm 4$ & $\leq 5.75$  \\
 & & & & & 2004-09-26 & $28.5 \pm0.7$ & $\leq 5.75$\\
T\,Dra & $197\pm14 $ & $66 \pm 3$ & $15.8 \pm0.9$ & $5.7\pm0.3$ & 2005-06-20 & $\leq 5.75$ & $7.93\pm1.1$ \\
 & & & & & 2007-07-12 & $32 \pm3$ & $\leq 5.75$\\
TU\,And & $20.5\pm 1.2$ & $7.6\pm0.6$ & $1.23\pm0.14$ & $\leq 1.41$ & 2004-10-02 & $22.8 \pm1.3$ & $\leq 5.75$ \\
 & & & & & 2003-10-07 & $8 \pm2$ & $\leq 5.75$ \\
UY\,Leo & $8.6\pm0.3$ & $3.9\pm0.3$ & $0.6\pm0.06 $ & $\leq 1.0$ & 2005-03-16 & $132 \pm4 $ & $24 \pm5 $ \\
 & & & & & 2006-02-06 & $167 \pm2 $ & $30 \pm1.5$\\
V\,Eri & $326\pm10$ & $184\pm 7$ & $24\pm2$ & $7.2\pm0.6 $ & 2005-11-03 & $142 \pm2 $ & $79 \pm 3$ \\
 & & & & & 2007-01-07 & $56 \pm5 $ & $54 \pm8 $ \\
VY\,Uma & $52\pm2$ & $14.0\pm0.7$ & $4.8\pm0.4$ & $4.9\pm0.4$ & 2004-01-21 & $93 \pm 7$ & $10\pm4 $ \\
 & & & & & 2004-01-21 & $84 \pm6$ & $8 \pm3$ \\
 & & & & & 2006-01-07 & $234 \pm3$ & $6.6\pm0.8 $ \\
W\,Peg & $218\pm9 $ & $94\pm4$ & $13.1\pm 1.2$ & $3.5\pm0.4$ & 2009-08-28 & $157.3 \pm1.7 $ &  \\
 & & & & & 2006-10-02 & $201 \pm 7$ & $13 \pm3$\\
Y\,Crb & $27.0\pm1.1$ & $10.8\pm0.5$ & $1.67\pm0.12$ & $0.62\pm0.09$ & 2005-07-28 & $9.2 \pm0.6 $ & $\leq 5.75$ \\
 & & & & & 2004-05-15 & $15 \pm 4$ & $\leq 5.75$  \\
Y\,Gem & $28.5\pm 1.7$ & $9.0\pm0.6$ & $1.23\pm0.11$ & $\leq 1.0 $& 2006-01-04  & $11100\pm 50$ & $10560 \pm 90$ \\
 & & & & & 2008-12-28 & $3690 \pm20 $ & $2230 \pm 30$\\
 & & & & & 2008-01-19 & $1613 \pm 7$ & $1380 \pm10 $\\
Z\,Cnc & $44\pm 2$ & $22.0\pm1.3$ & $3.0\pm0.3$ & $1.34\pm0.15$ & 2008-02-24 & $73.5 \pm1.7$ & $13.7 \pm1.3 $ \\
 & & & & & 2008-01-28 & $72 \pm 5$ & $11 \pm 3$ \\
\end{longtable}

\section{Extinction correction} \label{ext_corr}

We have estimated the extinction due to dust in the ISM toward our targets using the GALEXtin tool \citep[see][]{amores_2021}. This tool adopts the 3D dust map of the Galaxy from Bayestar19 \citep{Green_2019} and provides the corresponding ISM extinction considering the galactic coordinates and distance to a given source.
The values of the colour excess  $E(B-V)$ obtained from GALEXtin are smaller and more accurate than those provided by the GALEX MAST archive because the latter does not consider the distance to the target and estimates the integrated extinction along the line of sight across the whole Galaxy in a given direction \cite{Schlegel_1998}. 

The extinction-corrected bolometric luminosity of each source listed in Table~\ref{tab:sources} was estimated by numerical integration of its Spectral Energy distribution (SED). The SEDs were built using photometric data at different wavelengths (from 300\,nm to 300 $\mathrm{\mu m}$) using the VizieR Photometry viewer tool available at VizieR database \citep{vizier}. We applied the ISM extinction correction to the SEDs (see values on table~\ref{tab:sources}) adopting $A_{\lambda}/A_{V} \propto \lambda^{-1}$ and $\mathrm{A}_{V}=E(B-V)\mathrm{R}_{V}$, where $A_{\lambda}$ is the extinction in a certain wavelength and $E(B-V)$ is the redenning, which have been estimated from extinction in the johnson V band with the canonical value $\mathrm{R}_{V}=3.1$ of the extinction ratio.

The NUV and FUV magnitudes and fluxes of our sample are affected by extinction by both dust in the ISM and dust in the circumstellar envelope. A rough estimate of the circumstellar extinction of our targets has been obtained using the total H$_{2}$ column density deduced from our CO-based analysis (sect.~\ref{anal}) and adopting the canonical conversion ratio $\mathrm{N_{tot}/A}_{V}=1.8\times10^{21} \mathrm{mag^{-1}cm^{-2}}$. 
GALEX fluxes were corrected from both ISM and CSM extinction (see tables~\ref{tab:sources} and \ref{tab:parameters}) assuming $\mathrm{R_{NUV}}=7.81$ and $\mathrm{R_{FUV}}=6.30$ \citep[same as][]{Montez_2017}.

\section{Comparison of values of \mloss\ with previous estimates} \label{mloss_comp}

 For a subset of our sample of uvAGB stars (11 sources), we have compared our derived values of the mass-loss rate (\mloss) obtained from a first-order approximation using the population diagram method with previous CO-based estimates of mass-loss rates available in the literature (\mloss$_{\rm lit}$). For this comparison, aimed at evaluating the uncertainties associated with our simplified approach, we place greater emphasis on a few selected sources for which detailed radiative transfer models of more than one low-$J$ CO transition, typically the \doceuno, \docedos, and/or \docetres\ emission lines,  have been previously conducted.
  Details on previous CO observations and on the corresponding data analysis and mass-loss rate estimates available for some of our targets are given below. 
  To ensure a meaningful comparison of the mass-loss rates, we have rescaled the previous values by adopting the same distance ($D$), expansion velocity (\vexp), and fractional CO abundance (\xco) as used in this study (see tables~\ref{tab:astronomy} and \ref{tab:parameters}),  taking into account that \mloss$\propto$$D^2$\vexp/\xco.

\begin{figure*}[h!]
    \centering
    \includegraphics[width=\linewidth]{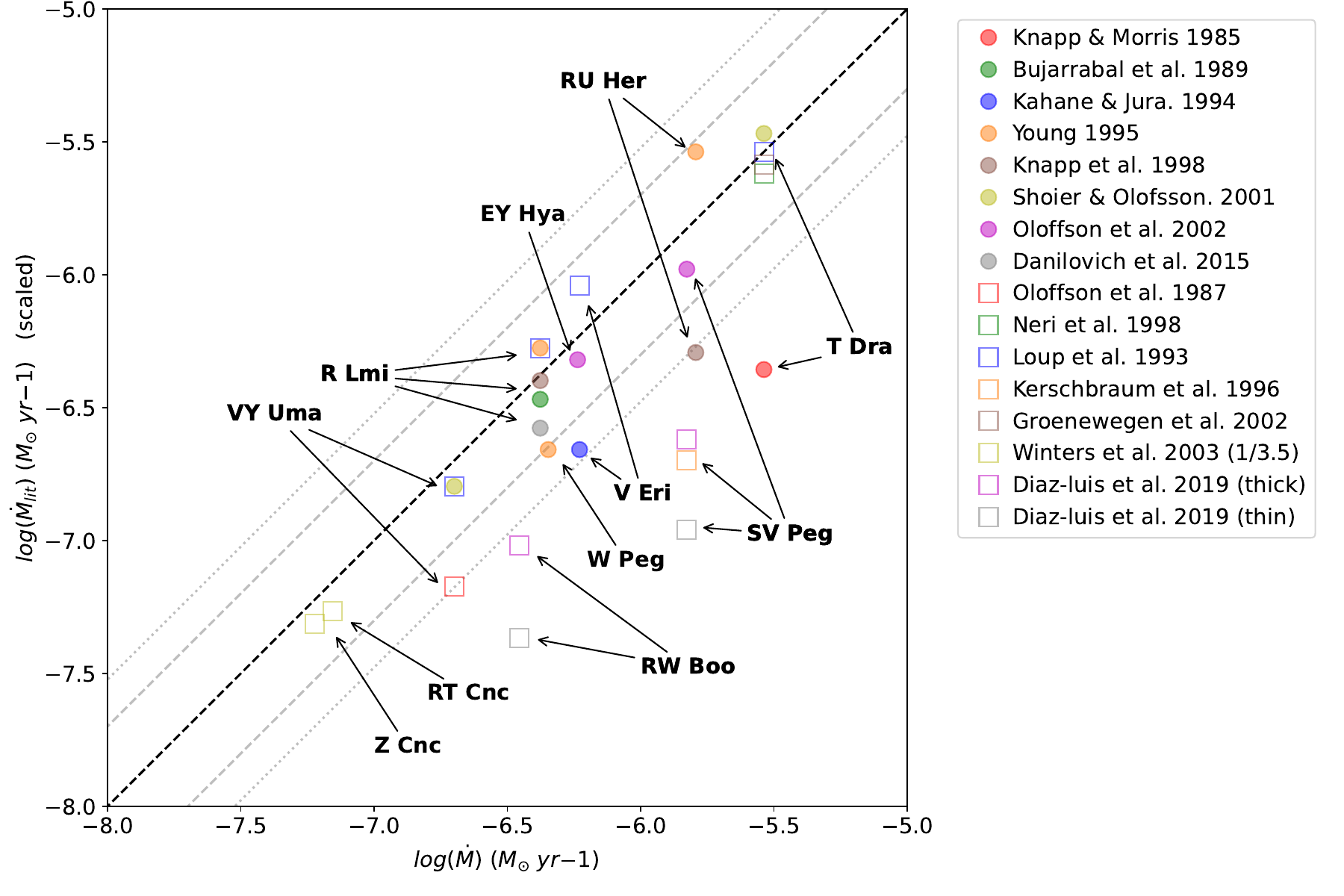}
    \caption{Comparison between mass-loss rates scaled values from the literature (\mloss$_{\rm lit}$) and \mloss \, obtained in this work fitted with $\rs=\rco/2.5$. The solid line indicates equality, the dashed lines relationships 1/2 and 2, and the dotted lines relationships 1/3 and 3. Filled circles and empty squares indicate previous studies with radiative transfer models and with empirical laws, respectively.}
    \label{fig:massloss}
\end{figure*}

Figure~\ref{fig:massloss} shows the properly scaled \mloss-to-\mloss$_{\rm lit}$ comparison. Overall, the obtained values of the mass-loss are in good agreement (within a factor 3) with those from the literature, specially those obtained from radiative transfer models and using more than one CO transitions, which are expected to provide the most reliable and accurate values. In a few cases, larger discrepancies are observed, which can be attributed to different factors, namely, marginal line(s) detection, calibration uncertainties, etc, which are discussed in some detail below. We can safely conclude that uncertainties in our mass-loss estimates are not lower than a factor $\approx$3, but probably not significantly larger considering the moderate scatter of points in Fig.\,\ref{fig:massloss}. This is comparable to uncertainties in most previous CO-based studies using different methods, even using radiative transfer models as shown in Fig.~\ref{fig:massloss}. The largest source of uncertainty in our estimate, and in many previous works, probably lies on the lack of a precise value of the characteristic radius of the CO envelope (\rs), which can only be improved by carrying out interferometric CO emission mapping.

In the following paragraphs we list in some detail previous CO detections and mass-loss rates estimates for individual targets, together with a brief description of some of their properties: 

\textbf{EY\,Hya} is a 180 day period O-rich semi-regular (SRa) variable. There are previous \docedos \, and \docetres \, detections and a CO-based mass-loss rate estimation.

\cite{Olofsson_2002} used the \docedos \, and \docetres \, detections presented in \cite{Kerschbaum_1999} to evaluate the mass-loss rate using a radiative transfer code. The value obtained is \mloss=2.5\ex{-7}\,\my\ ($D$=300\,pc, \vexp=11\,\kms, and \xco=2\ex{-4}), which scaled to the values of $D$, \vexp\ and \xco values used by us (tables~\ref{tab:astronomy} and \ref{tab:spectral_analysis}) is \mloss=4.8\ex{-7}\,\my.

\textbf{R\,Lmi} is a 370 day period O-rich Mira variable. There are previous \docedos \,, \docetres \, and \docecuatro \, detections and CO-based mass-loss rate estimations.

\cite{Bujarrabal_1989} reported a \doceuno \, detection. They used a Large Velocity Gradient model to fit the \docetres \,and SiO($J$=2-1) line profiles and evaluate the mass-loss rate. The value obtained is \mloss=5ex{-7}\,\my\ ($D$=350\,pc, \vexp=6.4\,\kms, and \xco=2\ex{-4}), which scaled to the values of $D$, \vexp\ and \xco values used by us (tables~\ref{tab:astronomy} and \ref{tab:spectral_analysis}) is \mloss=3.4\ex{-7}\,\my.

\cite{Loup_1993} used the \doceuno detection reported by \cite{Bujarrabal_1986} to evaluate the mass-loss rate using the formulation of \cite{knapp1985}. The value obtained is \mloss=2.8\ex{-7}\,\my\ ($D$=330\,pc, \vexp=7.0\,\kms, and \xco=5\ex{-4}), which scaled to the values of $D$, \vexp\ and \xco values used by us (tables~\ref{tab:astronomy} and \ref{tab:spectral_analysis}) is \mloss=5.3\ex{-7}\,\my.

\cite{Young_1995} reported a \docetres \, and \docecuatro \, detections. They used a Large Velocity Gradient model to fit the peak temperature of the \docetres \, line profile and evaluate the mass-loss rate. The value obtained is \mloss=1.6\ex{-7}\,\my\ ($D$=260\,pc, \vexp=8.7\,\kms, and \xco=3\ex{-4}), which scaled to the values of $D$, \vexp\ and \xco values used by us (tables~\ref{tab:astronomy} and \ref{tab:spectral_analysis}) is \mloss=5.3\ex{-7}\,\my.

\cite{Knapp_1998} reported a \docedos \, detection. These authors evaluate the mass-loss rate using a radiative transfer model. The value obtained is \mloss=1.6\ex{-7}\,\my\ ($D$=270\,pc, \vexp=7.8\,\kms, and \xco=5\ex{-4}), which scaled to the values of $D$, \vexp\ and \xco values used by us (tables~\ref{tab:astronomy} and \ref{tab:spectral_analysis}) is \mloss=4.0\ex{-7}\,\my.

\cite{Danilovich_2015} used a detailed radiative transfer model to fit the \docedos \, and \docetres line profiles and evaluate the mass-loss rate. The value obtained is \mloss=2.6\ex{-7}\,\my\ ($D$=330\,pc, \vexp=7.5\,\kms, and \xco=3\ex{-4}), which scaled to the values of $D$, \vexp\ and \xco values used by us (tables~\ref{tab:astronomy} and \ref{tab:spectral_analysis}) is \mloss=2.7\ex{-7}\,\my. 

\textbf{RT\,Cnc} is a 90 day period O-rich semi-regular (SRb) variable. There is a previous \docedos \, detection and a CO-based mass-loss rate estimation. 

\cite{Winters_2003} reported a \docedos \, detection, which they use to evaluate both the mass-loss rate using the formulation of \cite{Loup_1993} \citep[first introduced by][]{knapp1985}. The value obtained is \mloss=4.4\ex{-7}\,\my\ ($D$=340\,pc, \vexp=12\,\kms, and \xco=2\ex{-4}), which scaled to the values of $D$, \vexp\ and \xco values used by us (tables~\ref{tab:astronomy} and \ref{tab:spectral_analysis}) is \mloss=1.9\ex{-7}\,\my. \cite{Winters_2003} note that their values for \mloss\, are systematically larger by a factor 3.5 than those obtained by \cite{Olofsson_2002} for a number of objects in common in both works. Considering a correction by a factor 3.5, the value of \mloss\ derived \cite{Winters_2003} becomes 5.4\ex{-8}\,\my.

\textbf{RU\,Her} is a 480 day period Mira (M) variable. There are previous \docedos \,, \docetres \, and \docecuatro \, detections and CO-based mass-loss rate estimations.

\cite{Young_1995} reported a \docetres \, and \docecuatro \, detections.They used a Large Velocity Gradient model to fit the peak temperature of the \docetres \, line profile and evaluate the mass-loss rate. The value obtained is \mloss=7.5\ex{-7}\,\my\ ($D$=410\,pc, \vexp=8.9\,\kms, and \xco=3\ex{-4}), which scaled to the values of $D$, \vexp\ and \xco values used by us (tables~\ref{tab:astronomy} and \ref{tab:spectral_analysis}) is \mloss=2.9\ex{-6}\,\my.

\cite{Knapp_1998} reported a \docedos \, detection which they use to evaluate both the mass-loss rate using a radiative transfer model. The value obtained is \mloss=3.2\ex{-7}\,\my\ ($D$=400\,pc, \vexp=9.4\,\kms, and \xco=5\ex{-4}), which scaled to the values of $D$, \vexp\ and \xco values used by us (tables~\ref{tab:astronomy} and \ref{tab:spectral_analysis}) is \mloss=5.1\ex{-7}\,\my.

\textbf{RW\,Boo} is a 209 day period O-rich semi-regular (SRb) variable. There are previous \doceuno \, and \docedos \, detections and a CO-based mass-loss rate estimation.

\cite{diaz-luis_2019} reported \doceuno \, and \docedos \, detections and estimated the mass-loss rate using the formulation based on that presented in \cite{knapp1985}, in the case of optically thick envelope and in the case of optically thin. The values obtained are $\mloss_{thick}$=9.9\ex{-8}\,\my\ and $\mloss_{thin}$=4.4\ex{-8}\,\my\ ($D$=307\,pc, \vexp=17.29\,\kms, and \xco=3\ex{-4}), which scaled to the values of $D$, \vexp\ and \xco values used by us (tables~\ref{tab:astronomy} and \ref{tab:spectral_analysis}) is $\mloss_{thick}$=9.6\ex{-8}\,\my and $\mloss_{thin}$=4.3\ex{-8}\,\my.

\textbf{SV\,Peg} is a 145 day period O-rich semi-regular (SRb) variable. There are previous\doceuno, \docedos \, and \docetres \, detections and CO-based mass-loss rate estimations.

\cite{Kerschbaum_1996} reported \doceuno \, and \docedos \, detections which they use to evaluate both the mass-loss rate using the formulation of \cite{Kastner_1992}. The value obtained is \mloss=1.0\ex{-7}\,\my\ ($D$=230\,pc, \vexp=10.6\,\kms, and \xco=2\ex{-4}), which scaled to the values of $D$, \vexp\ and \xco values used by us (tables~\ref{tab:astronomy} and \ref{tab:spectral_analysis}) is \mloss=2.0\ex{-7}\,\my.

\cite{Olofsson_2002} used the \doceuno\,, \docedos \, and \docetres \, detections presented in \cite{Kerschbaum_1999} to evaluate the mass-loss rate using a radiative transfer code. The value obtained is \mloss=3.0\ex{-7}\,\my\ ($D$=190\,pc, \vexp=7.5\,\kms, and \xco=2\ex{-4}), which scaled to the values of $D$, \vexp\ and \xco values used by us (tables~\ref{tab:astronomy} and \ref{tab:spectral_analysis}) is \mloss=1.1\ex{-6}\,\my.

\cite{diaz-luis_2019} reported \doceuno \, and \docedos \, detections and estimated the mass-loss rate using the formulation based on that presented in \cite{knapp1985}, in the case of optically thick envelope and in the case of optically thin. The values obtained are $\mloss_{thick}$=1.1\ex{-6}\,\my\ and $\mloss_{thin}$=4.8\ex{-7}\,\my\ ($D$=890\,pc, \vexp=9.07\,\kms, and \xco=3\ex{-4}), which scaled to the values of $D$, \vexp\ and \xco values used by us (tables~\ref{tab:astronomy} and \ref{tab:spectral_analysis}) is $\mloss_{thick}$=2.4\ex{-7}\,\my and $\mloss_{thin}$=1.1\ex{-7}\,\my.

\textbf{T\,Dra} is a 420 day period C-rich Mira variable. There are previous \doceuno \, and \docedos \, detections and CO-based mass-loss rate estimations.

\cite{knapp1985} reported a tentative \doceuno \, detection with  low S/N. They used a radiative transfer model to fit the \doceuno \, line profile and evaluate the mass-loss rate. The value obtained is \mloss=1.3 \ex{-6}\,\my\ ($D$=525\,pc, \vexp=14.0\,\kms, and \xco=8\ex{-4}), which scaled to the values of $D$, \vexp\ and \xco values used by us (tables~\ref{tab:astronomy} and \ref{tab:spectral_analysis}) is \mloss=4.4\ex{-7}\,\my. 
The mass-loss rate discrepancy (with respect to our value of \mloss\ (by a factor $\sim$7) in this case can be mainly attributed to the marginal detection of the \doceuno\ line and the lack of a constraint for the excitation temperature in the work by \cite{knapp1985}. 

\cite{Loup_1993} used the \doceuno\ detection reported by \cite{knapp1985, Margulis_1990, Nyman_1992} to evaluate the mass-loss rate using the formulation of \cite{knapp1985}. The value obtained is \mloss=2.2\ex{-6}\,\my\ ($D$=910\,pc, \vexp=13.1\,\kms, and \xco=1\ex{-3}), which scaled to the values of $D$, \vexp\ and \xco values used by us (tables~\ref{tab:astronomy} and \ref{tab:spectral_analysis}) is \mloss=2.9\ex{-6}\,\my.

\cite{Neri_1998} reported \doceuno \, and \docedos \, interferometric maps obtined with the IRAM Plateau de Bure Interferometer (PdBI) as well as single-dish \iram\ spectra, which they use to evaluate both the mass-loss rate using the formulation of \cite{Loup_1993} \citep[first introduced by][]{knapp1985}. The value obtained is \mloss=2.9\ex{-6}\,\my\ ($D$=910\,pc, \vexp=14.3\,\kms, and \xco=1\ex{-3}), which scaled to the values of $D$, \vexp\ and \xco values used by us (tables~\ref{tab:astronomy} and \ref{tab:spectral_analysis}) is \mloss=2.4\ex{-6}\,\my.
They also estimated the source radius directly from the interferometric maps of $\simeq$8.7\arcsec, which is 
comparable to, but slightly larger than, 
the \rco/2.5 approximation used by us resulting in \rsou$\simeq$4.8\arcsec.

\cite{schoier_2001} used a detailed radiative transfer analysis in combination with an energy balance equation for the gas to model the \doceuno \, and \docedos \, detections reported by \cite{Olofsson_1993} and the \doceuno \,detection reported by \cite{Neri_1998}. The value obtained is \mloss=1.2\ex{-6}\,\my\ ($D$=610\,pc, \vexp=13.5\,\kms, and \xco=1\ex{-3}), which scaled to the values of $D$, \vexp\ and \xco values used by us (tables~\ref{tab:astronomy} and \ref{tab:spectral_analysis}) is \mloss=3.4\ex{-6}\,\my, which is in very good agreement with our estimate \mloss=[1.6-2.9]\ex{-6}\,\my\ (adopting Re=\rco\ and Re=\rco/2.5, respectively).

\cite{Groenewegen_2002b} used the \docedos\, detection reported by \cite{Groenewegen_2002a} to evaluate the mass-loss rate using the formulation of \cite{Olofsson_1993}. The value obtained is \mloss=2.4\ex{-6}\,\my\ ($D$=860\,pc, \vexp=16.8\,\kms, and \xco=8\ex{-4}), which scaled to the values of $D$, \vexp\ and \xco\ values used by us (tables~\ref{tab:astronomy} and \ref{tab:spectral_analysis}) is \mloss=2.6\ex{-6}\,\my.

\textbf{V\,Eri} is a 100 day period semi-irregular  (SRC) variable. There are previous \doceuno \, and \docedos\, detections and CO-based mass-loss rate estimations.

\cite{Loup_1993} used the \docedos\, detection reported by \cite{Zuckerman_1986} to evaluate the mass-loss rate using the formulation of \cite{knapp1985}. The value obtained is \mloss=1.1\ex{-6}\,\my\ ($D$=460\,pc, \vexp=13\,\kms, and \xco=5\ex{-4}), which scaled to the values of $D$, \vexp\ and \xco values used by us (tables~\ref{tab:astronomy} and \ref{tab:spectral_analysis}) is \mloss=9.1\ex{-7}\,\my.

\cite{Kahane_1994} reported \doceuno \, and \docedos\, detections. They used a Large Velocity Gradient model to fit both line profiles and evaluate the mass-loss rate. The value obtained is \mloss=2.9ex{-7}\,\my\ ($D$=250\,pc, \vexp=11.0\,\kms, and \xco=2\ex{-4}), which scaled to the values of $D$, \vexp\ and \xco values used by us (tables~\ref{tab:astronomy} and \ref{tab:spectral_analysis}) is \mloss=2.2\ex{-7}\,\my.

\textbf{VY\,Uma} is a C-rich irregular variable. There are previous \doceuno \, and \docedos \, detections and CO-based mass-loss rate estimations.

\cite{Olofsson_1987} reported a \doceuno \, detection which they use to evaluate both the mass-loss rate using the formulation of \cite{knapp1985}. The value obtained is \mloss=1.3\ex{-7}\,\my\ ($D$=520\,pc, \vexp=8.4\,\kms, and \xco=8\ex{-4}), which scaled to the values of $D$, \vexp\ and \xco values used by us (tables~\ref{tab:astronomy} and \ref{tab:spectral_analysis}) is \mloss=6.7\ex{-8}\,\my.

\cite{Loup_1993} used the \doceuno \, detection reported by \cite{Olofsson_1987} to evaluate the mass-loss rate using the formulation of \cite{knapp1985}. The value obtained is \mloss=4.8\ex{-7}\,\my\ ($D$=750\,pc, \vexp=7.9\,\kms, and \xco=1\ex{-3}), which scaled to the values of $D$, \vexp\ and \xco values used by us (tables~\ref{tab:astronomy} and \ref{tab:spectral_analysis}) is \mloss=1.6\ex{-7}\,\my.

\cite{schoier_2001} used a detailed radiative transfer analysis in combination with an energy balance equation for the gas to model the \doceuno \, and \docedos\, detections reported by \cite{Olofsson_1993}. The value obtained is \mloss=7\ex{-8}\,\my\ ($D$=330\,pc, \vexp=6\,\kms, and \xco=1\ex{-3}), which scaled to the values of $D$, \vexp\ and \xco values used by us (tables~\ref{tab:astronomy} and \ref{tab:spectral_analysis}) is \mloss=1.6\ex{-7}\,\my.

\textbf{W\,Peg} is a 350 day period Mira (M) variable. There is a previous \docetres \, detection and a CO-based mass-loss rate estimation.

\cite{Young_1995} reported a \docetres\, detection and used a Large Velocity Gradient model to fit the peak temperature of the \docetres \, line profile and evaluate the mass-loss rate. The value obtained is \mloss=9.8\ex{-8}\,\my\ ($D$=270\,pc, \vexp=7.7\,\kms, and \xco=3\ex{-4}), which scaled to the values of $D$, \vexp\ and \xco values used by us (Tables~\ref{tab:astronomy} and \ref{tab:spectral_analysis}) is \mloss=2.2\ex{-7}\,\my.

\textbf{Z\,Cnc} is a 95 day period O-rich semi-regular (SRb) variable. There is a previous \docedos \, detection and a CO-based mass-loss rate estimation.

\cite{Winters_2003} reported a \docedos \, detection, which they use to evaluate both the mass-loss rate using the formulation of \cite{Loup_1993} \citep[first introduced by][]{knapp1985}. The value obtained is \mloss=3\ex{-7}\,\my\ ($D$=389\,pc, \vexp=8\,\kms, and \xco=2\ex{-4}), which scaled to the values of $D$, \vexp\ and \xco values used by us (Tables~\ref{tab:astronomy} and \ref{tab:spectral_analysis}) is \mloss=1.8\ex{-7}\,\my. \cite{Winters_2003} note that their values for \mloss\, are systematically larger by a factor 3.5 than those obtained by \cite{Olofsson_2002} for a number of objects in common in both works. Considering a correction by a factor 3.5, the value of \mloss\ derived \cite{Winters_2003} becomes 5.1\ex{-8}\,\my.

\section{Population diagrams of CO} \label{rotdiagrams}

\begin{figure*}[h!] 
  \begin{subfigure}[b]{0.5\linewidth}
    \centering
    \includegraphics[width=0.95\linewidth]{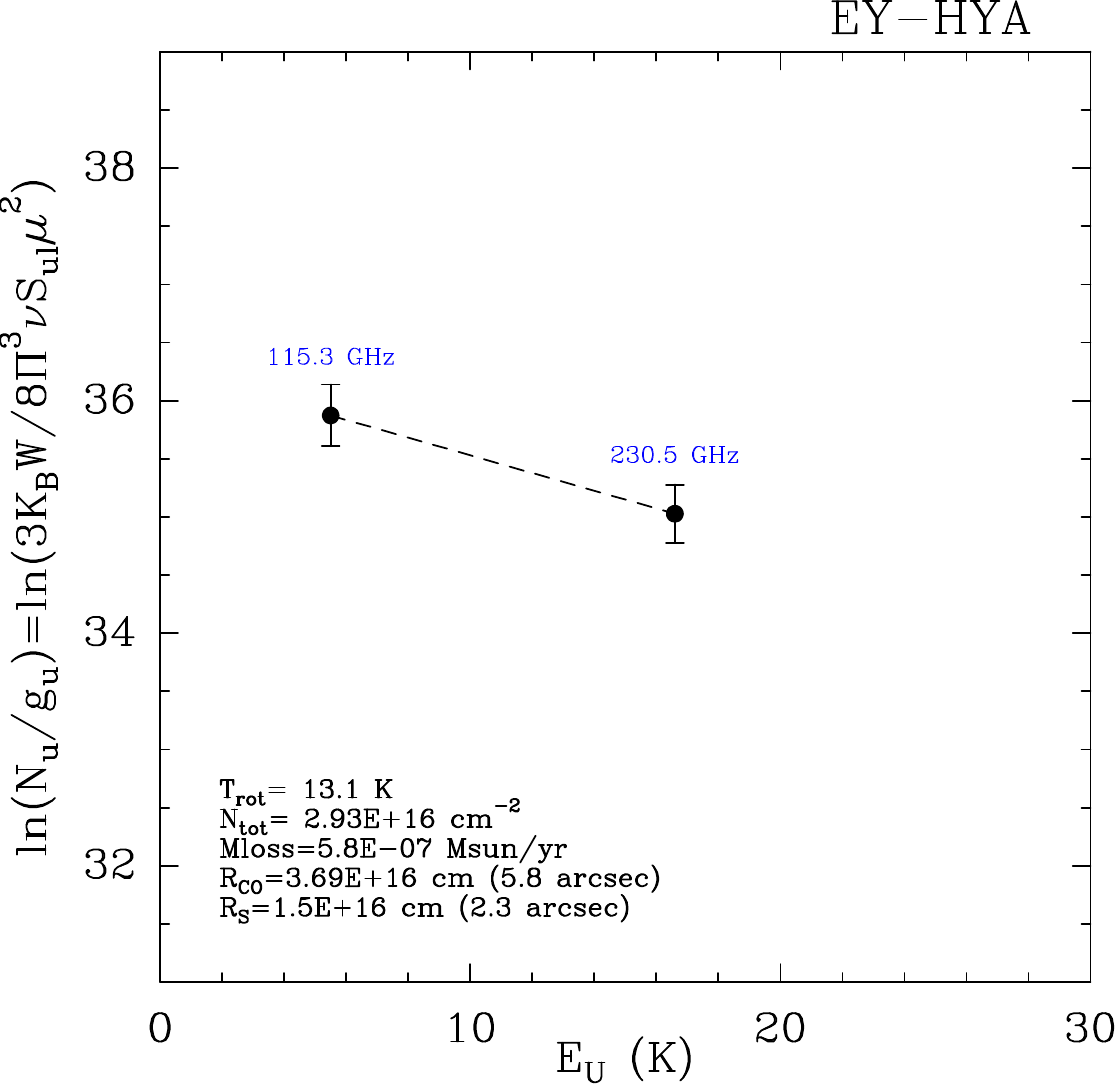} 
    \label{fig7:a} 
    \vspace{4ex}
  \end{subfigure}
  \begin{subfigure}[b]{0.5\linewidth}
    \centering
    \includegraphics[width=0.95\linewidth]{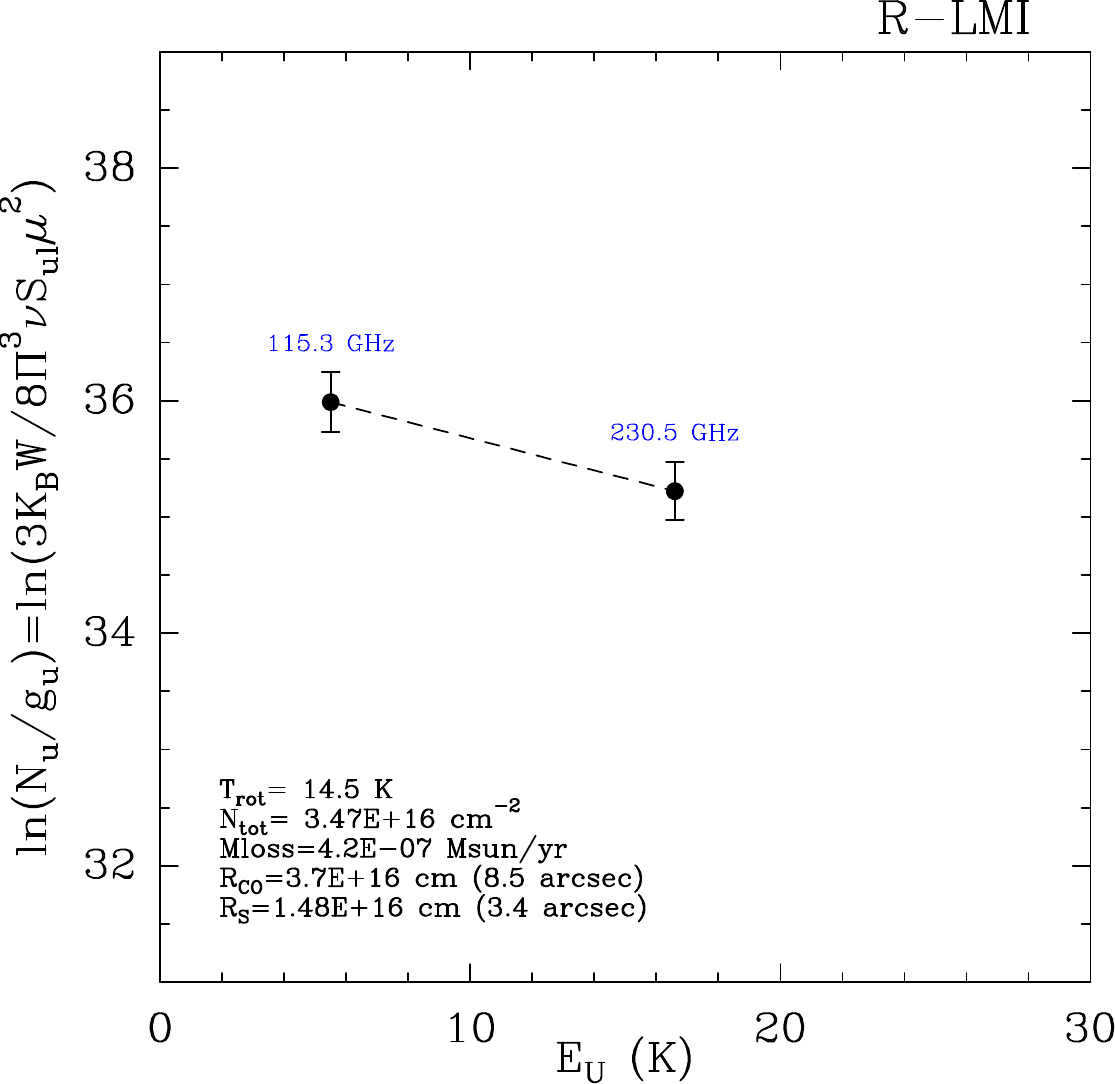} 
    \label{fig7:b} 
    \vspace{4ex}
  \end{subfigure} 
  \caption{Opacity-corrected population diagrams for uvAGB stars with CO detections (see Appendix~\ref{rotdiagrams}). Upper limits (3$\sigma$) for \doceuno\ non-detections are indicated by triangles. Values of the main envelope parameters deduced from the fits (dashed line) are indicated in the bottom left corner of the boxes.}
  \label{fig77}  
\end{figure*}

\begin{figure*}[h!] \ContinuedFloat
  \begin{subfigure}[b]{0.5\linewidth}
    \centering
    \includegraphics[width=0.95\linewidth]{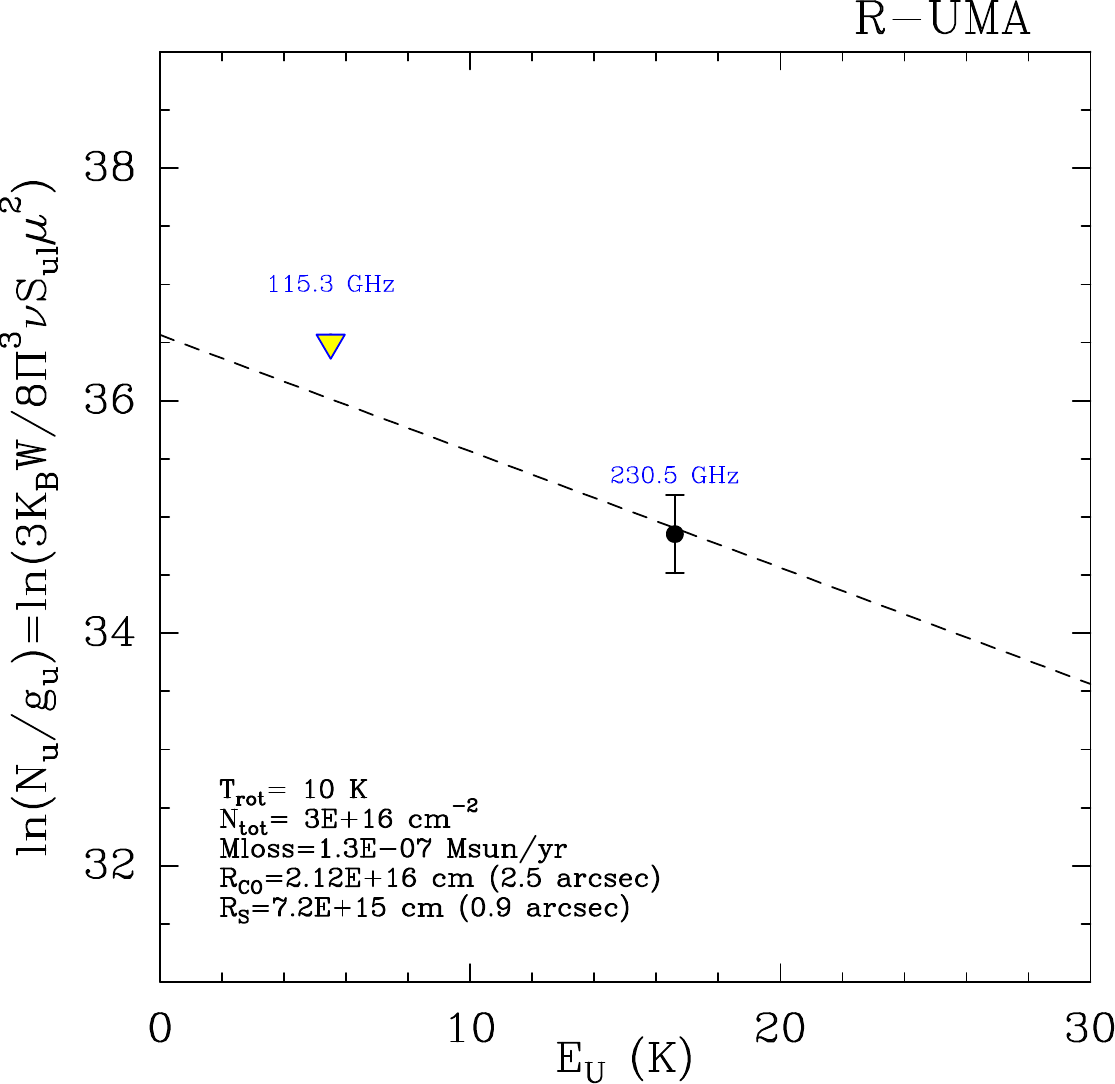} 
    \label{fig7:a} 
    \vspace{4ex}
  \end{subfigure}
  \begin{subfigure}[b]{0.5\linewidth}
    \centering
    \includegraphics[width=0.95\linewidth]{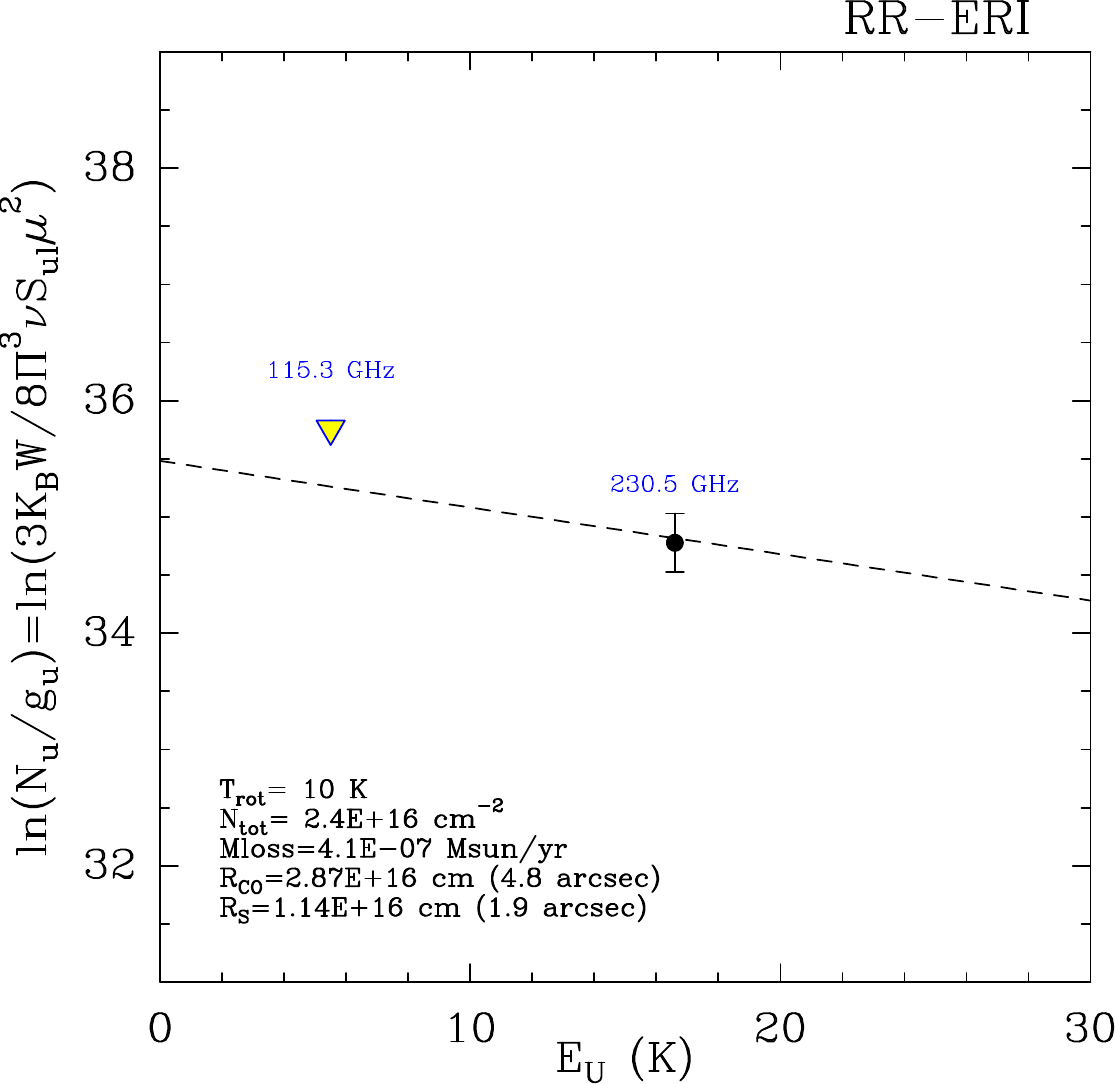} 
    \label{fig7:b} 
    \vspace{4ex}
  \end{subfigure} 
\end{figure*}

\begin{figure*}[h!] \ContinuedFloat
  \begin{subfigure}[b]{0.5\linewidth}
    \centering
    \includegraphics[width=0.95\linewidth]{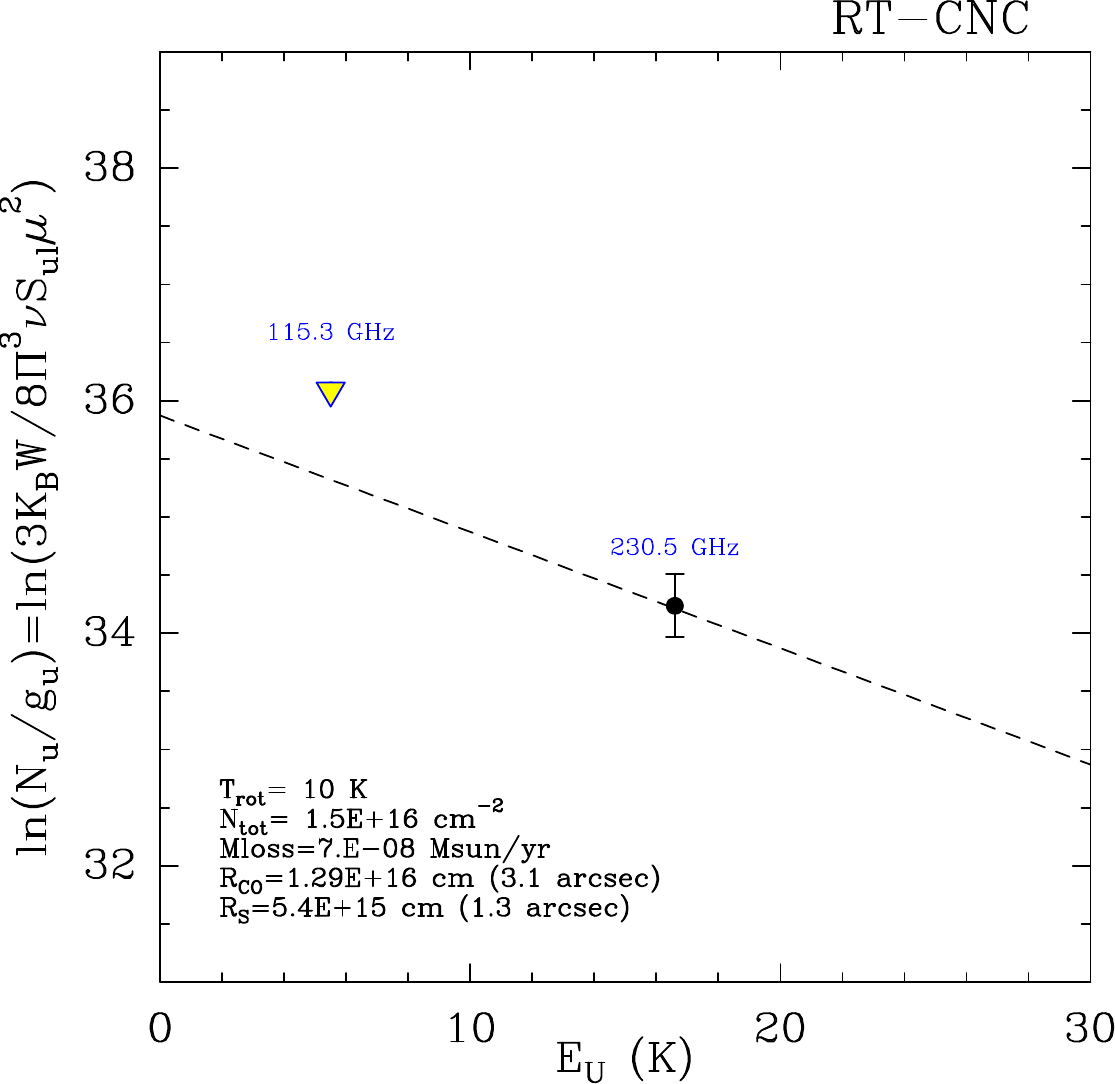} 
    \label{fig7:a} 
    \vspace{4ex}
  \end{subfigure}
  \begin{subfigure}[b]{0.5\linewidth}
    \centering
    \includegraphics[width=0.95\linewidth]{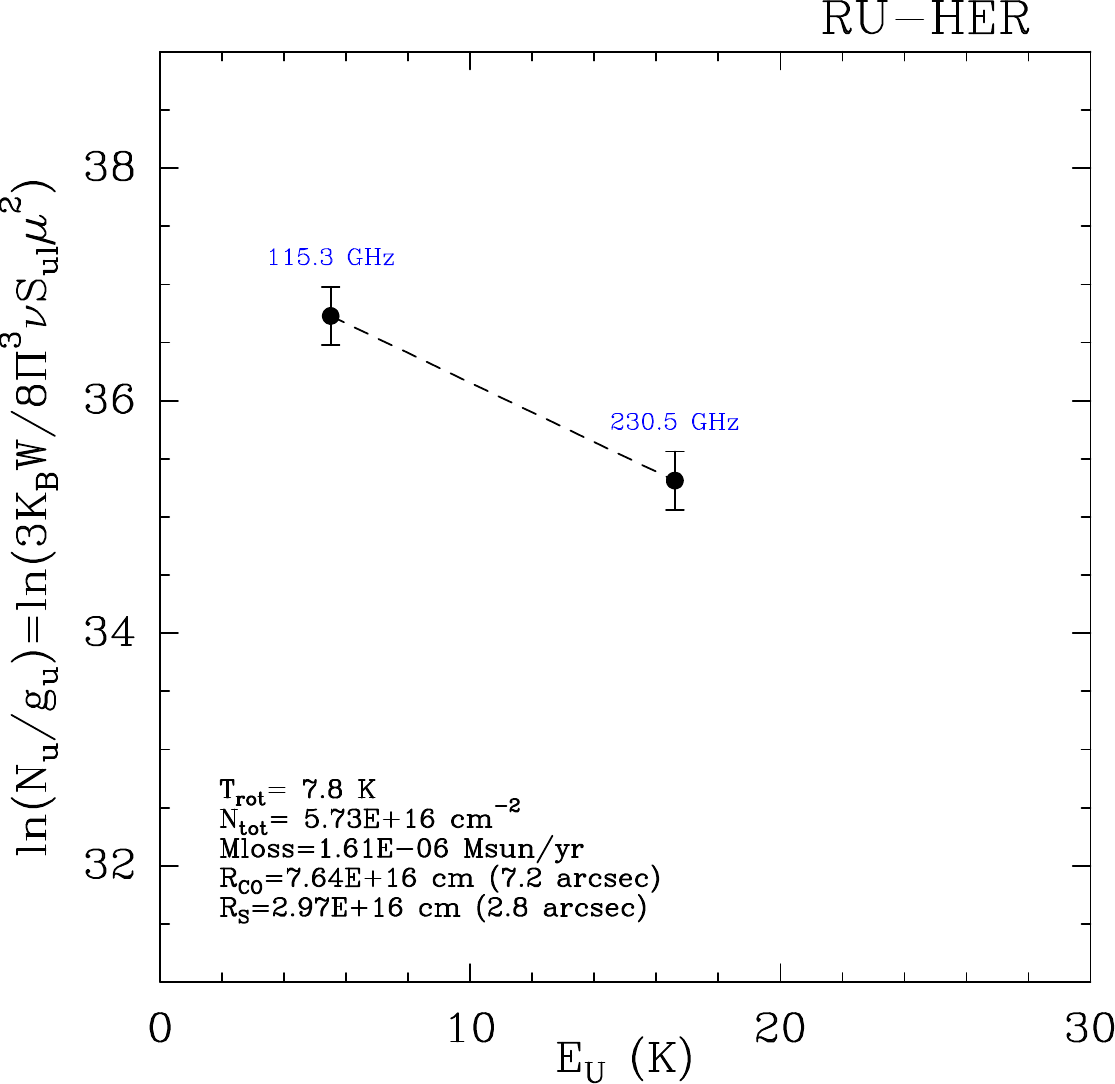} 
    \label{fig7:b} 
    \vspace{4ex}
  \end{subfigure}
  \caption*{Fig.~\ref{fig77} continued.} 
  \label{fig7} 
\end{figure*}

\begin{figure*}[h!] \ContinuedFloat
  \begin{subfigure}[b]{0.5\linewidth}
    \centering
    \includegraphics[width=0.95\linewidth]{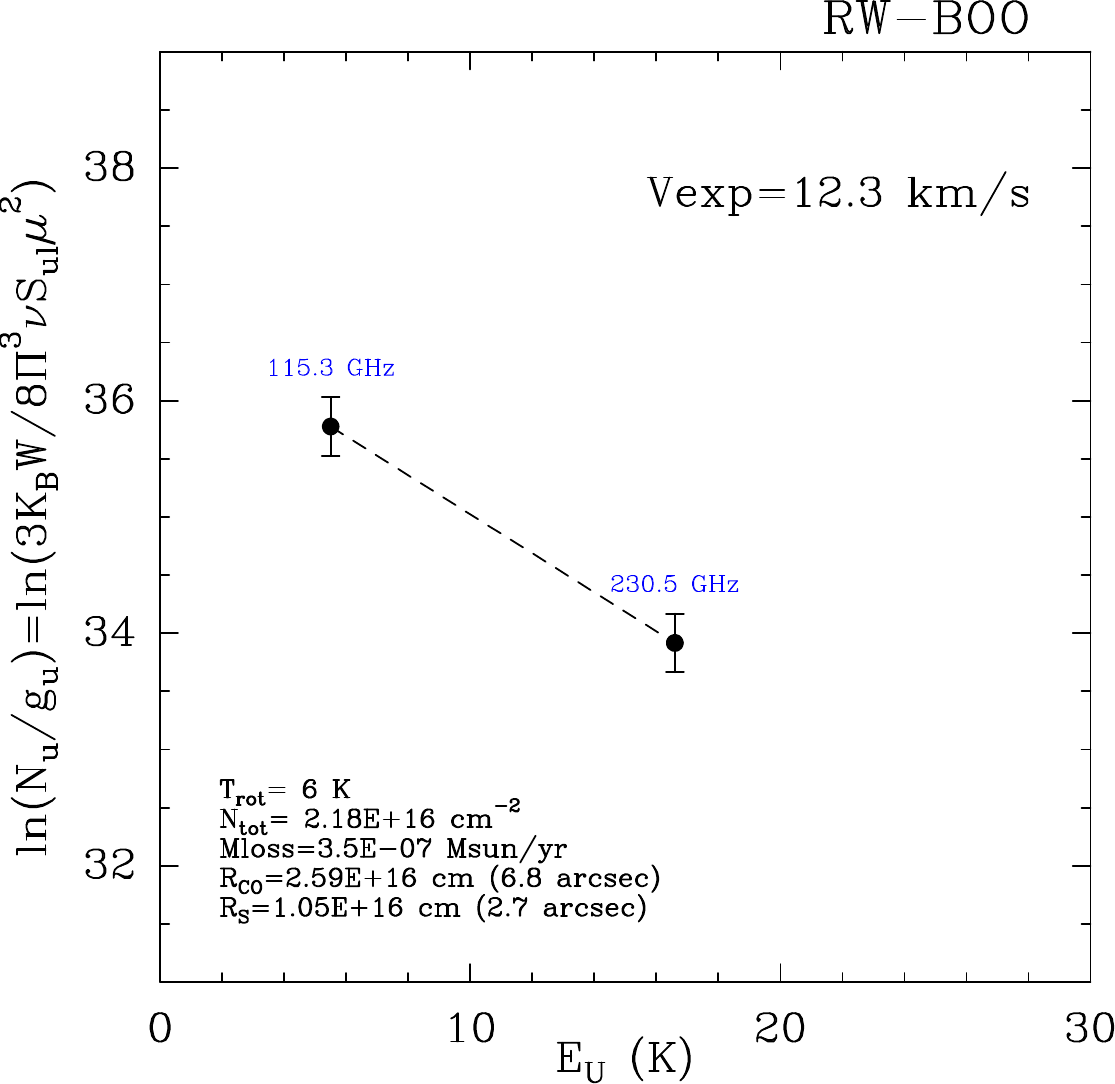} 
    \label{fig7:a} 
    \vspace{4ex}
  \end{subfigure}
  \begin{subfigure}[b]{0.5\linewidth}
    \centering
    \includegraphics[width=0.95\linewidth]{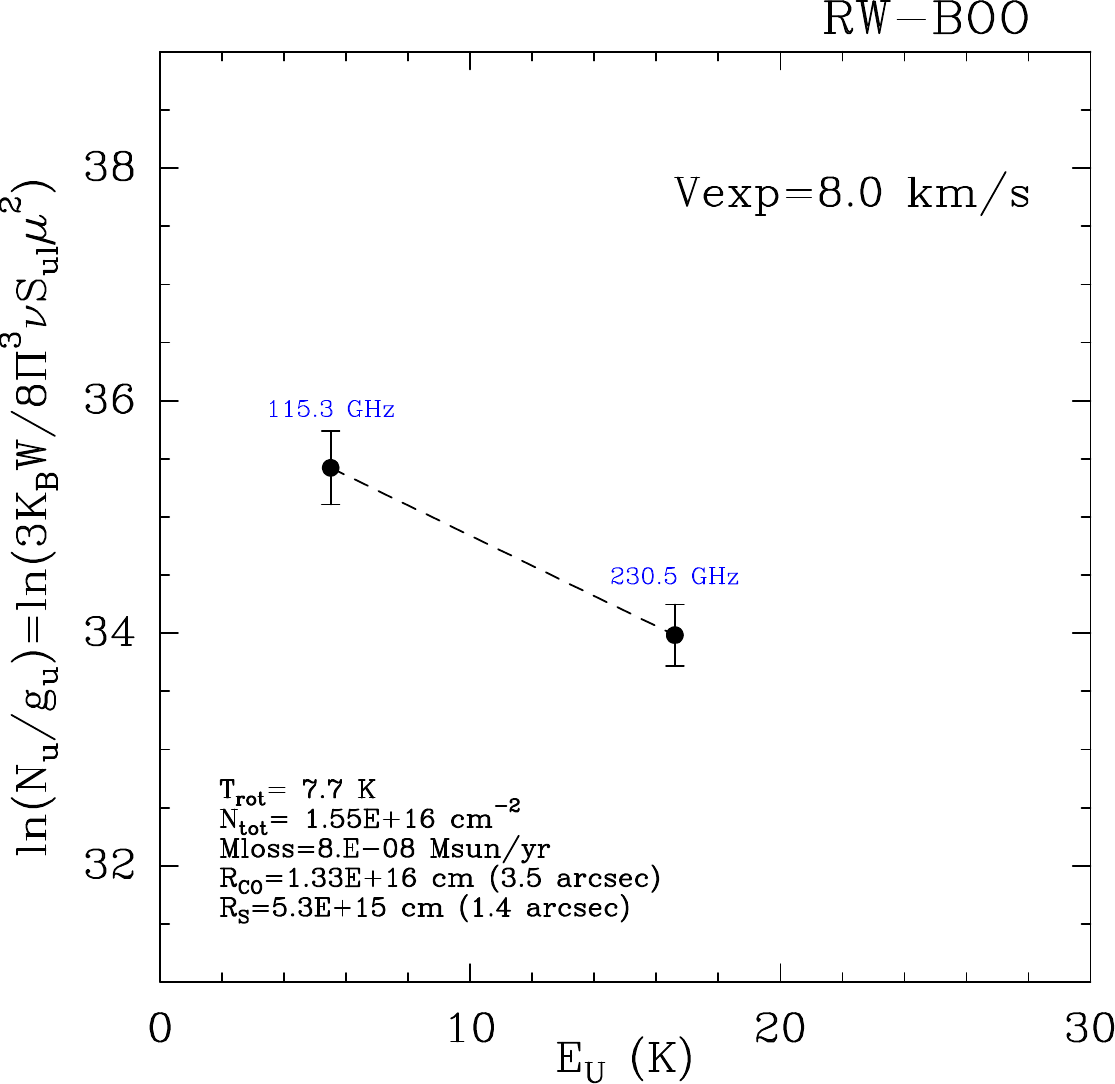}  
    \label{fig7:b} 
    \vspace{4ex}
  \end{subfigure}
  \label{fig7} 
\end{figure*}

\begin{figure*}[h!] \ContinuedFloat
  \begin{subfigure}[b]{0.5\linewidth}
    \centering
    \includegraphics[width=0.95\linewidth]{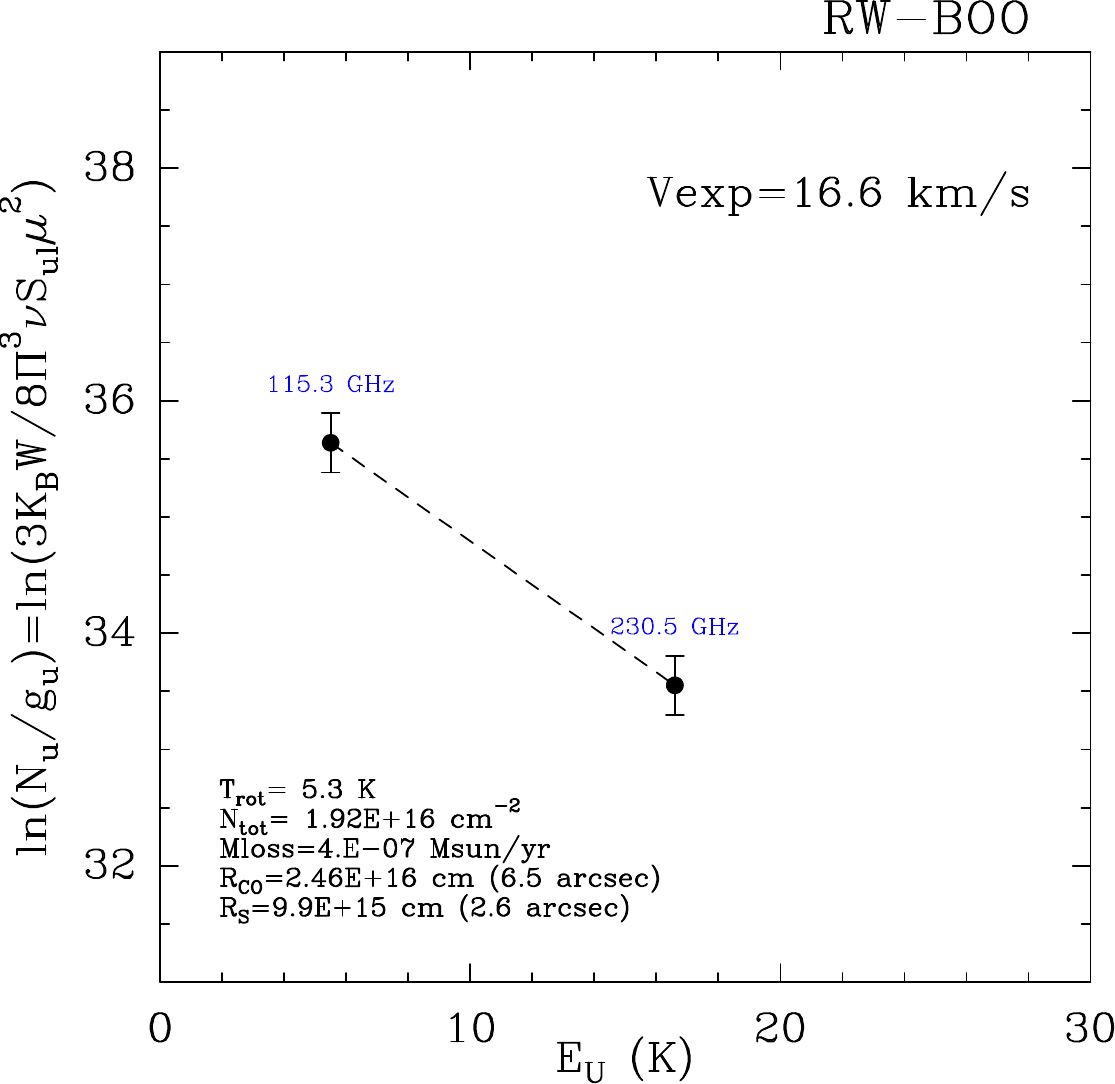} 
    \label{fig7:a} 
    \vspace{4ex}
  \end{subfigure}
  \begin{subfigure}[b]{0.5\linewidth}
    \centering
    \includegraphics[width=0.95\linewidth]{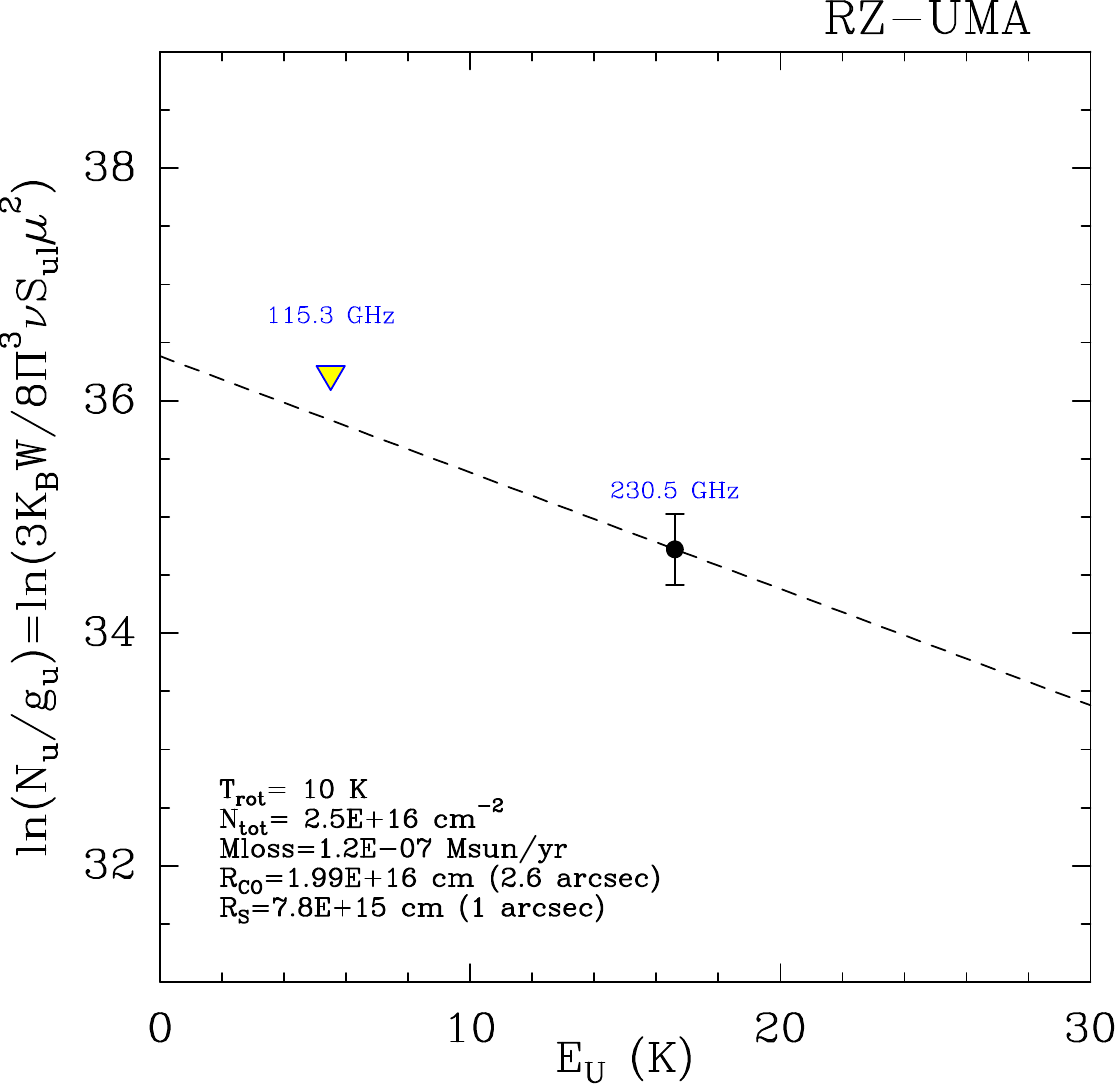} 
    \label{fig7:b} 
    \vspace{4ex}
  \end{subfigure} 
  \caption*{Fig.~\ref{fig77} continued.}
  \label{fig7} 
\end{figure*}

\begin{figure*}[h!] \ContinuedFloat
  \begin{subfigure}[b]{0.5\linewidth}
    \centering
    \includegraphics[width=0.95\linewidth]{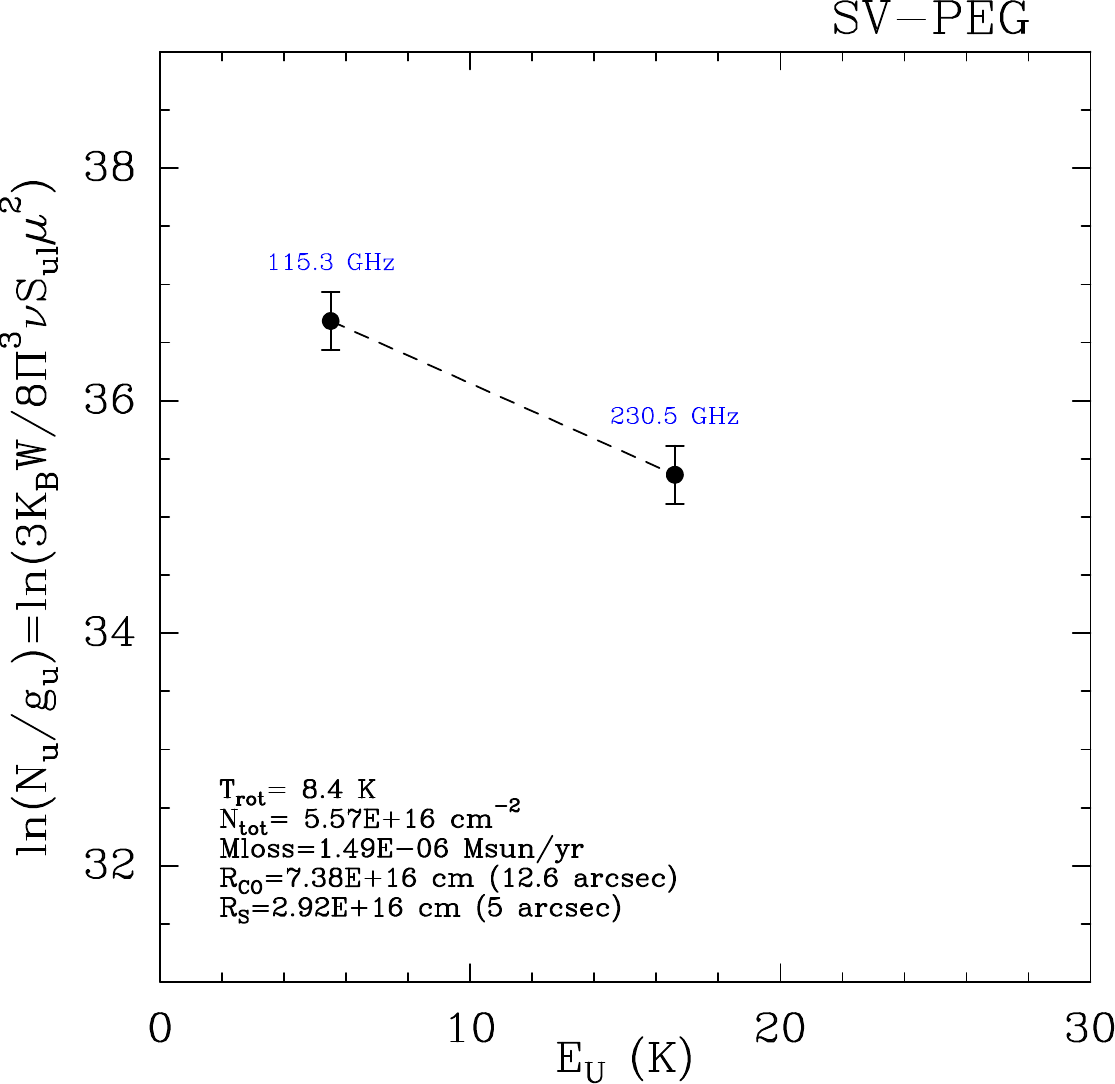} 
    \label{fig7:a} 
    \vspace{4ex}
  \end{subfigure}
  \begin{subfigure}[b]{0.5\linewidth}
    \centering
    \includegraphics[width=0.95\linewidth]{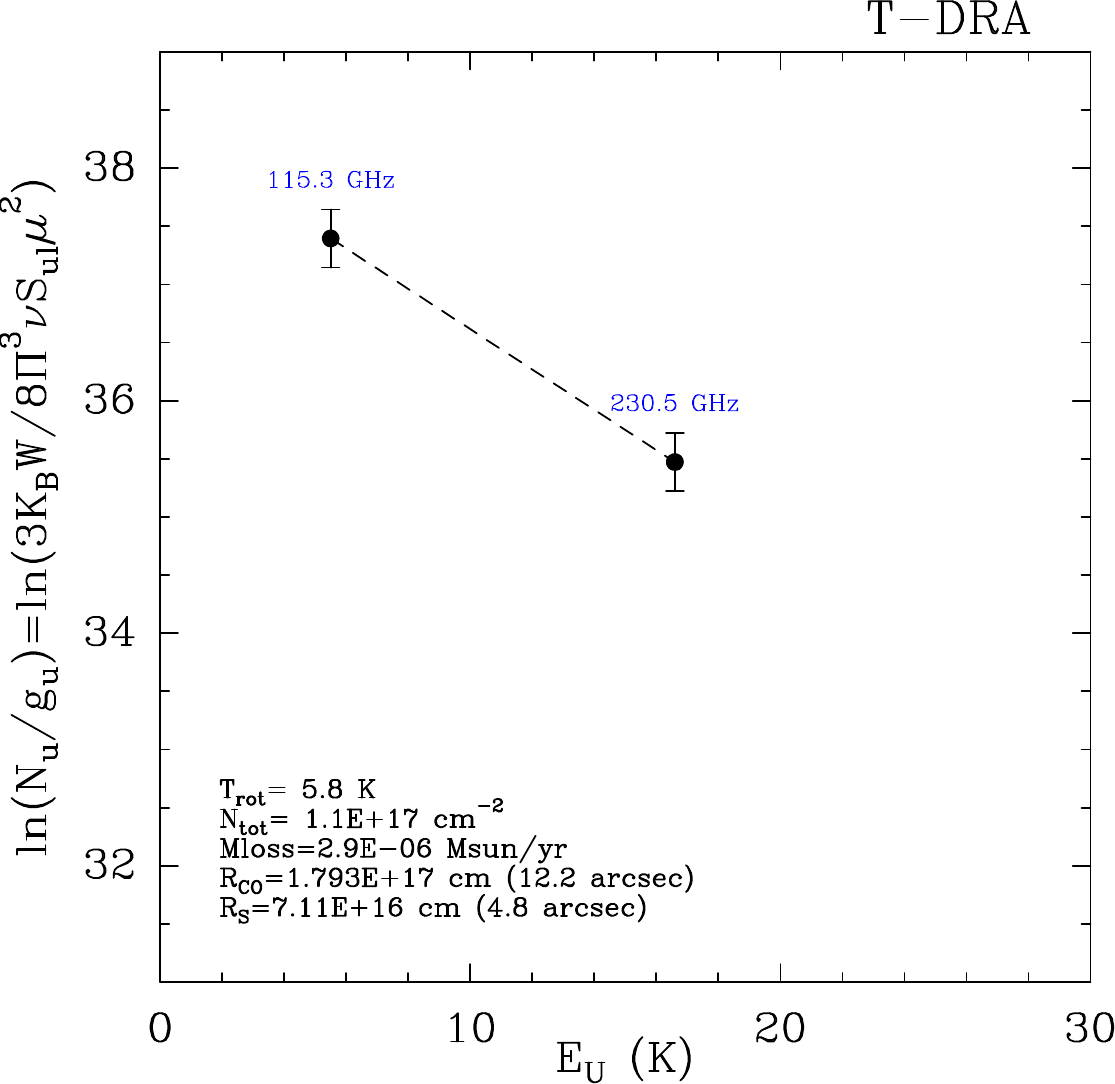} 
    \label{fig7:b} 
    \vspace{4ex}
  \end{subfigure} 
  \label{fig7} 
\end{figure*}

\begin{figure*}[h!] \ContinuedFloat
  \begin{subfigure}[b]{0.5\linewidth}
    \centering
    \includegraphics[width=0.95\linewidth]{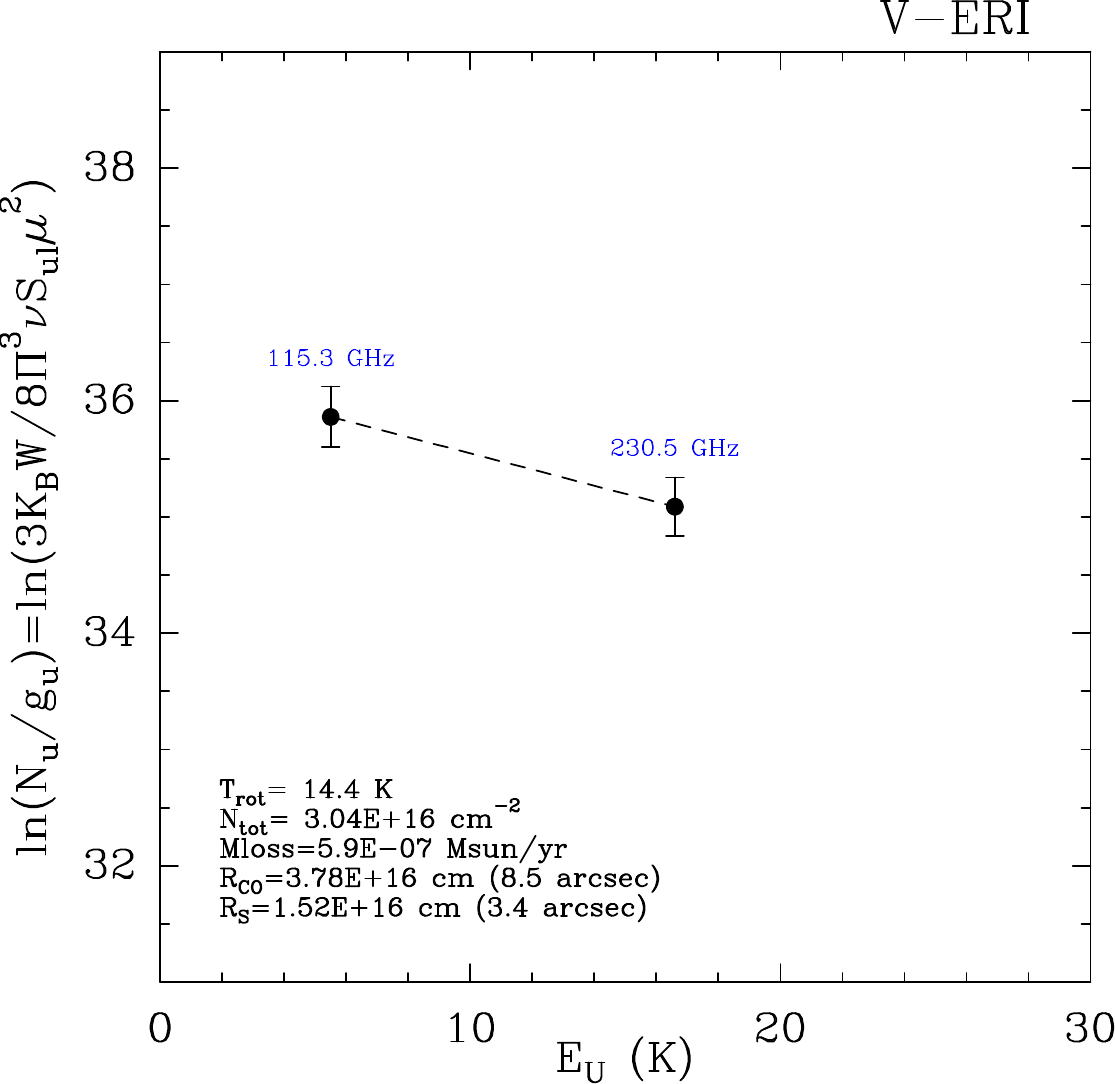} 
    \label{fig7:a} 
    \vspace{4ex}
  \end{subfigure}
  \begin{subfigure}[b]{0.5\linewidth}
    \centering
    \includegraphics[width=0.95\linewidth]{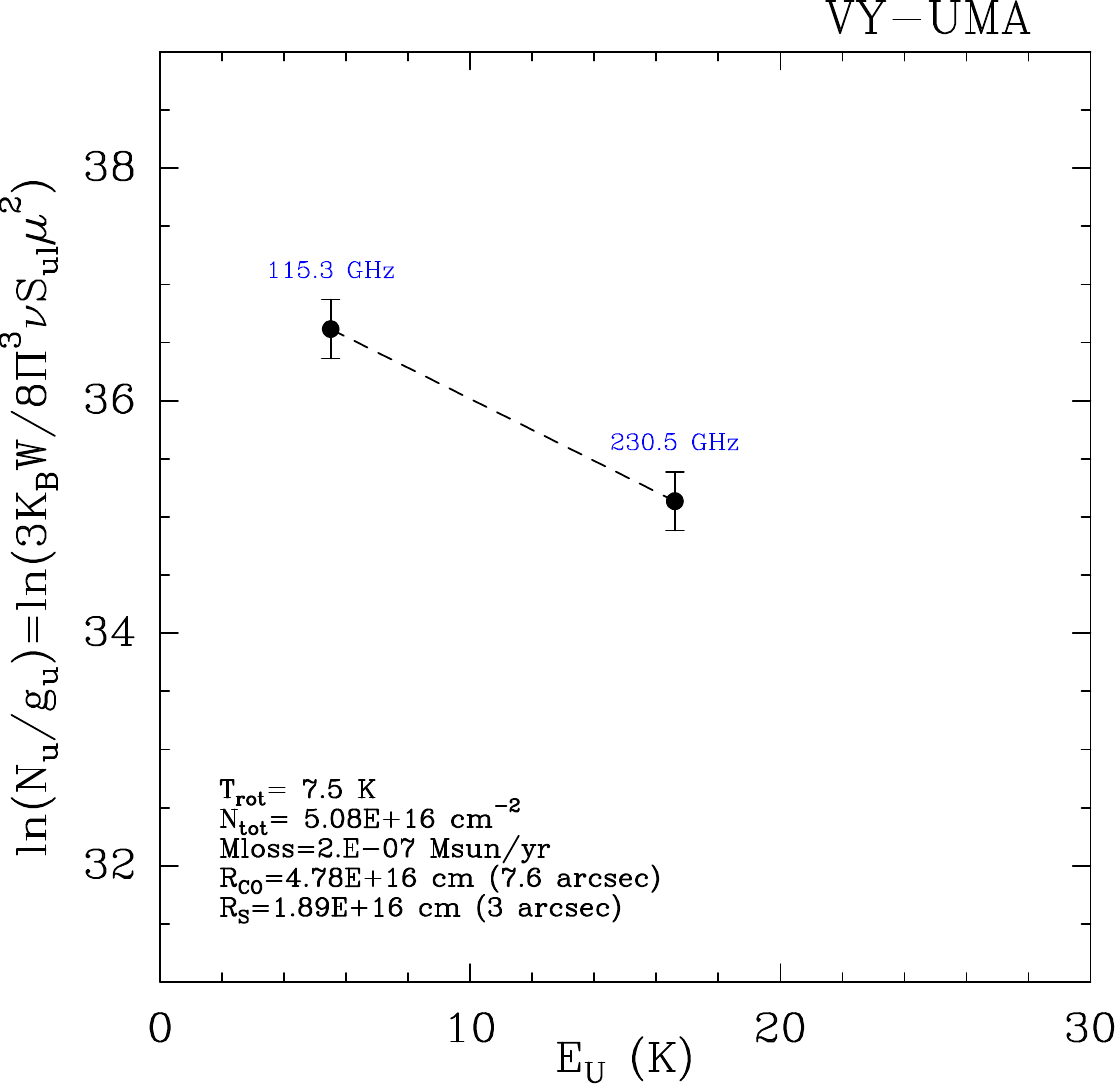} 
    \label{fig7:b} 
    \vspace{4ex}
  \end{subfigure} 
  \caption*{Fig.~\ref{fig77} continued.} 
  \label{fig7} 
\end{figure*}

\begin{figure*}[h!] \ContinuedFloat
  \begin{subfigure}[b]{0.5\linewidth}
    \centering
    \includegraphics[width=0.95\linewidth]{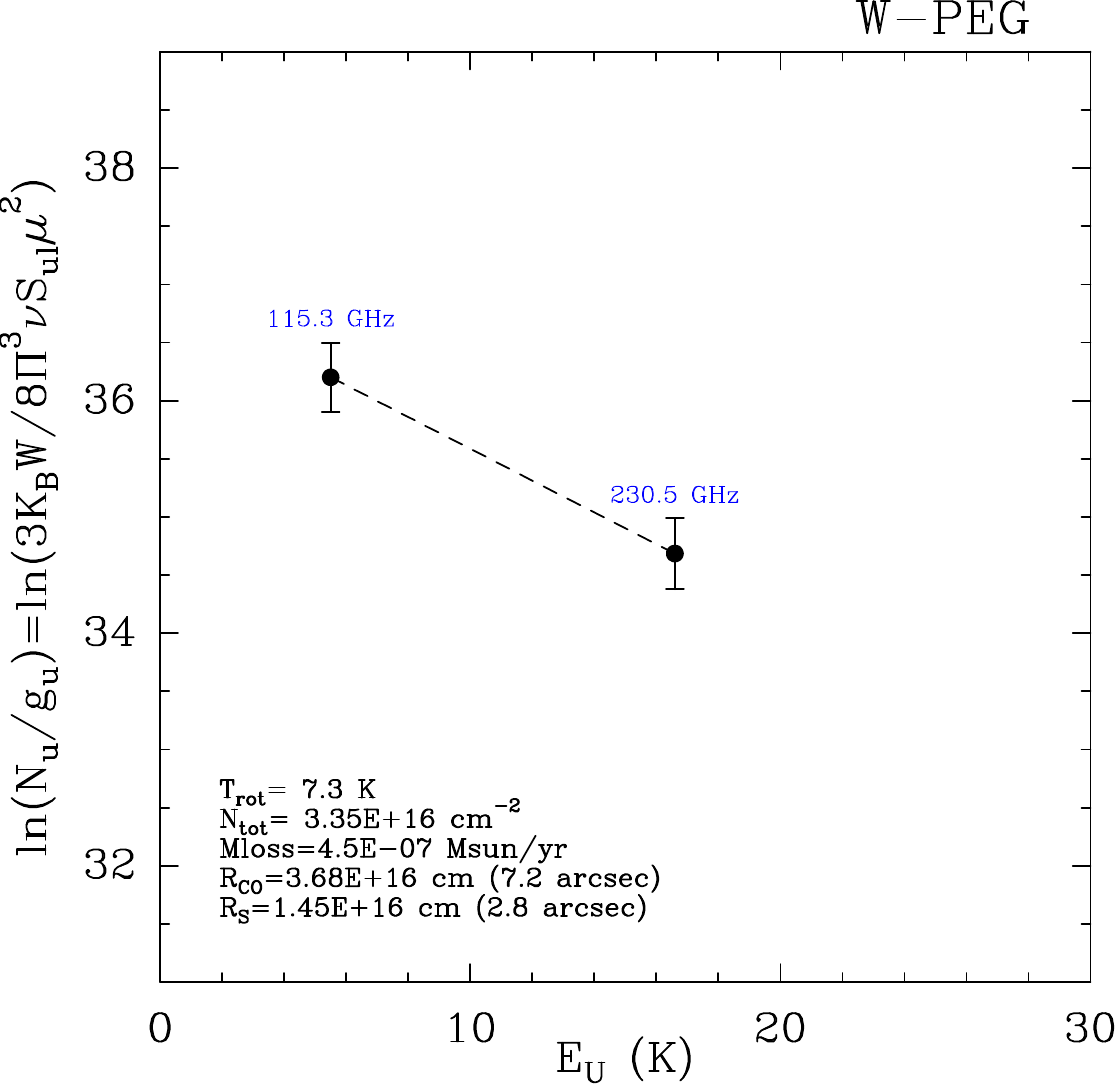} 
    \label{fig7:a} 
    \vspace{4ex}
  \end{subfigure}
  \begin{subfigure}[b]{0.5\linewidth}
    \centering
    \includegraphics[width=0.95\linewidth]{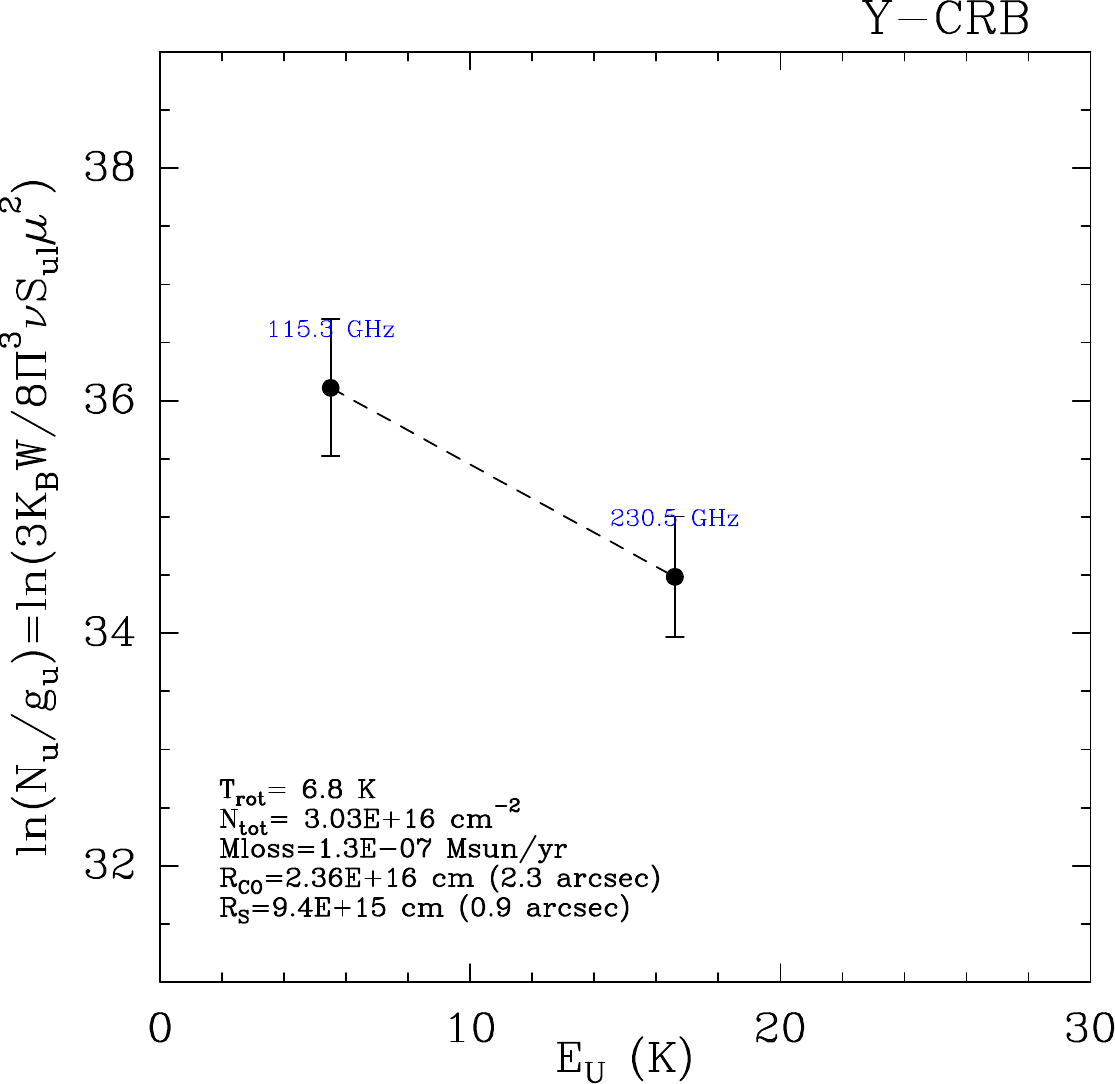} 
    \label{fig7:b} 
    \vspace{4ex}
  \end{subfigure} 
  \label{fig7} 
\end{figure*}

\begin{figure*}[h!] 
  \begin{subfigure}[b]{0.5\linewidth}
    \centering
    \includegraphics[width=0.95\linewidth]{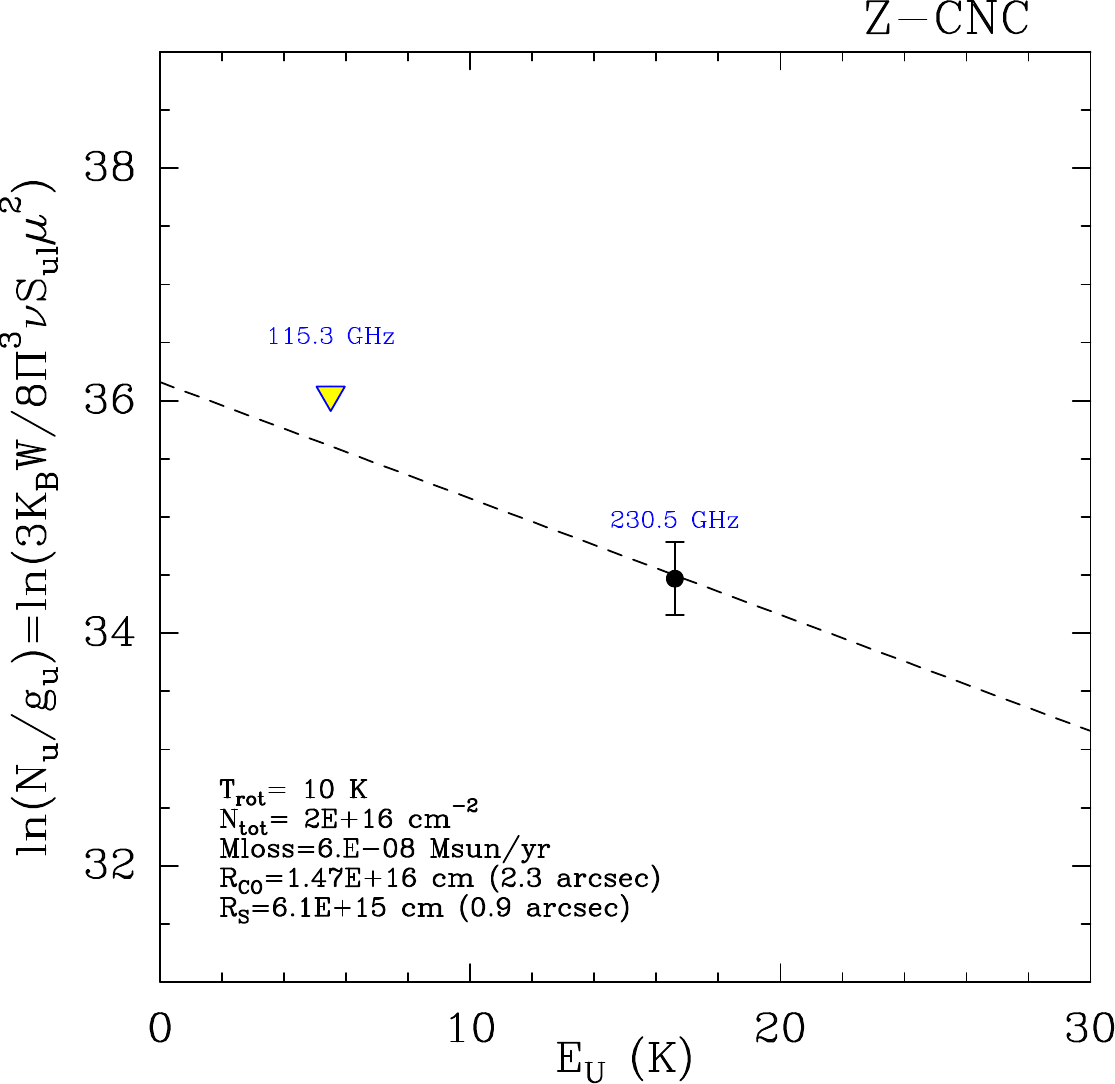} 
    \label{fig7:a} 
    \vspace{4ex}
  \end{subfigure}
  \begin{subfigure}[b]{0.5\linewidth}
    \centering
    \caption*{} 
    \label{fig7:b} 
    \vspace{4ex}
  \end{subfigure} 
  \caption*{Fig.~\ref{fig77} continued.} 
  \label{fig7} 
\end{figure*}

\end{appendix}

\end{document}